\documentclass[%
 reprint,
showpacs,preprintnumbers,
 amsmath,amssymb,
 aps,
]{revtex4-1}

\usepackage{graphicx}
\usepackage{dcolumn}
\usepackage{bm}

\begin{document}

\title{$ F_4$ , $E_6$ and $G_2$ Exceptional Gauge Groups in Vacuum Domain Structure Model}

\author{Amir Shahlaei}%
\email{amirshahlaei@gmail.com}
\author{Shahnoosh Rafibakhsh}
\email{rafibakhsh@srbiau.ac.ir}

\affiliation{
Department of Physics, Science and Research Branch, Islamic Azad University, Tehran 14665/678, Iran
}%

\date{\today}

\begin{abstract}
Using vacuum domain structure model, trivial static potentials in various representations of $F_4$, $E_6$ and $G_2$ exceptional groups are calculated by means of the unit center element. Due to the absence of the non-trivial center elements, the potential of every representation is screened at far distances. However, the linear part is observed at intermediate quark separations which is investigated by the decomposition of the exceptional group to its maximal subgroups. Comparing the group factor of the super-group with the corresponding one obtained from the non-trivial center elements of $SU(3)$ subgroup, shows that $SU(3)$ is not the direct cause of temporary confinement in any of the exceptional groups. However, the trivial potential obtained from the group decomposition to the $SU(3)$ subgroup is the same as the potential of the super-group itself. In addition, any regular or singular decomposition to the $SU(2)$ subgroup which produces the Cartan generator with the same elements as $h_1$, in any exceptional group, leads to the linear intermediate potential of the exceptional gauge groups. The other $SU(2)$ decompositions with the Cartan generator different from $h_1$, are still able to describe the linear potential if the number of $SU(2)$ non-trivial center element which emerge in the decompositions is the same. As a result,
it is the center vortices quantized in terms of non-trivial center element of the $SU(2)$ subgroup which give rise to the intermediate confinement in the static potentials.
\end{abstract}



\maketitle


\section{\label{sec:1}Introduction and Motivation}	

Quantum Chromodynamics is the theory of strong interactions. Quarks interact via gluons which are strong force carriers and attributed to the adjoint representation of the $SU(3)$ gauge group. The non-Abelian nature of gluons causes QCD to be fundamentally a non-perturbative theory in the infrared sector. To understand
the phenomena of the low energy regime, such as confinement, some topological field configurations, such as center vortices \cite{thooft,vin,cornwall,feynman,nielsen,ambjorn1,ambjorn2,ambjorn3,olesen,aharonov,mack1,mack2,mack3,mack4,tombolis}, are believed to play the key role in the non-trivial vacuum of QCD. They
assign a criterion to confinement through the area-law falloff of the Wilson loop which is one of the most efficient order parameters for investigating the large distance behavior of QCD.

In the center vortex model, confinement is the result of the interaction between center vortices and the Wilson loop. In fact, the Wilson loop running around the vortex, measures the vortex flux which is quantized in terms of the gauge group center. A center vortex which is topologically linked to a Wilson loop,
changes the Wilson loop by a group factor $z_n$:
\begin{equation}
W(C) \to (z_n)^k~W(C), 
\label{e1}
\end{equation}
where $z_n=\exp(\frac{2 \pi i n } {N}), n=1,2, \cdots , N-1$, and $k$ represents the N-ality of the representation $r$. This property implies a linear potential between static quarks which means confinement. 

The thick center vortex model developed by Faber et.~al. \cite{faber}, is aimed to study the potentials of higher representations of $SU(N)$ gauge groups. In this model, the quark anti-quark static potential behaves differently in three regions: At short distances, the interaction is determined by one-gluon exchange which leads to a Coulomb-like potential \cite{denis,rafi,ahmadi}. At intermediate distances, the string tension of the linear potential is proportional to Casimir scaling. At asymptotic region,
potentials of all representations with zero N-ality are screened. However, the potential of non-zero N-ality representations becomes parallel to the one of the lowest representation with the same N-ality \cite{faber,epjc,deldar-jhep,rafi2007,rafi2010}.

In 2007, J. Greensite et. al. \cite{greensite} claimed that there is no obvious reason to exclude the trivial center element from the model. In fact, even in $G_2$ gauge theory which only includes one trivial center element, one expects a linear potential from the breakdown of the perturbation theory to the onset of screening
while all asymptotic string tensions are zero. Monte Carlo numerical lattice calculations for $G_2$ \cite{olej,wipf1,wipf2} also confirm a confining potential, despite the absence of non-trivial center elements. In the new model, a vacuum domain is a closed tube of magnetic flux which unlike a vortex is quantized in
units corresponding to the gauge group trivial center. The string tensions are produced from random spatial variations of the color magnetic flux quantized in terms of unity. But, what accounts for the intermediate linear potential in such gauge groups? To answer this question and by using the idea of domain
structures, S. Deldar et.~al \cite{deldar1,deldar2,deldar3} showed that $SU(2)$ and $SU(3)$ subgroups of $G_2$ have an important role in the intermediate confinement of $G_2$. In fact, they were motivated by the two works done in refs. \cite{greensite} and \cite{holland}. Pepe et.~al \cite{holland} investigated that a scalar Higgs field in the fundamental representation of $G_2$ can break to $SU(3)$ representations. So, one is able to interpolate between exceptional and ordinary confinement. Moreover, Greensite et.~al \cite{greensite} used the abelian dominance idea to study $SU(3)$ and $SU(2)$ dominance in the $G_2$ gauge theory. Therefore, it seems interesting to investigate how confinement appears in a theory with exceptional gauge groups in the framework of the vacuum domain structure model.

In this paper, using the same method as in refs.~\cite{deldar1,deldar2,deldar3}, we present a general scheme to understand what kind of group decompositions lead to the temporary confinement of the exceptional gauge groups in the vacuum domain structure model. In the next section, the thick center vortex and vacuum domain structure models are discussed briefly. In Sec.~\ref{sec:3}, some properties
of exceptional groups are investigated. We apply $F_4$, $E_6$ and $G_2$ in the vacuum domain structure model and calculate the potentials in different representations in Sec.~\ref{sec:4}. The decomposition of these gauge groups to their subgroups is investigated as well.

\section{\label{sec:2}Thick Center Vortex Model and Vacuum Domain Structures}

A center vortex is a closed tube of magnetic flux which is quantized in terms of the non-trivial center elements of the gauge group. It might be considered as line-like (in three dimensions) or a surface-like (in four dimensions) object. In a pure non-Abelian gauge theory, the random fluctuations in the number of center vortices
which pierce the minimal area of the Wilson loop, give rise to the asymptotic string tension. In fact, a thin center vortex is capable of inducing the linear potential for the fundamental representation of the gauge group. Thickening the center vortices leads to a bigger piercing area and these topological objects
should be described by a profile function. Therefore, the gauge group centers in Eq.~\eqref{e1} should be replaced by a group factor:
\begin{equation} 
W(C) \to \mathcal{G}_r [\vec{\alpha}_C^n] W(C),
\label{W-thick}
\end{equation}
where the group factor is described as
\begin{equation}
 \mathcal{G}_{r} [\vec{\alpha}_C^n(x)]=\frac{1}{d_r} \, \textrm{Tr} \left[ \exp(i \,  \vec{\alpha}_C^n \cdot \vec{H}) \right],
 \label{e2}
\end{equation}
in which $ d_r$ depicts the dimension of the group representation and $ H_i$,{ $ i=1, \cdots , N-1 $}, are simultaneous diagonal generators of the group spanning the Cartan sub-algebra and $n$ represents the type of center vortices. Vortices of type $n$ and type $N-n$ are complex conjugate of one another and their
magnetic fluxes are in the opposite directions. Therefore,
\begin{equation} 
\mathcal{G}_{r} [\vec{\alpha}_C^n (x)]=\mathcal{G}_r^{\ast}[\vec{\alpha}_C^{N-n}(x)] 
\label{e3}
\end{equation}
The function $ \vec{\alpha}_C^n (x) $ is the vortex profile ansatz. It depends on the location of the vortex midpoint $x$, from the Wilson loop, the shape of the contour $C$ and the vortex type $n$. Mathematically, there are various candidates which can simulate a well-defined potential but all of them should obey the
following conditions:
\begin{enumerate}
\item{ As $ R \to 0 $ then $ \alpha \to 0 $.}

\item{When the vortex core lies entirely outside the minimal planar area enclosed by the Wilson loop, there is no interaction:
\begin{equation} 
\textrm{exp}[i \,  \vec{\alpha}_C^n \cdot \vec{H}]=\mathbb{I}
\Rightarrow  \vec{\alpha}_C^n  =0.
\label{e4}
\end{equation}
}
\item{Whenever the vortex core is completely inside the planar area of the Wilson loop,
\begin{equation}  
\exp [i \, \vec{\alpha}_C^n \cdot \vec{H}]=z_n \, \mathbb{I}
\Rightarrow \vec{\alpha}_C^n=\vec{\alpha}_{max}^n.
\label{e6}
\end{equation}
}
\end{enumerate}
Here, we have chosen the flux profile introduced in ref.~\cite{faber}:
\begin{equation} 
\vec{\alpha}_C^n (x)=\frac{\vec{\alpha}_{max}^n}{2} \left[ 1-\tanh \left( a \, y(x)+\frac{b}{R} \right) \right],
\label{e7}
\end{equation}
where $a$ and $b$ are free parameters of the model. The distance between the vortex midpoint and the nearest time-like leg of the Wilson loop is measured by $ y(x) $:
\begin{equation} 
y(x)=
\begin{cases}
x-R &\textrm{for}  \quad |R-x| \leq |x| \\
-x   & \textrm{for}  \quad |R-x|>|x| \\
\end{cases}
\label{e8}
\end{equation}
It seems changing the ansatz may have no effect on the extremum points of the group factor $ \mathcal{G}_{r} [\vec{\alpha}(x)] $ \cite{rafibakhsh}. Whereas, an alteration of $ \vec{\alpha}_C^n (x) $ is influential in the potential itself in a way that string tension ratios might be in more or less agreement with Casimir
scaling \cite{deldar-jhep,deldar3}.

Now we are able to write the Wilson loop for $SU(N)$ gauge groups:
\begin{equation} \label{e9}
\langle W(C) \rangle =\prod_{x} \left( 1-\sum_{n=1}^{N-1} f_n \left( 1-\mathcal{G}_{r}[\vec{\alpha}_C^n(x)] \right) \right) \langle W_0(C) \rangle,
\end{equation}
where, $f_n$ shows the probability that the midpoint of a center vortex is located at any plaquette in the plane of the Wilson loop. The probability of locating center vortices of any type at any two plaquettes are independent which is an over-simplification of the model. $<W_0(C)>$ is the Wilson loop expectation
value when no vortices is linked with the loop. It should be noted that in addition to the regions associated with the non-trivial center elements, the domains corresponding to unity center elements are also allowed in the vacuum. Therefore, the sum in Eq.~\eqref{e9}, should contain $n=0$ as well. Using the fact that
$f_n=f_{N-n}$, the static potential between a color and an anti-color sources induced by thick center vortices and vacuum domains is
\begin{small}
\begin{equation} 
V(R)=- \sum_{m=-\infty}^{m=+\infty} \ln \left\{  1- \sum_{n=0}^{N-1}  f_n \big( 1- \textrm{Re} \, \mathcal{G}_r [ \vec{\alpha}_C^n (x_m) ] \big) \right\}
\label{e10}
\end{equation}
\end{small}
where $n=0$ denotes a vacuum domain type vortex and $n=1, \cdots , N-1 $ indicates the type of center vortices.

\section{\label{sec:3}Some Properties of Exceptional groups}
The idea of symmetries and Lie exceptional groups has been always attractive in modern high energy physics. $G_2$ is the simplest exceptional gauge group which confirms the chance of having confinement without the center \cite{holland}. $G_2$ gauge theory is a theoretical laboratory in which $SU(N)$ subgroups are embedded. This
provides us with an understanding not only about the exceptional $G_2$ confinement but also about the $SU(3)$ confinement which happens in nature. In this section, we briefly explain some properties of the exceptional groups applied in this article including their subgroups and Dynkin diagrams.

\begin{figure}
\centering
\includegraphics[width=\linewidth]{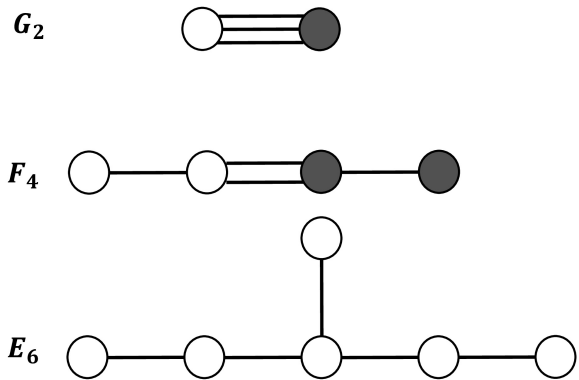}
\caption{Dynkin diagrams of  $ G_2$, $ F_4 $ and $ E_6 $ exceptional groups.}
\label{fig:dynkin}       
\end{figure}

In general, there are five distinguishable exceptional groups named $G_2$, $F_4$, $E_6$, $E_7$ and $ E_8 $. The subscripts point out the rank of the groups. The number of simple roots and simultaneous diagonal generators of simple Lie groups are equivalent to their rank. One may draw the whole root diagram by having
simple roots and the angles between them in a simple Lie group. The angle between simple roots in a Dynkin diagram is always obtuse. Three, two, one or no line between simple roots measures their mid angles which are $ 150^{\circ} $, $ 135^{\circ} $, $ 120^{\circ} $ or $90^{\circ} $, respectively \cite{das}. Fig.~\ref{fig:dynkin}
depicts Dynkin diagrams of the exceptional groups used in this research.
Filled circles represent shorter roots and empty ones show longer roots in terms of their length. 

Using Dynkin diagrams, one is able to find the subgroups of every lie group. There are three different sorts of maximal subgroups \cite{feger}:
\begin{itemize}
\item{Regular maximal non-semisimple subgroups},
\item{Regular maximal semisimple subgroups},
\item{Singular (special) maximal subgroups}.
\end{itemize}
The sum of the rank of the regular subgroups equals to the rank of their super-group. However, this is not true for the singular subgroups. It should be noted that if a factor $ U(1) $ appears in a subgroup, it makes the subgroup as a non-semisimple one. 

\begin{figure}
\centering
\includegraphics[width=\linewidth]{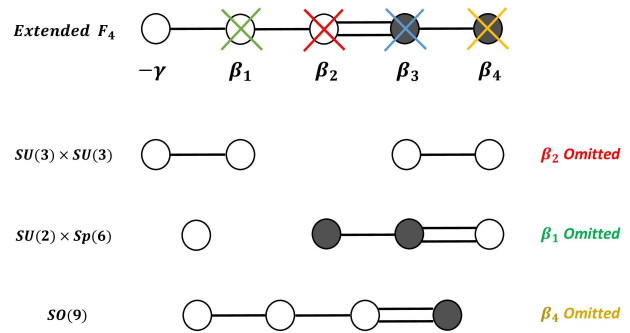}
\caption{Three different regular maximal subgroups of $F_4$ obtained from its extended Dynkin diagram by omitting the original simple roots one by one.}
\label{fig:dynkin-f4}       
\end{figure}

In this article, we briefly discuss how to derive the subgroups of $F_4$ and use the same method for other exceptional groups. The extended Dynkin diagram is structured by adding the most negative root ($-\gamma$) to the set of simple roots (Fig.~\ref{fig:dynkin-f4}). Then, by omitting the original $\beta_i$ roots,
regular subgroups will emerge one by one. For example, in Fig.~\ref{fig:dynkin-f4}, eliminating the root $ \beta_2 $ leads to the $ SU(3) \times SU(3) $ subgroup of $ F_4 $.
Moreover, when the root $\beta_4$ is omitted, $SO(9)$ subgroup of $F_4$ is obtained. It should be pointed out that omitting the root $\beta_3$ gives off the $ SU(2) \times SU(4) $ subgroup. In some references, it has been claimed as a direct subgroup of the $ F_4 $ \cite{van} and in some others it is not
\cite{sorba,slansky}. However, it is a direct subgroup of $ SO(9) $. Therefore, it might be, at least, considered as an indirect subgroup of the $F_4$. To achieve singular maximal subgroups of exceptional groups, a determined method does not exist and each subgroup has to be extracted individually \cite{feger}. All
maximal subgroups of the $ F_4 $ exceptional group have been presented in Tab.~\ref{tab2}.
\begin{table*}
\begin{ruledtabular}
\caption{Maximal subgroups of some exceptional groups \cite{slansky}. [R] and [S] represent regular and singular subgroups of each group, respectively.}
\begin{tabular}{ c c c }
 $\mathbf{E_6}$ & $\mathbf{F_4}$ & $\mathbf{G_2}$\\
\colrule
$ SU(3) \times SU(3) \times SU(3) $  [R] & $ SU(3) \times SU(3)$  [R] &  $SU(3)$ [R] \\

$ SU(2) \times SU(6) $  [R]  &   $ SU(2) \times Sp(6) $  [R] &  $ SU(2) \times SU(2)$ [R]\\

$ SO(10) \times U(1) $  [R]  & $ SO(9)$  [R]  &  $ SU(2) $ [S] \\

$ SU(3) \times G_2 $  [S] &  $ SU(2) \times G_2 $  [S] &  \\

$ SU(3) $  [S]  &  $ SU(2) $  [S]  & \\

$ Sp(8) $  [S] &    &\\

$ G_2 $  [S] &   & \\

$ F_4 $  [S]  &   & \\
\end{tabular}
\label{tab2}
\end{ruledtabular}
\end{table*}

‌Based on the branching rules, an irreducible representation of a group can be decomposed to the irreps of its subgroup as the following \cite{sorba}:
\begin{equation}  
R(G)=\bigoplus_{i} m_i \, R_i(g),
\label{e11}
\end{equation}
where $ R(G) $ is an irrep of the super-group $G$ and $ R_i(g) $ is the irrep of the subgroup $g$.  $ m_i $ is the degeneracy of the representation $ R_i(g) $ in the decomposition of representation $ R(G) $ \cite{slansky,sorba}. To be more precise, we consider one of the regular subgroups of $F_4$: 
\begin{equation} 
F_4 \supset SU(3) \times SU(3).\nonumber
\end{equation}
Using the the branching rules, one might write \cite{sorba,slansky,gellmann}:
\begin{equation}
26 = (8 , 1) \oplus (3 , 3) \oplus (\bar{3} , \bar{3}).
\label{e12}
\end{equation}
From Eq.~(\ref{e11}), it is obvious that the first numbers in each parenthesis could be considered as the degeneracy of the second representation emerging in the decomposition:
\begin{equation}
26 = 8(1)+3(3)+3(\bar{3}).
\label{e13}
\end{equation}
Therefore, an $F_4$ ``quark" is made of of three $SU(3)$ quarks, three anti-quarks and one singlet.

\section{\label{sec:4}Confinement Without a Center}
\subsection{\label{subsection-f4}$F_4$ exceptional group}
$ F_4 $ exceptional group has rank four and contains four Cartan generators. The diagonal generators for the fundamental 26-dimensional representation of the $F_4$ are \cite{bincer,patera}:
\begin{equation} 
\begin{aligned}
h_1 =N_1 \ (& D_5^5+D_6^6-D_7^7+D_8^8-D_9^9-D_{10}^{10}), \\
h_2 =N_2 \ (& D_3^3+D_4^4-D_5^5-D_6^6+D_{10}^{10}-D_{11}^{11}), \\
h_3 = \frac{N_3}{2} \ \big(& D_2^2-2D_3^3-D_4^4+D_6^6-D_8^8+D_9^9-D_{10}^{10}+ \\
&D_{11}^{11}-D_{12}^{12} \big), \\
h_4 = \frac{N_4}{2} \ \big(& - 2 D_2^2+D_3^3-D_4^4+D_5^5-D_6^6+D_7^7-D_9^9+ \\
  & D_{12}^{12}-D_{13}^{13} \big),
\end{aligned}
\label{e14}
\end{equation}
where 
\begin{equation}
D_a^b=I_{ab}-I_{\overline{ba}},
\label{e15}
\end{equation}
and $I_{ab}$ are $26 \times 26$ matrices with the following matrix elements
\begin{equation}
(I_{ab})_{jk}= \delta_{aj} \, \delta_{bk}.
 \label{e16}
\end{equation}
Subscripts $j$ and $k$ take on the same values as $a$ and $b$ such that $ a , b : -13 \leq j , k \leq 13 $ with zero excluded.

Using the standard normalization condition
\begin{equation} \label{e17}
\textrm{Tr}[h_a , h_b]= \frac{1}{2} \, \delta_{ab},
\end{equation}
the normalization factors are calculated as follows:
\begin{eqnarray} 
N_1 &=&N_2=\frac{1}{2 \sqrt{6}}, \nonumber\\
N_3 &=&N_4=\frac{1}{2 \sqrt{3}}.
\label{e18}
\end{eqnarray}

\begin{figure}
\centering
\includegraphics[width=\linewidth]{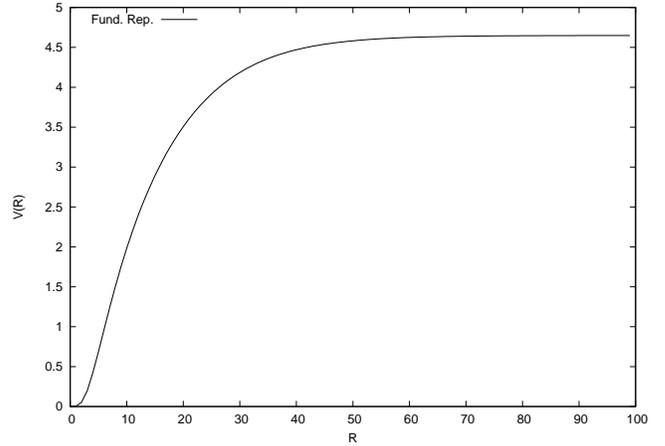}
\caption{The Potential between two static sources in the fundamental representation of the $ F_4 $ for $R \in [1,100]$ . Screening is clearly observed at large quark separations while the potential is linear at intermediate distances. The free parameters of the model have been chosen to be $a=0.05$, $b=4$ and $f=0.1$.}
\label{fig:26f4}       
\end{figure}

To find the maximum amount of the domain structure flux, we use Eq.~\eqref{e6}, using the fact that the $F_4$ gauge group includes only one trivial center element:
\begin{equation} \label{e19}
\textrm{exp}[i \, \vec{\alpha} \cdot \vec{H}]=\mathbb{I},
\end{equation}
and we find:
\begin{equation} \label{e20}
\begin{aligned}
\alpha_1^{max}=2 \pi \sqrt{24}, \\
\alpha_2^{max}= 6 \pi \sqrt{24}, \\
\alpha_3^{max}=4 \pi \sqrt{48}, \\
\alpha_4^{max}=2 \pi \sqrt{48}.
\end{aligned}
\end{equation}
Now, one can calculate the static potential of Eq.~\eqref{e10} for the fundamental representation of the $F_4$ exceptional gauge group. This potential has been pictured in Fig.~\ref{fig:26f4} for $R \in [1,100]$ . The free parameters of the model are chosen to be $a=0.05$, $b=4$ and $f=0.1$ in every calculation of this article.

In Fig.~\ref{fig:26f4}, the linear potential is demonstrably located in the approximate range of $ R \in [2 , 9] $. In addition, at large distances where the vacuum domain is entirely located inside of the Wilson loop, a flat potential is induced. Hence, one can deduce that in groups without a non-trivial center,
static potentials of all representations behave like a $ SU(N) $ representation with zero N-ality.

The adjoint representation of the $F_4$ is $52$ dimensional. As a consequence, like any gauge group, the ``bosonic" gluons of the $ F_4 $ exceptional group are made of the adjoint representation. Thus, mathematically one can derive the way of screening of color sources in any representation from tensor products of that
representation with the adjoint one i.e. when a singlet emerges, it means screening. So, for the fundamental representation, one may write:
\begin{equation} \label{e21}
26 \times 52=26 \oplus 273 \oplus 1053.
\end{equation}
Therefore, the fundamental representation of $F_4$ can not be screened just by one set of gluons. Energetically speaking, color sources in the fundamental representation of the $F_4$ are not screened until the potential reaches that extent where four sets of gluons pop out of the vacuum:
\begin{equation} \label{e22}
26 \times \overbrace{52 \times \cdots \times 52}^{4 \, \textrm{times}}=\mathbf{1} \oplus 46 (26) \oplus 10 (52) \oplus \cdots .
\end{equation}
These tensor products have been calculated by LieART project on the Mathematica \footnote{http://www.hepforge.org/downloads/lieart/}. The numbers out of the parentheses are the degeneracy of the representations being repeated in the tensor product. Therefore, four $F_4$ ``gluons" are able to screen an $F_4$ ``quark" to create a color singlet hybrid $qGGGG$. Moreover, two ``quarks" form a singlet:
\begin{equation} \label{e23new}
26 \times 26=1 \oplus 26 \oplus 52 \oplus 273 \oplus 324.
\end{equation}
As in $SU(N)$ gauge groups, three $F_4$ ``quarks" can create a baryon:
\begin{equation} \label{e24new}
\begin{aligned}
26 \times 26 \times 26=1 \oplus 5(26) \oplus 2(52) \oplus 4(273) \oplus 3(324) \oplus \\3(1053) \oplus 1274 \oplus 2652 \oplus 2(4096).
\end{aligned}
\end{equation}

Evidently, the function $ \textrm{Re} \, \mathcal{G}_r [\vec{\alpha}_C^n(x)] $ looks predominant in the potential formula in Eq.~\eqref{e10}. It shows that the group factor varies between $1$ and $\exp (\frac{2 \pi i n k}{ N} ) $ corresponding to the N-ality=$k$ of the representation and the vortex type $n$. An unaffected
Wilson loop which has not been pierced by any vortex means $ \textrm{Re} \, \mathcal{G}_r [\vec{\alpha}_C^n(x)]=1 $. When the vortex is linked to the Wilson loop, the group factor deviates from $1$. Hence, to investigate what happens to the $ F_4 $ potentials, one may study the behavior of the group factor.

In reference \cite{rafibakhsh}, it has been proven that the third Cartan generator of the $ SU(4) $ gauge group i.e. $H_3=\frac{1}{2\sqrt{6}} \textrm{diag} [1 , 1 , 1 , -3] $ can produce the total potential individually. In $ F_4 $ exceptional group, one might use only $ h_1 $ or $ h_2 $ Cartan generators or both of
them together to find the same group factor and also the same potential as if we apply all four Cartan generators in our calculations. This property will be helpful in the decomposition of the $F_4$ representations into its subgroups. In fact, when the identical diagonal generators are constructed, the same potentials
as the $F_4$ itself will be achieved.

\begin{figure}
\centering
\includegraphics[width=\linewidth]{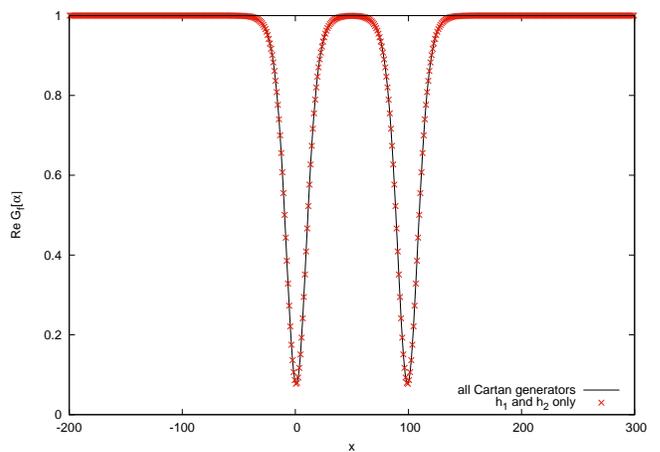}
\caption{The real part of the group factor versus $x$ -the location of the vacuum domain midpoint- for the fundamental representation of the $F_4$ exceptional gauge group in the range $x \in [-200,300]$ by applying $h_1$ and $h_2$ Cartan generators (stars) and all diagonal generators
(solid line). It is clear that two sets of data are the same. The distance $R$ between color and anti-color sources equals to $100$.}
\label{fig-2}       
\end{figure}

\begin{figure}
\centering
\includegraphics[width=\linewidth]{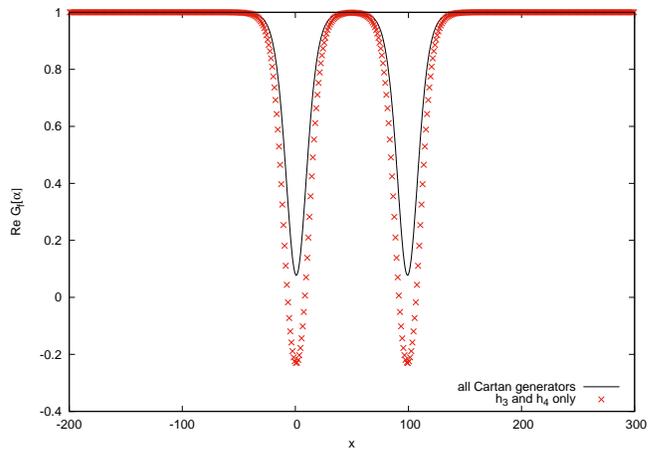}
\caption{The same as Fig.~\ref{fig-2} but in comparison with the group factor when only $ h_3 $ or $ h_4 $ is used.}
\label{fig:gr-h34}       
\end{figure}

In  Fig.~\ref{fig-2}, the real part of the group factor versus the location $x$ of the vacuum domain midpoint has been plotted, for $R=100$ and in the range $ x \in [-200,300] $, by considering all generators and also by utilizing only $ h_1 $ and $ h_2 $. 
It is clear that both diagrams in Fig.~\ref{fig-2} are identical and the group factor reaches the minimum amount of $\approx 0.076 $ at $x=0$ and $x=100$.  To confirm our conclusion, we have plotted the similar diagram in Fig.~\ref{fig:gr-h34} but by using only $ h_3 $ or $ h_4 $ diagonal generators.
In this Figure, the group factor reaches the minimum amount equal to $ \approx  -0.23 $ that is way less than the minimum amount of the $F_4$ group factor. 

It has been shown that \cite{rafibakhsh} the group factor reaches the minimum points where $50\%$ of the vortex maximum flux enters the Wilson loop. These points are responsible for the intermediate linear potential. We aim to show that the $SU(N)$ subgroups of the exceptional groups might be the reason for the
appearance of the linear potential at intermediate distances. It means that the minimum points of the group factor could be explained by the group decomposition to the subgroups.

\hfill\\
\bm{$SU(3) \times SU(3)$} \textbf{decomposition}\\
Using the decomposition in Eq.~\eqref{e13}, we are able to reconstruct Cartan generators of the $ F_4$ with respect to its $ SU(3)$ subgroup:
\begin{equation}
\begin{aligned}
H_a^{26}=\frac{1}{\sqrt6} \textrm{diag} \Big[ \overbrace{0 , \cdots , 0}^{8 \, \textrm{times}} , \lambda_{a}^3 , \lambda_{a}^3 , \lambda_{a}^3 , -(\lambda_{a}^3)^{\ast} ,  -(\lambda_{a}^3)^{\ast} , \\ -(\lambda_{a}^3)^{\ast} \Big],
\end{aligned}
\label{e23}
\end{equation}
where $ \lambda_{a}^3 , a=3,8 $ are Cartan generators of the $ SU(3) $ in the fundamental representation:
\begin{equation} 
\begin{aligned}
\lambda_3^3 &=\frac{1}{2} \, \textrm{diag} [ 1 , -1 , 0],  \\
\lambda_8^3 &=\frac{1}{2\sqrt{3}} \, \textrm{diag} [ 1 , 1 , -2].
\end{aligned}
 \label{e24}
\end{equation}
Meanwhile, the matrices of Eq.~\eqref{e23} are normalized using the normalization conditions in Eq.~\eqref{e17}. If the matrix $H_3^{26}$ in Eq.~\eqref{e23} is considered, its components are identical with the ones for the Cartan generators $ h_1 $ and $ h_2 $ of Eq.~\eqref{e14}. But this is not the case with $H_8^{26}$. To examine the results coming out of these two matrices, one is supposed to establish the same normalization condition as in Eq.~\eqref{e19}:
\begin{equation}
\exp(i \alpha_{{max}_1}^{26} H_3^{26}+i\alpha_{{max}_2}^{26} H_8^{26})=\mathbb{I}.
\label{e25}
\end{equation}
In this case, we have six distinctive equations and find:
\begin{equation} 
\begin{aligned}
\alpha_{{max}_1}^{26} &=2 \pi \sqrt{6}, \\
\alpha_{{max}_2}^{26} &=6 \pi \sqrt{2}.
\end{aligned}
\label{e26}
\end{equation}
Now, the potential of Eq.~\eqref{e10} could be calculated using Eqs.~\eqref{e23} and \eqref{e26}. This potential is identical with Fig.~\ref{fig:26f4}, despite the difference between $H_8^{26}$ and $h_1$ or $h_2$. To investigate this matter, one might manually estimate the value of $ \textrm{Re} \, \mathcal{G}_r [\alpha] $ when the vacuum domain is completely inside the Wilson loop:
\begin{equation} 
\begin{aligned}
 \textrm{Re} \, \mathcal{G}_r^{1} &\left[ {\alpha} \right]_{SU_3 \times SU_3} =\\&\frac{1}{26}  \times 
\textrm{Re}\left( \textrm{Tr} \left[ \textrm{exp} \big( i \, \alpha_{{max}_1}^{26} \cdot  H_3^{26} \big) \right] \right) 
\approx  0.076 
\end{aligned}
\label{e27}
\end{equation}
\begin{equation} \label{e28}
\begin{aligned}
 \textrm{Re} \, \mathcal{G}_r^{2} &\left[ {\alpha} \right]_{SU_3 \times SU_3} =\\&\frac{1}{26}  \times 
 \textrm{Re} \left( \textrm{Tr} \left[ \textrm{exp} \big( i \, \alpha_{{max}_2}^{26} \cdot  H_8^{26} \big) \right] \right)
\approx  0.076
\end{aligned}
\end{equation}
It is clear that both group factor functions earned by either $ \alpha_{{max}_1}^{26} $ or $ \alpha_{{max}_2}^{26} $ reach the same amount which is the minimum amount of the $F_4$ group factor in Fig.~\ref{fig-2} as well. Consequently, based on the analogy of the group factor functions acquired by both $H_3^{26}$ and $H_8^{26}$, we claim that, although the second matrix of $ SU(3) \times SU(3) $ has different components but it has the same effect as the Cartan $ h_1 $ on the $ F_4 $ group. Then, the trivial static potential of the $ F_4 $ exceptional group is similar to the potential gained by its $ SU(3) \times SU(3) $ subgroup. Therefore, it seems that this decomposition could be generalized for higher representations to find the corresponding potential. 

The decomposition of the 52-dimensional adjoint representation of the $ F_4$  is \cite{slansky,sorba,gellmann}:
\begin{equation} 
\begin{aligned}
52=& (8 , 1) \oplus (1 , 8) \oplus (\bar{6} , 3) \oplus (6 , \bar{3}), \\
52=& 8 (1) + 1 (8) + 6 (3) + 6 (\bar{3}).
\end{aligned}
\label{e29}
\end{equation}
This shows that $F_4$ ``gluons" are made of the usual $SU(3)$ gluons (representation $8$) and some additional ``gluons" consist of $SU(3)$ quarks (representation $3$) and anti-quarks (representation $\bar 3$) and also eight singlets. It is clear that these representations have different trialities. 

Using Eq.~\eqref{e29}, the Cartan generators of the $ F_4 $ in the adjoint representation might be reconstructed as the following
\begin{equation} 
\begin{aligned}
 H_a^{52}= \frac{1}{\sqrt{18}} \textrm{diag} \Big[ \overbrace{0 , \cdots , 0}^{8 \, \textrm{times}}  , \lambda_a^8 , \overbrace{ \lambda_a^3 , \cdots  \lambda_a^3}^{6 \, \textrm{times}},  \\
\overbrace{ -(\lambda_a^3)^{\ast} , \cdots,  -(\lambda_a^3)^{\ast}}^{6 \, \textrm{times}} \Big],
\end{aligned}
\label{e30}
\end{equation}
where $ \lambda_{a}^3 $, $a=3,8$, are the same generators as in Eq.~\eqref{e24} and $ \lambda_{a}^8 $ are simultaneous diagonal generators of the $ SU(3) $ gauge group in the adjoint 8-dimensional representation. Using Eq.~\eqref{e19}, the maximum values of the vortex flux for the adjoint representation of the $ F_4
\supset SU(3) \times SU(3) $ decomposition are
\begin{equation} \label{e31}
\begin{aligned}
\alpha_{{max}_1}^{52} &=6 \pi \sqrt{2}, \\
\alpha_{{max}_2}^{52} &=6 \pi \sqrt{6}.
\end{aligned}
\end{equation}

\begin{figure}
\centering
\includegraphics[width=\linewidth]{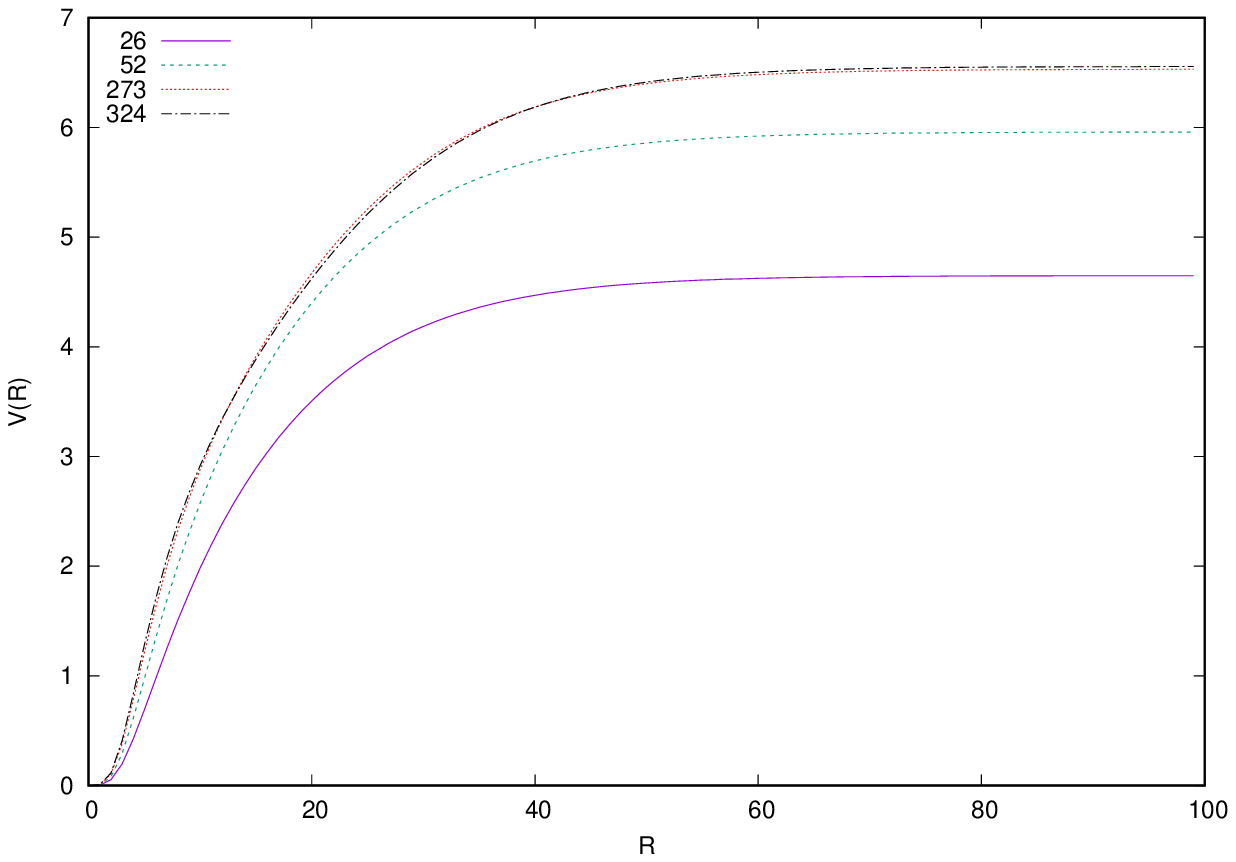}
\includegraphics[width=\linewidth]{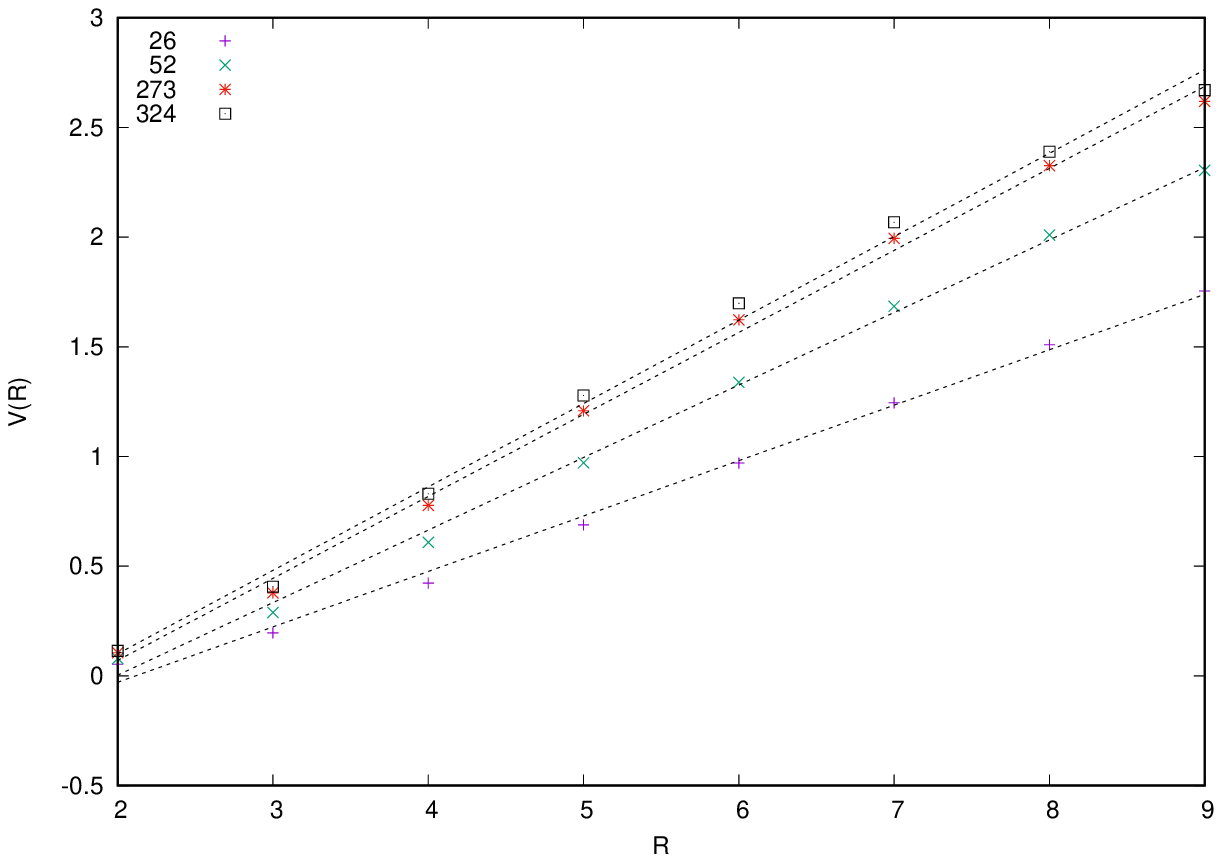}
\caption{Upper digram: The potential between static color sources in the fundamental, adjoint, 273 and 324-dimensional representations of the $ F_4 $ exceptional gauge group in the range $R \in [1,100]$. All potentials are screened at far distances, while linearity is evident at intermediate parts. Lower diagram: The same as the upper diagram but in the range $R \in [2,9]$. The slope of the potentials have been given in the forth row of Tab.~\ref{tab4}. Potentials are in agreement with Casimir scaling qualitatively. }
\label{fig-8}       
\end{figure}

The potential between static sources in the fundamental and adjoint representations of the $ F_4 $ exceptional gauge group has been plotted in Fig.~\ref{fig-8} along with the higher representations in the range $R \in [1,100]$. The decomposition of the higher representations and the corresponding Cartan generators have been presented in the
Appendix~\ref{appa}.
\begin{table}
\caption{Casimir numbers and Casimir ratios of different representations of the $F_4$ exceptional group has been presented in the second and third columns, respectively \cite{feifer} \footnote{It should be noted that Casimir scaling of the representation 273 has not been reported in this reference.}. The slope of the linear potentials of Fig.~\ref{fig-8} and the potential ratios have been given in the forth and fifth columns, respectively. The numbers in the parentheses show the fit error.}
\begin{center}
{ \setlength{\extrarowheight}{3pt}
\begin{tabular}{ |c|c|c|c|c| }
\hline
 \footnotesize{\textbf{Rep.}} &  \footnotesize{\textbf{Casimir number}}  &  $ \mathbf{\frac{C_r}{C_F}} $ & \footnotesize{\textbf{Potential slope}} & $\mathbf{\frac{k_r}{k_F}} $\\
\hline
26  &  $ \frac{2}{3}  $  &   1  & 0.252(7) & 1\\ 
\hline
52  &  $ 1 $  &   1.5  & 0.331(7) & 1.31(1)\\
\hline
273 & $2$  & 2 & 0.374(8) & 1.48(1)\\
\hline
324 & $ \frac{13}{9} $ & 2.16 & 0.38(1) & 1.5(1)\\
\hline
\end{tabular}
}
\end{center}
\label{tab4}
\end{table} 
In Fig.~\ref{fig-8}, screening is observed for the potentials of every representation at far distances. Since $F_4$ does not own any non-trivial center element, all representations act like $SU(N)$ representations with zero N-ality. Hence, screening was anticipated. Another reason for this phenomenon is the creation of gluons in the QCD vacuum which are able to screen the initial static color charges and produce a flat potential at high levels of energy:
\begin{equation} \label{e32}
\begin{aligned}
&52 \times 52 = \mathbf{1} \oplus 52 \oplus 324 \oplus \cdots ,\\
&273 \times 52 \times 52 \times 52 = \mathbf{1} \oplus 15 (26) \oplus \cdots ,\\
&324  \times 52 \times 52 = \mathbf{1}  \oplus 26 \oplus 3 (52) \oplus \cdots .\\
\end{aligned}
\end{equation}
Furthermore, in Fig.~\ref{fig-8}, there are linear parts at intermediate
distance scales for all representations which are situated at the interval $ R \in [2 , 9] $, approximately. The linear potentials have been depicted in the lower diagram of Fig.~\ref{fig-8}. The slope of the linear potentials of different representations have been given in the forth column of Tab.~\ref{tab4}, as well as the potential ratios ($\frac{k_r}{k_F}$) in the last column. It is observed that potential ratios are qualitatively in agreement with Casimir scaling ($\frac{C_r}{C_F}$) which has been presented in the third column of Tab.~\ref{tab4}. However, Casimir scaling has not been proved numerically for $F_4$. 

\begin{figure}
\centering
\includegraphics[width=\linewidth]{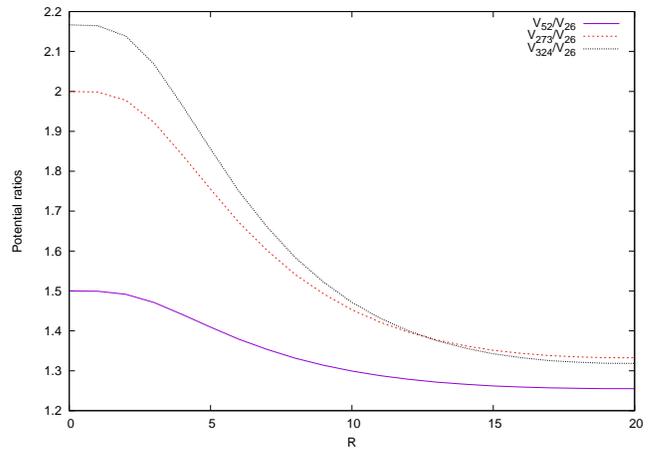}
\caption{Potential ratios of the $ F_4 $ representations to the fundamental one in the range $R \in [0,20]$. These ratios start up at the values of the corresponding Casimir ratios presented in Tab.~\ref{tab4}. }
\label{fig-9}       
\end{figure}

Fig.~\ref{fig-9} presents the point by point ratio of the potential of each representation to the fundamental one in the range $R \in [1,20]$ . These ratios start up at the ratios of the corresponding Casimirs. However, they abruptly decline at intermediate intervals. The inclination becomes more pronounced as the
dimension of the representations grows, e.g. the deviation from the exact Casimir scaling is more significant for representations 273 and 324. 

To investigate whether the linear potentials of the $ F_4$ exceptional group are caused by the non-trivial center elements of the $ SU(3) \times SU(3) $ subgroup or not, one may plot the group factor function $ \textrm{Re} \, \mathcal{G}_r [\vec \alpha ] $ with respect to the non-trivial center elements of $ SU(3) $. Using
the same method as in refs.~\cite{deldar1,deldar2,deldar3} and Eqs.~\eqref{e13} and \eqref{e29}, we are able to compose matrices containing center elements of the $ SU(3) $ depending on the N-ality of each representation. Thus,
\begin{equation} 
\begin{aligned}
\mathbb{Z}_{SU(3)}^{26}= \textrm{diag} \Big[& 1 , 1 , 1 , 1 , 1 , 1 ,1 , 1 ,  z \, \mathbb{I}_{3 \times 3} , 
  z \,  \mathbb{I}_{3 \times 3} ,  z \, \mathbb{I}_{3 \times 3} ,\\ &z^{\ast} \,  \mathbb{I}_{3 \times 3} , z^{\ast} \,  \mathbb{I}_{3 \times 3} , z^{\ast} \, \mathbb{I}_{3 \times 3} \Big],\\
\mathbb{Z}_{SU(3)}^{52}= \textrm{diag} \Big[& 1 , 1 , 1 , 1 , 1 , 1 ,1 , 1 , \mathbb{I}_{8 \times 8} , 
  z \, \mathbb{I}_{3 \times 3} ,  z \,  \mathbb{I}_{3 \times 3} , \\  
&z \, \mathbb{I}_{3 \times 3} , z \, \mathbb{I}_{3 \times 3} ,  z \, \mathbb{I}_{3 \times 3} , 
  z \, \mathbb{I}_{3 \times 3} , z^{\ast} \, \mathbb{I}_{3 \times 3} ,\\& z^{\ast} \,  \mathbb{I}_{3 \times 3} ,   z^{\ast} \, \mathbb{I}_{3 \times 3} ,  z^{\ast} \,  \mathbb{I}_{3 \times 3} , z^{\ast} \,  \mathbb{I}_{3 \times 3} , z^{\ast} \, \mathbb{I}_{3 \times 3} \Big],
\end{aligned}
\label{e33}
\end{equation}
where $z=\exp(\pm \frac{2\pi i}{3})$  is the $SU(3)$ non-trivial center element. We previously mentioned that $z$ and $z^{\ast}$ vortices carry the same magnetic fluxes but in the opposite directions. Interestingly, the number of $z$ and $z^{\ast}$ vortices which appear in the above decompositions of Eq.~\eqref{e33} is the same. Therefore, one might conclude that $F_4$ vacuum domain consists of the $SU(3)$ center vortices. For our purpose, we use the normalization condition as the following:
\begin{equation} \label{e34}
\exp[i \, \vec{\alpha} \cdot \vec{H}^{26  \, \textrm{or} \,  52}]=\mathbb{Z}_{SU(3)}^{26  \, \textrm{or} \,  52} \, \mathbb{I},
\end{equation}
where $H^{26}$ and $H^{52}$ are the generators depicted in Eqs.~\eqref{e23} and \eqref{e30}, respectively. Solving the corresponding equations results in:
\begin{equation}  
\begin{aligned}
& \alpha_{max_1}^{26-\textrm{non}}= 2 \pi \sqrt{6}, \\
& \alpha_{max_2}^{26-\textrm{non}}= 2 \pi \sqrt{2}, \\
\end{aligned}
\label{e35}
\end{equation}
and
\begin{equation}  
\begin{aligned}
& \alpha_{max_1}^{52-\textrm{non}}= 6 \pi \sqrt{2}, \\
& \alpha_{max_2}^{52-\textrm{non}}= 2 \pi \sqrt{6}, \\
\end{aligned}
\label{e36}
\end{equation}
where the term ``non" denotes non-trivial center element. It should be mentioned that an unusual normalization condition has been applied in Eq.~\eqref{e34}. Therefore, neither the $G_2$ potentials are expected nor the $SU(3)$ ones. However, as the Cartan generators of Eq.~\eqref{e23} are taken, we expect the potentials obtained from Eqs.~\eqref{e35} and \eqref{e36} to be parallel to the corresponding ones in Fig.~\ref{fig-8}, in some range of $R$. To study this fact more accurately, we study the group factor function.

The minimum points of the group factor function which happen at the positions where half of the vortex flux enters the Wilson loop, are responsible for the intermediate linear potential. Therefore, we compare the group factors of different representations of the $F_4$ obtained from the trivial center element with the
ones calculated from the decomposition to the $SU(3)$ subgroup.

\begin{figure}
\centering
\includegraphics[width=\linewidth]{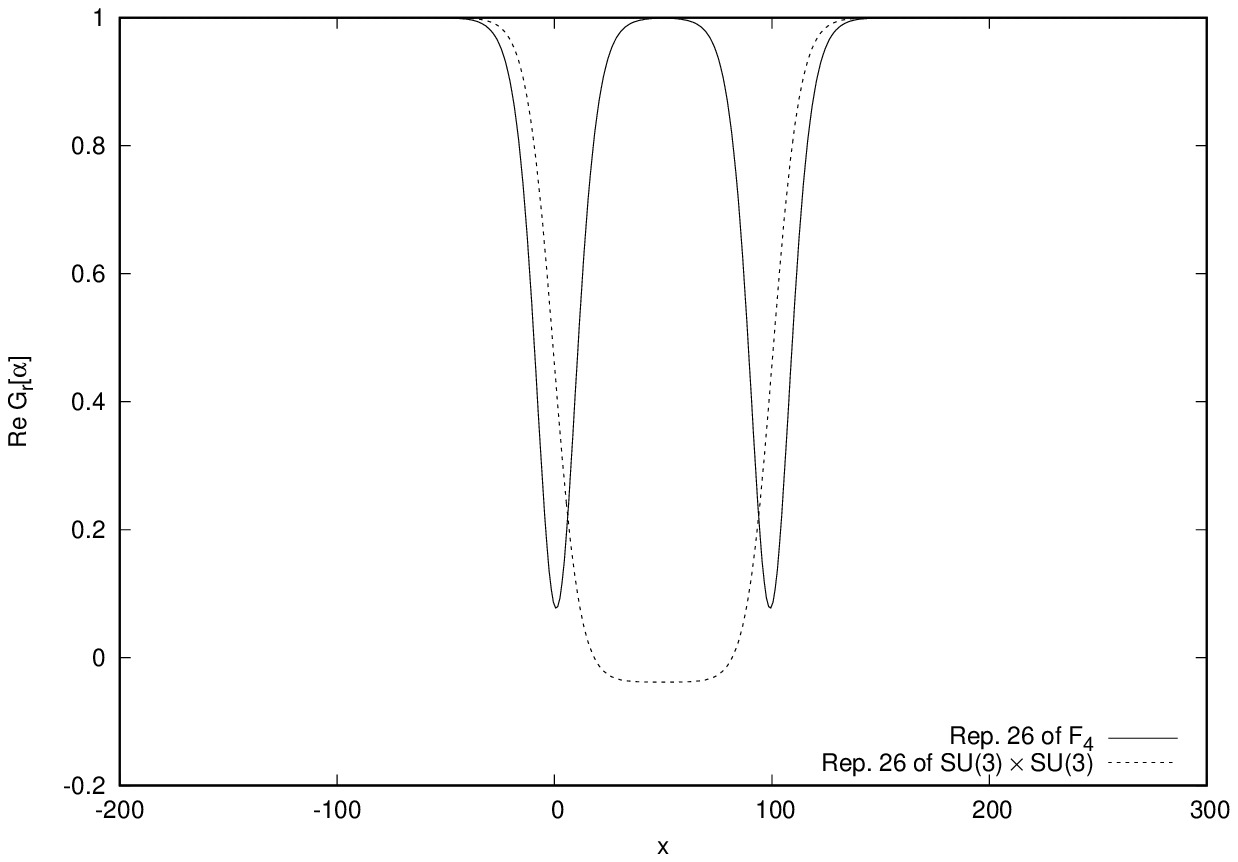}
\includegraphics[width=\linewidth]{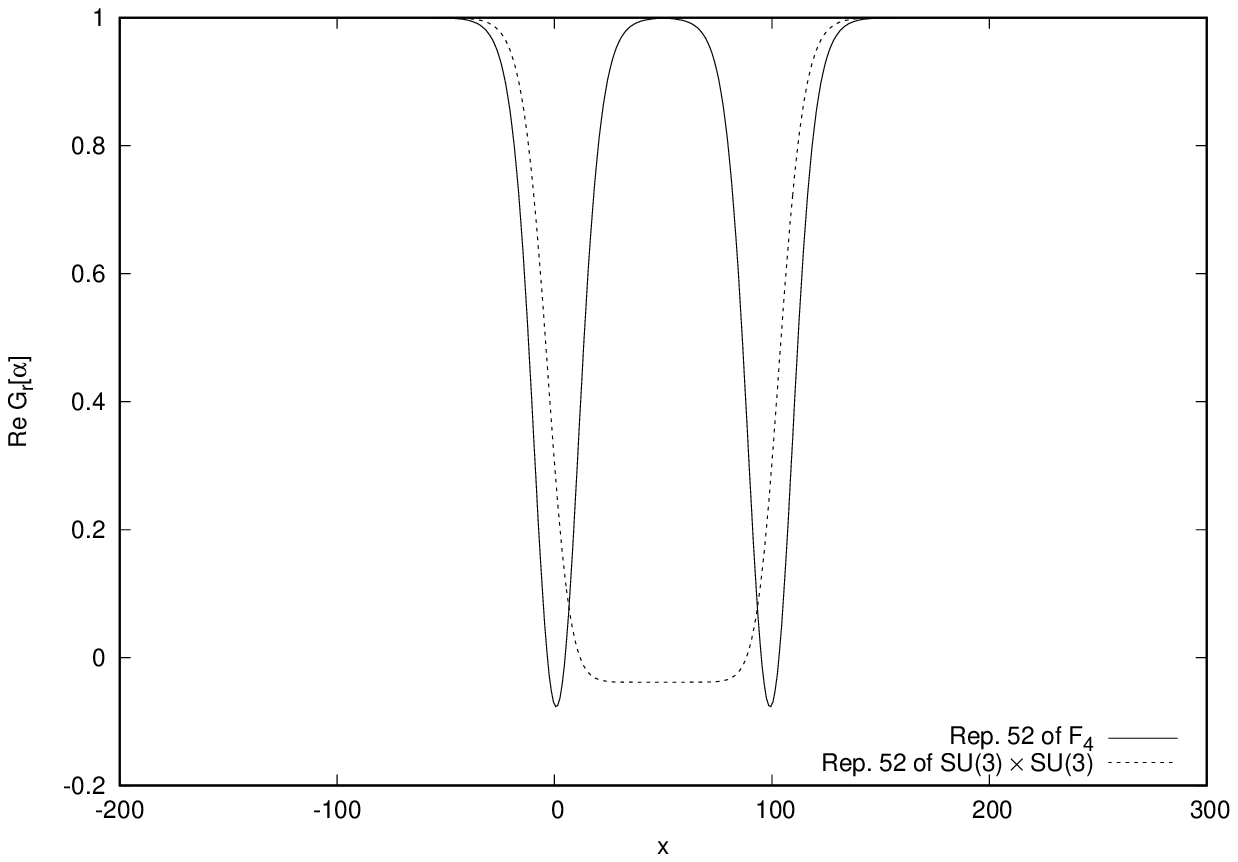}
\caption{The real part of the group factor versus the location $x$ of the vacuum domain midpoint, for $R=100$ and in the range $x \in [-200,300]$, for the fundamental and adjoint representation of the $F_4$ (solid lines) in comparison with the corresponding ones obtained from $ SU(3) \times SU(3) $ decomposition using its non-trivial center elements (dashed lines). The minimum value of the $F_4$ group factor for the fundamental and adjoint representations are $0.076$ and $-0.076$, respectively. It is clear that the minimum value of the group factors for the $SU(3) \times SU(3)$ decomposition is not identical with the corresponding ones for the $F_4$. }
\label{fig:f4-su3}       
\end{figure}

Fig.~\ref{fig:f4-su3} shows the real part of the group factor function for both fundamental and adjoint representations of the group $F_4$ and the $ SU(3) $ subgroup using its non-trivial center elements.
The discrepancies in the minimum amounts of the group factors in these figures are undeniable. As a result, one might say that the center elements of the $ SU(3) $ subgroup are not the direct factors for the confinement of the $F_4$ static potentials. The same reason is applicable for the higher representations of the
$ F_4$ exceptional group. The calculations of the higher representations have been presented in the Appendix~\ref{appa}.

So far, we have shown that the decomposition of the $ F_4 $ representations to the $ SU(3) $ subgroup leads to the Cartan generators that give the exact potential of the $F_4$ and Casimir scaling is also achieved. However, the $SU(3)$ non-trivial center elements are not responsible for the linearity observed in the
potentials of the $ F_4 $ representations. So, what accounts for the temporary confining potential? To answer this question, we study other subgroups of $F_4$. Greensite et.~al \cite{greensite} found that $SU(3)$ and $Z_3$-projected lattices are successful in reproducing the asymptotic string tension of $G_2$ gauge theory. However, no correlation between the gauge invariant Wilson loops and the $SU(3)$ and $Z_3$-projected loops is observed. They conclude that the results of $SU(3)$ and $Z_3$ projections in $G_2$ gauge theory are misleading. Therefore, they look for the smallest subgroup of $G_2$ i.e. $SU(2)$ and the Wilson loop imposing a ``maximal $SU(2)$" gauge is calculated. It is observed that the potential of the full $G_2$ theory is approximately parallel to the one obtained from the $SU(2)$-only Wilson loop. However, $SU(2)$ projection also appears to be  problematic. 

\hfill\\
\bm{$SU(2) \times Sp(6)$} \textbf{subgroup}\\
We try to achieve pure $SU(N)$ subgroups of $F_4$ out of this decomposition. One may focus on the $Sp(6)$ to branch it to its $ SU(2) $ subgroups, using the following process:
\begin{equation}  \label{e37}
 Sp(6) \supset SU(2) \times Sp(4)  \qquad (R)\\
\end{equation}
then,
\begin{equation} \label{e38}
 Sp(4) \supset SU(2) \times SU(2) \qquad (R)\\
\end{equation}
Therefore, $F_4$ branches to a pure $SU(2)$ subgroup. For the fundamental and adjoint representations of the $ F_4$, one may have \cite{slansky,sorba},
\begin{equation} 
\begin{aligned}
F_4 &\supset SU(2) \times Sp(6) \\
26 &=(2 , 6 ) \oplus (1 , 14 ),\\
52 &=(3 ,1) \oplus (1 , 21 ) \oplus (2 , 14^{\prime}).
\end{aligned}
\label{e39}
\end{equation}
In the next step,
\begin{equation} 
\begin{aligned}
 Sp(6) &\supset SU(2) \times Sp(4) \\
 6 &=(2 , 1) \oplus (1 , 4 ), \\
 14 &=(1 , 1) \oplus (1 , 5) \oplus (2 , 4),\\
 14^{\prime} &=(1 , 4) \oplus (2 , 5), \\
 21 &=(3 , 1) \oplus (1 , 10) \oplus (2 , 4).
\end{aligned}
\label{e40}
\end{equation}
Then,
\begin{equation}  \label{e41}
\begin{aligned}
 Sp(4) &\supset SU(2) \times SU(2) \\
 4 &=(2 , 1 ) \oplus (1 , 2),\\
 5 &=(1 , 1) \oplus (2 , 2),\\
 10 &=(3 , 1) \oplus (1 , 3) \oplus (2 , 2).
\end{aligned}
\end{equation}
Ultimately, for the $ F_4$ exceptional group, pure $ SU(2) $ subgroups are formed:
\begin{equation}  
\begin{aligned}
26 =&(2 , 1) \oplus (2 , 1) \oplus (1 , 2) \oplus (2 , 1) \oplus  (2 , 1) \oplus (1 , 2)\oplus \\
&(1 , 1) \oplus (1 , 1) \oplus  (2 , 2) \oplus (2 , 1) \oplus (1 , 2) \oplus (2 , 1) \oplus\\
& (1 , 2),\\
52 =&(3 , 1) \oplus (3 , 1) \oplus (3 , 1) \oplus  (1 , 3) \oplus  (2 , 2) \oplus (2 , 1) \oplus\\
& (1 , 2) \oplus  (2 , 1) \oplus  (1 , 2) \oplus (2 , 1) \oplus (1 , 2) \oplus  (1 , 1) \oplus \\
& (2 , 2)  \oplus  (1 , 1) \oplus (2 , 2) \oplus  (2 , 1) \oplus  (1 , 2) \oplus (1 , 1) \oplus \\
& (2 , 2)  \oplus (1 , 1) \oplus (2 , 2).
\end{aligned}
\label{e42}
\end{equation}
The Cartan generators extracted out of these decompositions are
\begin{equation}  
\begin{aligned}
H_{SU(2)}^{26} =\frac{1}{\sqrt{6}} \, \textrm{diag} \Big[ 0 , 0 , 0 , 0 ,  \sigma_3^2  ,  
0 , 0 , 0 , 0 ,  \sigma_3^2  , 0 , 0 , \sigma_3^2  , 
\sigma_3^2 ,\\
  0 , 0 ,  \sigma_3^2  ,  0 , 0 ,  \sigma_3^2 \Big],\\
 H_{SU(2)}^{52}=\frac{1}{3 \sqrt{2}} \, \textrm{diag} \Big[ 0 , 0 , 0 , 0 , 0 , 0 , 0 , 0 , 0 ,  \sigma_3^3 , \sigma_3^2  , \sigma_3^2  , 0 , 0 , \\
\sigma_3^2 ,  0 , 0 ,  \sigma_3^2  , 0 , 0 ,  \sigma_3^2  , 0 , \sigma_3^2 , 
 \sigma_3^2 ,  0 , \sigma_3^2 , \sigma_3^2 , 0 , 0 , \sigma_3^2 ,  0 , \sigma_3^2 ,  \sigma_3^2  ,\\
  0 , \sigma_3^2  ,  \sigma_3^2 \Big]
\end{aligned}
\label{e43}
\end{equation}
where $ \sigma_3^2=\frac{1}{2} \, \textrm{diag}[1 , -1] $ and $ \sigma_3^3= \textrm{diag}[1 , 0 , -1] $ are diagonal generators of the $ SU(2) $ gauge group in the fundamental and adjoint representations, respectively. Hence, with respect to the $SU(2) $ subgroup, we can reconstruct just one diagonal generator. It should be recalled that matrices of Eq.~\eqref{e43}, are normalized with the normalization condition in Eq.~\eqref{e17}. Considering normalization coefficient, it is obvious that the components of these matrices are identical with $H_3^{26}$ and $H_3^{52}$ in Esq.~\eqref{e23} and \eqref{e30}, respectively. Therefore, one expects
the potential between static color sources in the fundamental and adjoint representations to be the same as in Fig.~\ref{fig-8}, using the maximum flux values below:
\begin{equation} \label{e44}
\begin{aligned}
\alpha_{max}^{26} &=4 \pi \sqrt{6},\\
\alpha_{max}^{52} &=12 \pi \sqrt{2}.
\end{aligned}
\end{equation}

The next step is to investigate if the non-trivial center element of $SU(2)$, i.e. $ z_1^{SU(2)}=e^{i \pi} $ is responsible for the confinement of the $F_4$. So, we calculate the maximum flux values from Eq.~\eqref{e6}:

Similar to what we did for the $SU(3)$ subgroup, we are going to develop matrices containing the non-trivial center element of the $SU(2) $ subgroup corresponding to the decomposition of Eq.~\eqref{e42}:
\begin{equation} 
\begin{aligned}
 \mathbb{Z}_{SU(2)}^{26} = \textrm{diag} \Big[ 1 , 1 , 1 , 1 ,  z_1 \, \mathbb{I}_{2 \times 2} , 1 , 1 , 1 , 1 ,  
 z_1 \, \mathbb{I}_{2 \times 2} , 1 , 1 , \\
 z_1 \, \mathbb{I}_{2 \times 2} ,  z_1 \, \mathbb{I}_{2 \times 2} , 1 , 1 ,  z_1 \, \mathbb{I}_{2 \times 2} , 
 1 , 1 ,  z_1 \, \mathbb{I}_{2 \times 2}  \Big],\nonumber\\
\end{aligned}
\end{equation}
\begin{equation}
\begin{aligned}
\mathbb{Z}_{SU(2)}^{52} = \textrm{diag} \Big[ 1 , 1 , 1 , 1 ,  1 , 1 , 1 , 1 , 1 , \mathbb{I}_{3 \times 3} , 
  z_1 \, \mathbb{I}_{2 \times 2} ,  z_1 \, \mathbb{I}_{2 \times 2} ,\\ 1 , 1 ,   z_1 \, \mathbb{I}_{2 \times 2} ,  
 1 , 1 ,   z_1 \, \mathbb{I}_{2 \times 2} , 1 , 1 ,  z_1 \, \mathbb{I}_{2 \times 2} , 1  ,  z_1 \, \mathbb{I}_{2 \times 2} ,  
 z_1 \, \mathbb{I}_{2 \times 2} , 1 ,  \\ z_1 \, \mathbb{I}_{2 \times 2} ,  z_1 \, \mathbb{I}_{2 \times 2} , 1 , 1 ,  z_1 \, \mathbb{I}_{2 \times 2}, 1 ,  z_1 \, \mathbb{I}_{2 \times 2}  ,  z_1 \, \mathbb{I}_{2 \times 2}  ,  1 ,  z_1 \, \mathbb{I}_{2 \times 2}  , \\ z_1 \, \mathbb{I}_{2 \times 2}  \Big],
\end{aligned}
\label{e45}
\end{equation}
where $ \mathbb{I}_{ n \times n} $ are square identity matrices. As representations $2$ and $\bar 2$ are the same in the $SU(2)$, the vortices $z_1$ and $z_1^{\ast}$ are the same in this gauge group. It is observed that there are even number of $z_1$ vortices in the decompositions of Eq.~\eqref{e45}. Therefore, the vacuum domain or the trivial vortex might be thought to have these center vortices inside. 

The maximum flux condition of Eq.~\eqref{e6} could be written as follows:
\begin{equation}
\textrm{exp}[i \, \vec{\alpha} \cdot \vec{H}_{SU(2)}^{(26) \, \textrm{or} \,  (52)}]=\mathbb{Z}_{SU(2)}^{(26) \, \textrm{or} \,  (52)} \quad \mathbb{I},
 \label{e46}
\end{equation}
which leads to the amounts below:
\begin{equation} 
\begin{aligned}
\alpha_{max}^{26-\textrm{non}}&=2 \pi \sqrt{6},\\
\alpha_{max}^{52-\textrm{non}}&=6 \pi \sqrt{2}.
\end{aligned}
\label{e47}
\end{equation}

\begin{figure}
\centering
\includegraphics[width=\linewidth]{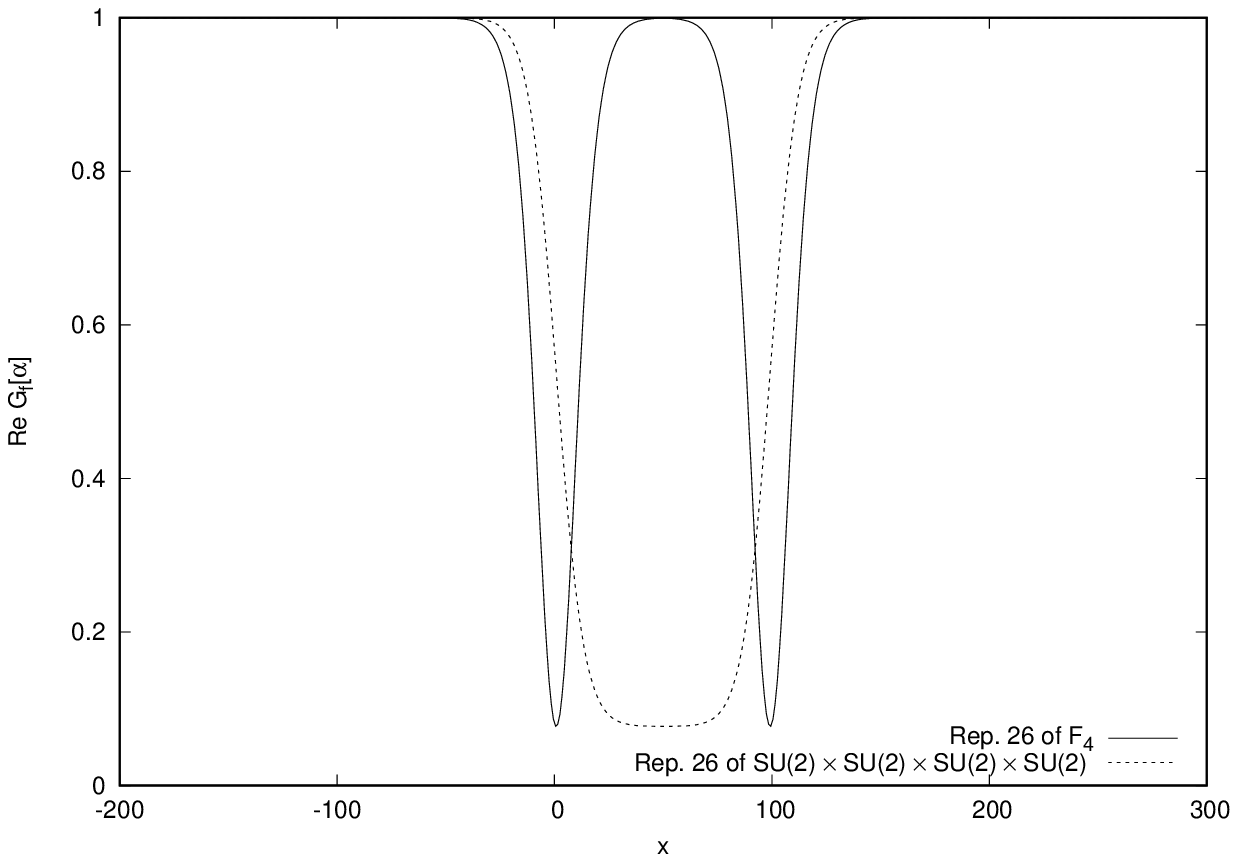}
\includegraphics[width=\linewidth]{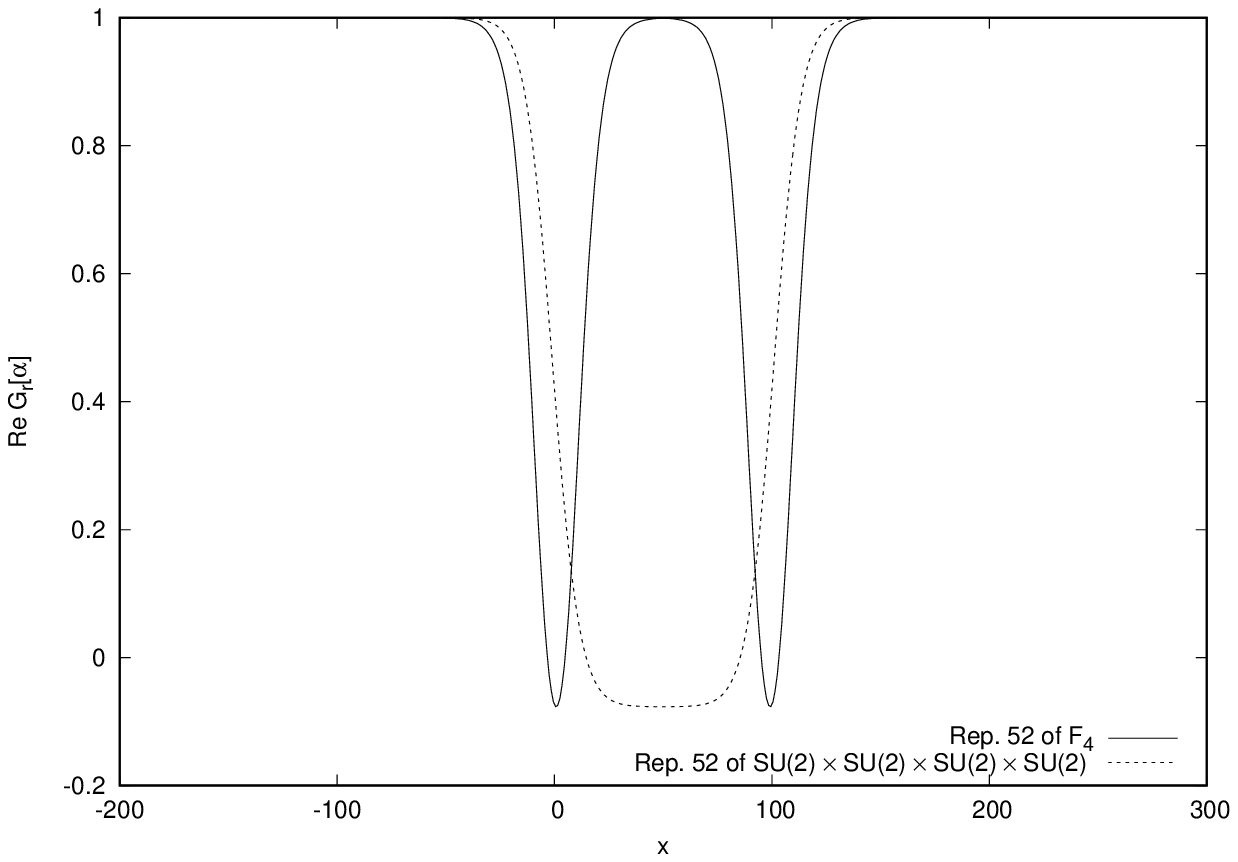}
\caption{The same as Fig.~\ref{fig:f4-su3} but the dashed lines represent the group factor for the $SU(2) \times SP(6) \supset SU(2) \times SU(2) \times SP(4) \supset SU(2) \times SU(2) \times SU(2) \times SU(2)$ decomposition. In each diagram, the minimum values of the $F_4$ group factor are the same as the corresponding ones obtained from the $SU(2)$ subgroup. These values are $0.076$ and $-0.076$ for
the fundamental and adjoint representations, respectively. }
\label{fig-10}       
\end{figure}

To compare the extremums of the group factor function of the $ F_4 $ exceptional group with its $ SU(2) $ subgroup, the real part of this function has been plotted versus the location of the vortex midpoint for the fundamental and adjoint representations in Fig.~\ref{fig-10}.
It is observed that the group factors corresponding to this decomposition reach the amounts $0.076$ and $-0.076$ at $x=50$ for the fundamental and adjoint representations, respectively. These amounts are identical with the corresponding ones for the $F_4$ which occur at $x=0$ and $x=100$. Since the extremums at $x=50$ imply the complete interaction between vortices and the Wilson loop that results in a linear asymptotic potential, it is the center element of the $ SU(2) $ subgroup which gives rise to the intermediate linear potential of $F_4$.

The other chain of possible decomposition is
\begin{equation} \label{e48}
\begin{aligned}
& F_4 \supset SU(2) \times Sp(6)  \quad (R) \\
& Sp(6) \supset SU(2) \times SU(2) \quad (S)
\end{aligned}
\end{equation}
This decomposition generates matrices similar to $ h_3 $ and $ h_4 $. According to our previous discussion in Fig.~\ref{fig:gr-h34}, when $h_3$ or $h_4$ is applied, the minimum points of the group factor function are not identical with the ones for the $F_4$ gauge group. Using this fact,
one might conclude that this decomposition is not responsible for the confinement in $F_4$.

\hfill\\
\bm{$SO(9)$} \textbf{subgroup}\\
Another regular maximal subgroup of $F_4$ group is $SO(9)$. To extract a pure $SU(N)$ subgroup out of this subgroup, one might choose the following decomposition process:
\begin{equation} \label{e49}
SO(9) \supset SU(2) \times SU(4)\quad (R)
\end{equation}
We present three methods to decompose the $SU(4)$ to the $SU(2)$ subgroups:
\begin{itemize}
\item{
A regular decomposition as follows:
\begin{equation} \label{e50}
SU(4) \supset SU(2) \times SU(2) \times U(1)
\end{equation}
There is a $U(1)$ factor in this decomposition which makes it a non-semisimple maximal subgroup. In fact, the $ U(1) $ which appears in some branching rules is a trivial abelian Lie group composed by all $ 1 \times 1 $ matrices of $ e^{ i \phi } $ with real $ \phi $ \cite{king}. This factor is excluded in branching
rules \cite{sorba,von}. In our case, we ignore it since it has no effect on our calculations. If we evade the $ U(1) $ factor in Eq.~\eqref{e50}, the same results as Eq.~\eqref{e42}, where the $F_4$ has been decomposed to its pure $ SU(2) $ subgroups, are achieved. Hence, this decomposition could be responsible for the intermediate linear potentials, as well. As the results for the fundamental and adjoint representations are the same as in Fig.~\ref{fig-10}, we just present the detailed calculations for the higher representations in Appendix~\ref{appb}.}
\item{$SU(4)$ has a singular subgroup:
\begin{equation}
 SU(4) \supset Sp(4) \quad (S)
\label{e51}  
\end{equation}
and it could be decomposed as the following:
\begin{equation}
 Sp(4) \supset SU(2) \times SU(2) \quad (R)
\label{e52}
\end{equation} 
The exact decompositions as in Eq.~\eqref{e42} and Cartan generators of Eq.~\eqref{e43} are obtained. Therefore, the same results are achieved. Consequently, singular maximal subgroups are able to bring out the same potentials as $F_4$ as well. Furthermore, this decomposition could also describe the linear potential of the $F_4$ correctly.
}
\item{Another chain of breaking to $SU(N)$ subgroups could be a singular decomposition:
\begin{equation} 
SU(4) \supset SU(2) \times SU(2) \quad (S)
\label{e53}
\end{equation}
Despite that this decomposition seems to be similar to Eq.~\eqref{e50}, due to the branching rules, representations decompose in a way that they compose exact matrices as $ h_3 $ and $ h_4 $ of Eq.~\eqref{e14} in the fundamental representation of $ F_4$. We previously learned that induced potentials by these matrices have different
behaviors according to Fig.~\ref{fig:gr-h34}.
}
\end{itemize}

So far, we can conclude that, in order to determine the subgroups whose Cartan decompositions result in a well-defined potential, one has to compare reconstructed Cartan matrices produced by means of the subgroups with the ones of the main exceptional group itself. In the $ F_4 $ case,
the potential out of applying all of its Cartan generators is the same as the case where just $h_1$ or $h_2$ is used. Therefore, any regular or singular subgroup which is able to reconstruct the same diagonal matrices as one of these two generators, produces the same potential as the $F_4$ itself.

\hfill\\
\bm{$SU(2) \times G_2$} \textbf{subgroup}\\
Ultimately, we are going to investigate the results of a direct singular maximal subgroup of the $F_4$ exceptional group, i.e. $ SU(2) \times G_2 $, because it shows a different behavior. To make a pure $ SU(N) $ subgroup out of this singular subgroup, one may choose to break $G_2$ into its $ SU(3) $ subgroup:
\begin{equation} 
 G_2 \supset SU(3)  \quad (R) 
\label{e54}
\end{equation}
It is obviously not a pure subgroup because it contains both $SU(2)$ and $SU(3)$ subgroups. However, if one considers the representations of the $SU(2)$ as degeneracies of the irreducible representations of the $SU(3)$ in the branching rules, the result will be the same as $F_4 \supset SU(3) \times SU(3) $
decomposition. We have argued that this decomposition is not responsible for the $F_4$ confinement.

A more challenging procedure is the following decomposition:
\begin{equation} \label{e55}
\begin{aligned}
 F_4 &\supset SU(2) \times G_2 \quad (S)  \\
 26 &=(5 , 1) \oplus (3 , 7),  \\
 52 &=(3 , 1) \oplus (1 , 14) \oplus (5 , 7).\\
\end{aligned}
\end{equation}
$G_2$ might be  decomposed as
\begin{equation} \label{e56}
\begin{aligned}
 G_2 &\supset SU(2) \times SU(2) \quad (R)  \\
 7 &=(2 , 2) \oplus (1 , 3), \\
 14 &=(3 , 1) \oplus (2 , 4) \oplus (1 , 3).  \\
\end{aligned}
\end{equation}
Finally,
\begin{equation} \label{e57}
\begin{aligned}
 26 =&(5 , 1) \oplus (2 , 2) \oplus (1 , 3) \oplus  (2 , 2) \oplus  (1 , 3) \oplus (2 , 2) \oplus \\
     &(1 , 3) \\ 
 52 =&(3 , 1) \oplus (3, 1) \oplus (2 , 4) \oplus  (1 , 3) \oplus (2 , 2) \oplus (1 , 3) \oplus \\
     &(2 , 2) \oplus   (1 , 3) \oplus  (2 , 2) \oplus (1 , 3) \oplus (2 , 2) \oplus  (1 , 3) \oplus \\
     &(2 , 2) \oplus (1 , 3).  
\end{aligned}
\end{equation}
Using these decompositions, we can reproduce Cartan generators in the fundamental and adjoint representations as the following:
\begin{equation} \label{e58}
\begin{aligned}
 H_{SU(2) \times G_2}^{26}= \frac{1}{3 \sqrt{2}} \, \textrm{diag} \Big[ 0 , 0 , 0 , 0 , 0 ,  
 \sigma_3^2 , \sigma_3^2 , \sigma_3^3 , \sigma_3^2 , \sigma_3^2 ,  \sigma_3^3 ,\\ \sigma_3^2 , \sigma_3^2 , \sigma_3^3  \Big],\\
 H_{SU(2) \times G_2}^{52}= \frac{1}{3 \sqrt{6}} \, \textrm{diag} \Big[ 0 , 0 , 0 , 0 , 0 , 0 , 
 \sigma_3^4 , \sigma_3^4 , \sigma_3^3 , \sigma_3^2 , \sigma_3^2 ,  \\ \sigma_3^3 , \sigma_3^2 , \sigma_3^2 , \sigma_3^3  , 
  \sigma_3^2 , \sigma_3^2 , \sigma_3^3  ,   \sigma_3^2 , \sigma_3^2 , \sigma_3^3 ,  \sigma_3^2 , \sigma_3^2 , \sigma_3^3   \Big]
\end{aligned}
\end{equation}
In these matrices, the normalization coefficients have been computed from Eq.~\eqref{e17}. $\sigma_3^2$, $\sigma_3^3$ and $\sigma_3^4$ are Cartan generators of the $SU(2)$ gauge group in the fundamental, adjoint and 4-dimensional representations, respectively. After an initial review, the elements of these
matrices are not fully the same as the corresponding ones in Eqs.~\eqref{e23} and \eqref{e30} or \eqref{e43}. Accordingly, the trivial static potentials, when we consider trivial center element of the $SU(2)$ subgroup, are not identical with the $F_4$ exceptional group itself. We have investigated this subgroup in ref.~\cite{greece}. However, it is another
aspect of this decomposition which is appealing.

\begin{figure}
\centering
\includegraphics[width=\linewidth]{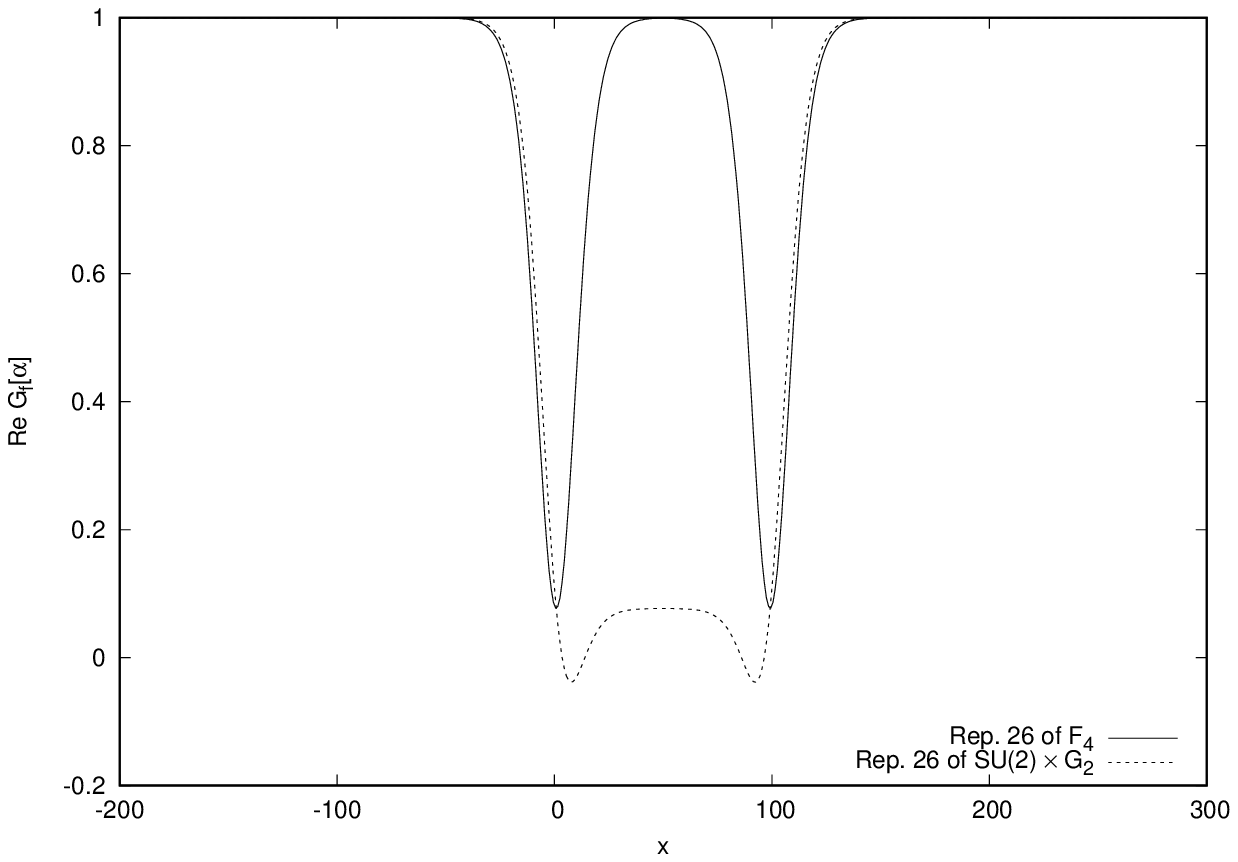}
\includegraphics[width=\linewidth]{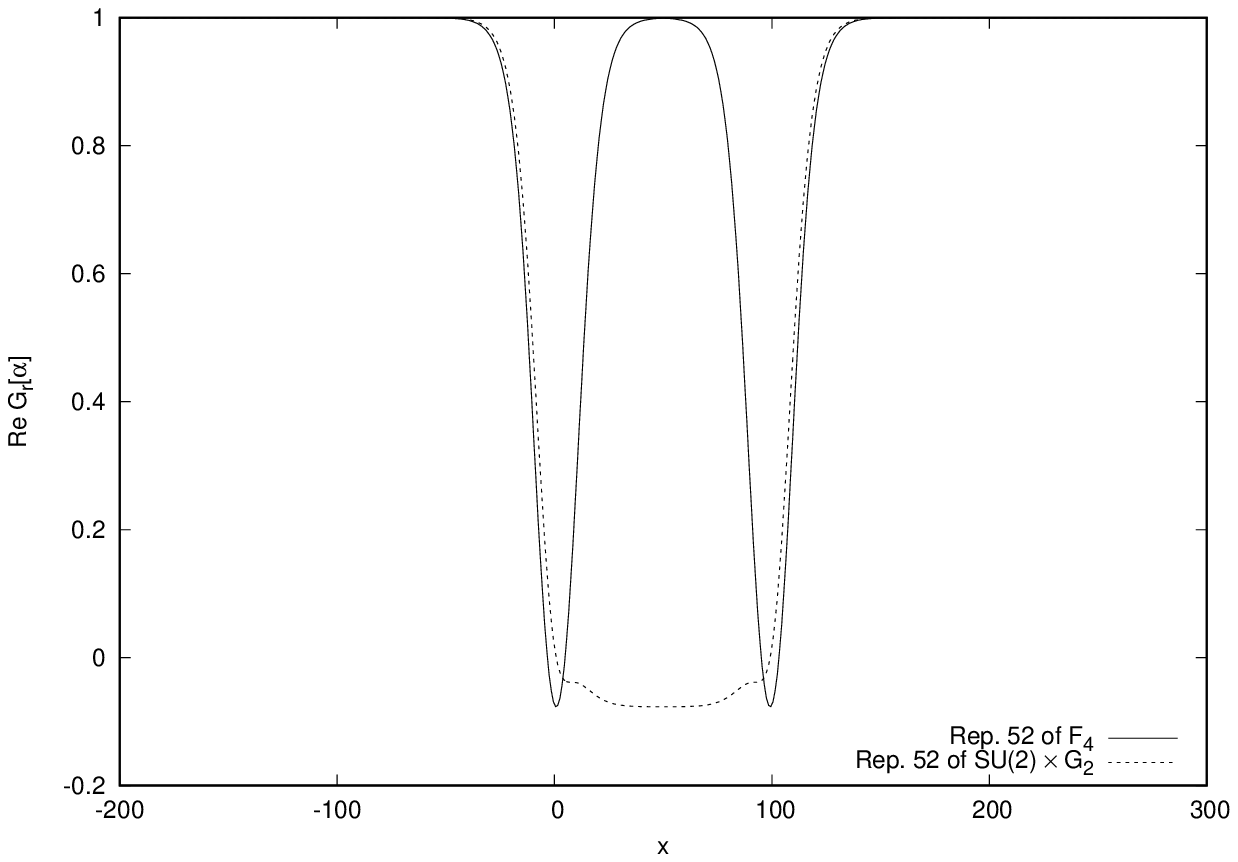}
\caption{The same as Fig.~\ref{fig-10} but for the $SU(2) \times G_2 $ subgroup. In both figures, the minimum values of the $F_4$ group factor are identical with the amount of the group factor corresponding to the $SU(2) \times G_2$ decomposition at $x=50$. Therefore, the slope of the intermediate linear of the $F_4$ is the same as the asymptotic one for the $SU(2) \times G_2$ subgroup.}
\label{fig-12}       
\end{figure}

The center element matrices of the $SU(2) \times G_2$ subgroup of $F_4$ in the fundamental and adjoint representations are given by
\begin{equation}  \label{e59}
\begin{aligned}
 \mathbb{Z}_{SU(2) \times G_2}^{26} = \textrm{diag} \Big[ 1 , 1 , 1 , 1 , 1 ,  z_1 \, \mathbb{I}_{2 \times 2} ,  
  z_1 \, \mathbb{I}_{2 \times 2} ,  \mathbb{I}_{3 \times 3} , \\ 
  z_1 \, \mathbb{I}_{2 \times 2} ,  z_1 \, \mathbb{I}_{2 \times 2} , \mathbb{I}_{3 \times 3} ,  \\
 z_1 \, \mathbb{I}_{2 \times 2} ,  z_1 \, \mathbb{I}_{2 \times 2} , \mathbb{I}_{3 \times 3}  \Big],\\
 \mathbb{Z}_{SU(2) \times G_2}^{52}= \textrm{diag} \Big[ 1 , 1 , 1 , 1 ,  1 , 1 , z_1 \, \mathbb{I}_{4 \times 4}  , 
  z_1 \, \mathbb{I}_{4 \times 4} , \mathbb{I}_{3 \times 3} ,\\
   z_1 \, \mathbb{I}_{2 \times 2} ,  z_1 \, \mathbb{I}_{2 \times 2} , \mathbb{I}_{3 \times 3} , 
  z_1 \, \mathbb{I}_{2 \times 2} ,  z_1 \, \mathbb{I}_{2 \times 2} , \mathbb{I}_{3 \times 3} ,  z_1 \, \mathbb{I}_{2 \times 2} ,  \\
 z_1 \, \mathbb{I}_{2 \times 2} , \mathbb{I}_{3 \times 3} ,   z_1 \, \mathbb{I}_{2 \times 2} ,  z_1 \, \mathbb{I}_{2 \times 2} ,  \\
 \mathbb{I}_{3 \times 3} ,  z_1 \, \mathbb{I}_{2 \times 2} ,  z_1 \, \mathbb{I}_{2 \times 2} , \mathbb{I}_{3 \times 3}  \Big].
\end{aligned}
\end{equation}
Now, putting the above matrices in Eq.~\eqref{e6}, the maximum flux values are calculated as follows:
\begin{equation}  \label{e60}
\begin{aligned}
 \alpha_{\textrm{max}}^{26 \textrm{-non}} &= 6 \pi \sqrt{2},  \\
 \alpha_{\textrm{max}}^{52 \textrm{-non}} &= 6 \pi \sqrt{6}.
\end{aligned}
\end{equation}
Fig.~\ref{fig-12} shows the group factor function versus the vortex midpoint $x$ for the fundamental and adjoint representations, respectively. 
As it is observed, the group factor has a totally different behavior in comparison with Fig.~\ref{fig-10}. The minimum points of the $F_4$ representations occur at $x=0$ and $x=100$ where half of the vortex flux enters the Wilson loop. These points are responsible for the intermediate linear potentials of $F_4$. Now, we focus on the decomposition of the representations to $SU(2) \times G_2 $ subgroup.
When the vortex midpoint is located at $x=50$, it means that the vortex is completely inside the Wilson loop. The non zero value of the group factor at this point, results in a linear potential at large distances. As the value of the group factor at this point equals to the corresponding one for the $F_4$, the slope of this linear potential seems to be identical with the intermediate linear potentials of the $F_4$.
Therefore, one might say that $SU(2) \times G_2 \supset SU(2) \times SU(2) \times SU(2)$ subgroup of the $F_4$ is responsible for the intermediate confinement of this exceptional group.

An interesting point here is that Cartan generators of this decomposition are different from $h_1$ and $h_2$. However, the minimum points of the $F_4$ group factor can still be investigated correctly via this decomposition. The question is why this happens. In fact, N-ality of the $SU(N)$ representations or the center element matrix obtained from the group decompositions has
predominant responsibility here. The representations could be classified by their N-ality. This means that the Wilson loop of the representations with the same N-ality is affected by a vortex type $n$ in the same manner. To make it more clear, we investigate center element matrices of the fundamental representation in Eq.~\eqref{e45}, obtained from the $SU(2) \times Sp(6)$ subgroup, and also in Eq.~\eqref{e59}, by means of $ SU(2) \times G_2 $ subgroup, which have different elements. In the former one, there exists fourteen $1$'s and six $ z_1 \, \mathbb{I}_{2 \times 2} $'s while in the latter one, in Eq.~\eqref{e59}, there exists five $1$'s, six $ z_1 \, \mathbb{I}_{2 \times 2} $'s and three $  \mathbb{I}_{3 \times 3} $'s. The number of $ z_1 \, \mathbb{I}_{2 \times 2} $ center elements is the same in both matrices which is corresponding to the fundamental representation with 2-ality$=1$. Elements $1$ and $  \mathbb{I}_{3 \times 3} $, which is corresponding to the $3$ dimensional representation with zero 2-ality, have no effect on the Wilson loop. So, the other elements of these two matrices do not affect the Wilson loop. As a result, although the matrices of Eqs.~\eqref{e43} and \eqref{e58} have different elements and the potentials out of these two generators behave differently, the number of center vortices which emerge in both decompositions is the same.  Thus, the group factor reaches the same minimum amount in both of them. Regarding 52-dimensional adjoint representation, the same process comes about.

\subsection{\label{sec:E6} $E_6$ exceptional group}
\begin{figure}
\centering
\includegraphics[width=\linewidth]{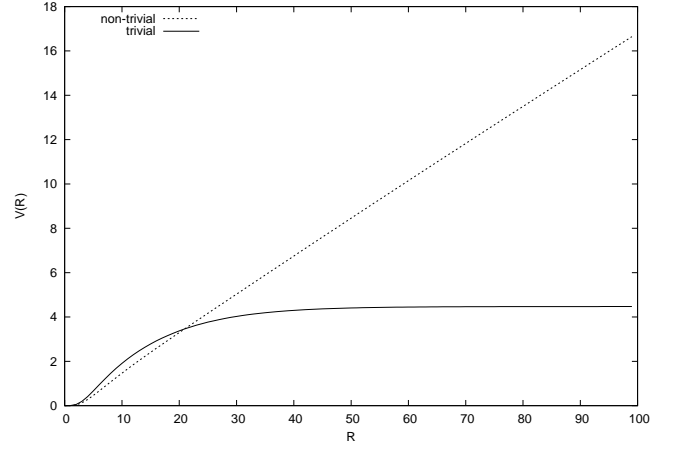}
\caption{Potential of the fundamental representation of $E_6$ using trivial and non-trivial center elements for $R \in [1,100]$.}
\label{fig:e6-both}       
\end{figure}

$E_6$ is the third exceptional group in terms of largeness. It makes a 78-dimensional space with 78 generators and, similar to $SU(3)$, has $\mathbb{Z}_3$ as its group center \cite{wipf2}. Here, we mostly focus on its trivial center to investigate the static potential behavior. As the rank of $E_6$ is six, it possesses 6 diagonal
matrices which are shown as following for the fundamental representation\cite{ekins}:
\begin{equation}
\begin{aligned}
h_1^{27} = N_1 \ \textrm{diag} \Big[ -1 , +1 , 0 , 0 , 0 , 0 , 0 , 0 , 
 0 , -1 ,  0 , 0 , 0 , -1 , \\+1 ,-1 , -1 , +1 , +1 ,  
 +1 ,  0 , -1 , +1 , 0 , 0 , 0 , 0 \Big], \nonumber
\end{aligned}
\end{equation}
\begin{equation} 
\begin{aligned}
h_2^{27}= N_2 \ \textrm{diag} \Big[0 , -1 , +1 , 0 , 0 , 0 , 0 , 0 , 
 -1 , +1 , 0 , -1 , -1 , \\+1 , 0 ,+1 , 0 ,  0 , 
  0 , -1 , +1 , 0 , -1 , +1 , 0 , 0 , 0 \Big], \nonumber
\end{aligned}
\end{equation}
\begin{equation} 
\begin{aligned}
h_3^{27}= N_3 \ \textrm{diag} \Big[0 , 0 , -1 , +1 , 0 , 0 , 0 , -1 , 
 +1 , 0 , -1 , +1 , 0 , 0 , \\0 ,-1 , +1 , 0 , 
 -1 , +1 , 0 , 0 , 0 , -1 , +1 , 0 , 0 \Big], \nonumber
\end{aligned}
\end{equation}
\begin{equation} 
\begin{aligned}
h_4^{27}= N_4 \ \textrm{diag} \Big[ 0 , 0 , 0 , -1 , +1 , 0 , -1 , +1 , 
 0 , 0 , 0 , -1 , +1 ,\\ -1 , 0 ,+1 , 0 , -1 ,
 +1 , 0 , 0 , 0 , 0 , 0 , -1 , +1 , 0 \Big],  \nonumber
\end{aligned}
\end{equation}
\begin{equation} 
\begin{aligned}
h_5^{27}= N_5 \ \textrm{diag} \Big[ 0 , 0 , 0 , 0 , -1 , +1 , 0 , -1 , 
-1 , -1 , +1 , +1 , 0 , \\+1 , -1 , 0 , 0 , 
 +1 , 0 , 0 , 0 , 0 , 0 , 0 , 0 , -1 , +1 \Big], \nonumber
\end{aligned}
\end{equation}
\begin{equation} 
\begin{aligned}
h_6^{27}= N_6 \ \textrm{diag} \Big[ 0 , 0 , 0 , -1 , -1 , -1 , +1 , 
+1 , 0 , 0 , +1 , 0 , 0 , 0 ,\\ 0 ,  0 , -1 , 0 , 0 ,  
 -1 ,  -1 , +1 , +1 , +1 , 0 , 0 , 0 \Big]. 
\end{aligned}
\label{e61}
\end{equation}
These matrices are normalized and their normalization coefficients could be calculated from Eq.~\eqref{e17}: 
\begin{equation}  \label{e62}
N_1= \cdots =N_6=\frac{1}{2 \sqrt{6}}.
\end{equation}
It should be noted that, due to the similarity of these matrices, one can use only $h_1^{27}$ to calculate the potential of the fundamental representation of $E_6$. The maximum flux values for the domain structures, calculated from the condition in Eq.~\eqref{e19}, are 
\begin{equation} \label{e63}
\alpha_1^{max}= \cdots =\alpha_6^{max}=2 \pi \sqrt{24}.
\end{equation}
On the other hand, if we include the non-trivial flux condition in Eq.~\eqref{e6}, the maximum amounts for the vortices fluxes are
\begin{equation} \label{e64}
\begin{aligned}
& \alpha_{{max}_{1}}^{\textrm{non}}=\alpha_{{max}_{4}}^{\textrm{non}}= \mp \frac{4 \pi }{3} \sqrt{6}, \\
& \alpha_{{max}_{3}}^{\textrm{non}}=\alpha_{{max}_{6}}^{\textrm{non}}= \mp 4 \pi  \sqrt{6}, \\
& \alpha_{{max}_{2}}^{\textrm{non}}=\alpha_{{max}_{5}}^{\textrm{non}}= \mp \frac{8 \pi }{3} \sqrt{6}. \\
\end{aligned}
\end{equation}
where ``non" indicates the answer pertaining to the non-trivial center elements. Negative answers have been gained by vortices type one when Eq.~\eqref{e6} equals to $z_1=\exp(\frac{2\pi i}{3})$ and positive answers are corresponding to the vortices type two ($z_2=\exp(\frac{-2\pi i}{3})$). Static potentials obtained
by both of these maximum trivial and non-trivial flux values in Eqs.~\eqref{e63} and \eqref{e64} have been depicted in Fig.\ref{fig:e6-both}.
As it could be predicted, at far distances, the potential obtained from the trivial center element of $E_6$ has been screened while the potential corresponding to the non-trivial center element is linear. This fact could be investigated by tensor products of the $E_6$ ``quark" and ``gluons": 
\begin{equation} \label{e64new}
27 \times 78 = 27 \oplus  351    \oplus 1728.
\end{equation}
It is seen that $E_6$ ``gluons" are not able to screen the potential of the $E_6$ ``quarks". Similar to $SU(N)$ gauge groups, one ``quark" and one ``anti-quark" can join to create a meson:  
\begin{equation} \label{e65-new}
27 \times \overline{27} = 1 \oplus  78   \oplus 650,
\end{equation}
and three ``quarks" form a baryon:
\begin{equation} \label{e66new}
27 \times 27 \times 27 = 1 \oplus  2(78)   \oplus 3(650) \oplus 2925 \oplus \overline{3003} \oplus 2\overline{5824}.
\end{equation}

Now, we aim to put on the same procedure applied for $F_4$, to explain what actually accounts for the temporary confinement in trivial static potential of $E_6$ exceptional group. The main question is, which kind of center vortices have filled the $E_6$ QCD vacuum which give rise to the intermediate confining potential
obtained by the trivial center element? In general, we have three candidates:
\begin{itemize}
\item{Non-trivial center elements of the $E_6$ exceptional group};
\item{Non-trivial center elements of its $SU(3)$ maximal subgroup};
\item{Non-trivial center element of its maximal $SU(2) $ subgroup}.
\end{itemize}

\begin{figure}
\centering
\includegraphics[width=\linewidth]{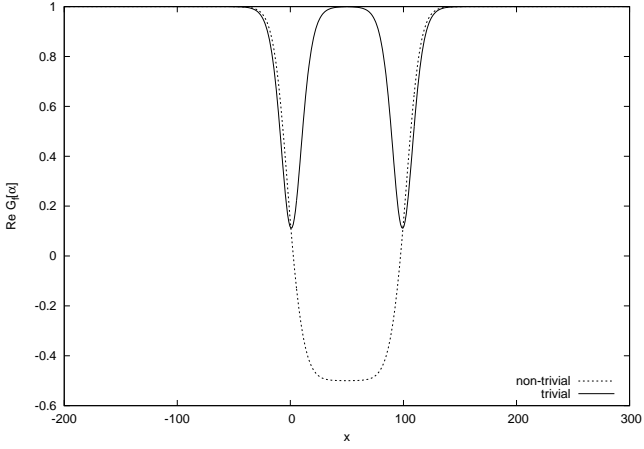}
\caption{Solid line represents the group factor of the fundamental representation of the $E_6$ versus the location $x$ of the vacuum domain midpoint, using the trivial center element, for $R=100$ in the range $x \in [-200,300]$. The dashed line shows the same function versus the location $x$ of the non-trivial center vortex.}
\label{fig:gre6-both}       
\end{figure}

To answer this question properly, the group factor function using Eq.~\eqref{e61} for both trivial and non-trivial center elements of the $E_6$ exceptional group have been demonstrated in Fig.~\ref{fig:gre6-both}.
Consequently, non-trivial center elements of $E_6$ are not the direct reason of the intermediate linear part in the trivial potential of $E_6$. Then, we go for some of the maximal subgroups of $E_6$ which have been mentioned in Tab.~\ref{tab2}.

\begin{figure}
\centering
\includegraphics[width=\linewidth]{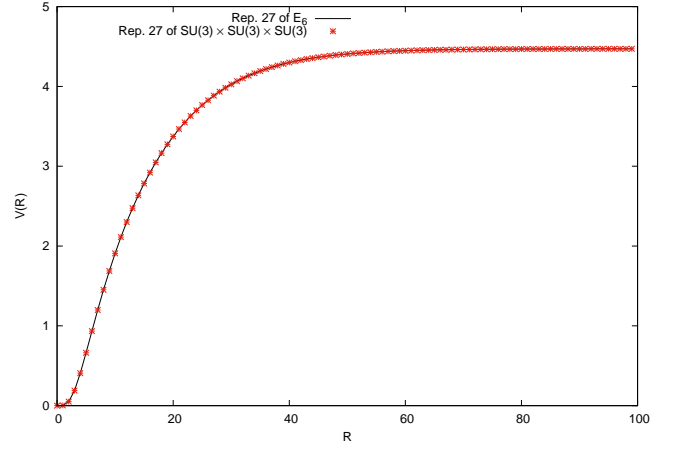}
\caption{The potential of the fundamental representation of $E_6$ using all Cartan generators (solid line) and the one corresponding to the $SU(3) \times SU(3) \times SU(3) $ decomposition (stars) in the range $R \in [1,100]$. The two sets of data are the same.}
\label{fig:su3-e6-2pi}       
\end{figure}

\hfill\\
\bm{$SU(3) \times SU(3) \times SU(3) $} \textbf{subgroup}\\
 The fundamental representation decomposes as \cite{sorba,slansky,gellmann}
\begin{equation}  \label{e65}
27=(3 , \bar{3} , 1) \oplus (1 , 3 , 3) \oplus (\bar{3} , 1 , \bar{3}),
\end{equation}
Thus, if we assume the first two representations in the parentheses in Eq.~\eqref{e65}, as degeneracies step by step, one might have:
\begin{equation}  \label{e67new}
27=9(1) \oplus 3(3) \oplus 3 (\bar 3).
\end{equation}
Although $E_6$ has a non-trivial center element, the method of decomposing its representations leads to the $SU(3)$ representations with different trialities. Therefore, an $E_6$ ``quark" could be decomposed to three $SU(3)$ quarks, three anti-quarks and nine singlets. Now, two Cartan generators with regard to this subgroup are reconstructed from Eq.~\eqref{e67new}:
\begin{equation} \label{e66}
\begin{aligned}
 H_a^{27}= \frac{1}{\sqrt{6}} \, \textrm{diag} \Big[ 0 , 0 , 0 , 0 , 0 , 0 , 0 , 0 , 0 ,  
 \lambda_{a}^3 , \lambda_{a}^3 , \lambda_{a}^3 ,\\ -(\lambda_a^3)^{\ast} ,  -(\lambda_a^3)^{\ast} , -(\lambda_a^3)^{\ast}   \Big], \\
\end{aligned}
\end{equation}
where $a=3,8$. Now, one can consider the condition in Eq.~\eqref{e19} and find
\begin{equation}  \label{e67}
\begin{aligned}
\alpha_{{max}_{1}}^{27}=2 \pi \sqrt{6}, \\
\alpha_{{max}_{2}}^{27}=6 \pi \sqrt{2}. 
\end{aligned}
\end{equation}
The potential between the fundamental sources of the $E_6$ using the six Cartan generators in Eq.~\eqref{e61} has been presented in Fig.~\ref{fig:su3-e6-2pi} which overlaps completely with the data obtained from the above values in Eq.~\eqref{e67} and Cartan generators of Eq.~\eqref{e66}. This fact is the result of the the identical components of $H_3^{27}$ and $h_1^{27}$. Although $H_8^{27}$ has different matrix elements, it has the same effect as $H_3^{27}$ on the $E_6$ potentials. The similar discussion has been given in the $F_4 \supset SU(3) \times SU(3)$ decomposition. Therefore,, one might use $SU(3)$ subgroup decomposition to find the $E_6$ adjoint potential.

Reconstruction of Cartan generators in the adjoint representation of $E_6$ with respect to its $SU(3)$ subgroup is possible only when we want to calculate the trivial static potential as they are identical. This method is not applicable for the potentials obtained by the non-trivial center elements. 

\begin{figure}
\centering
\includegraphics[width=\linewidth]{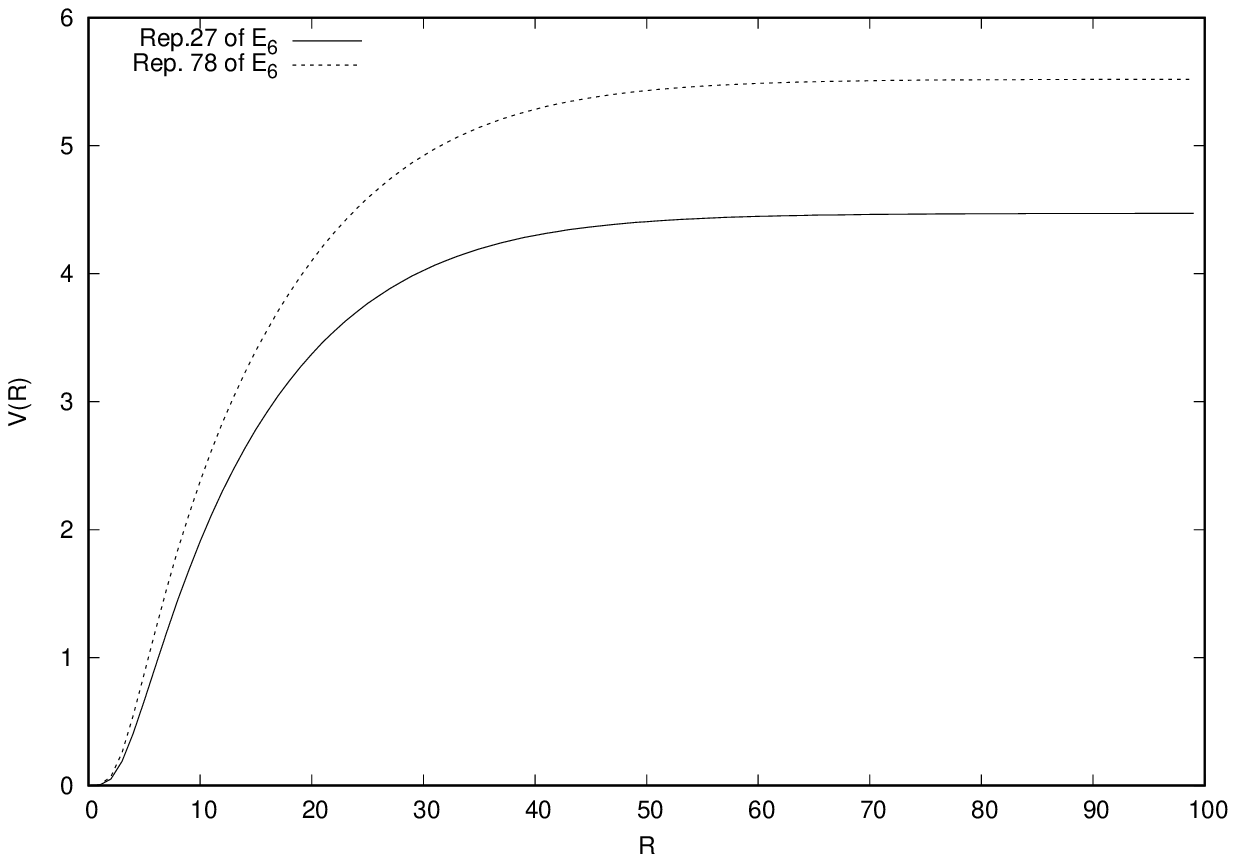}
\includegraphics[width=\linewidth]{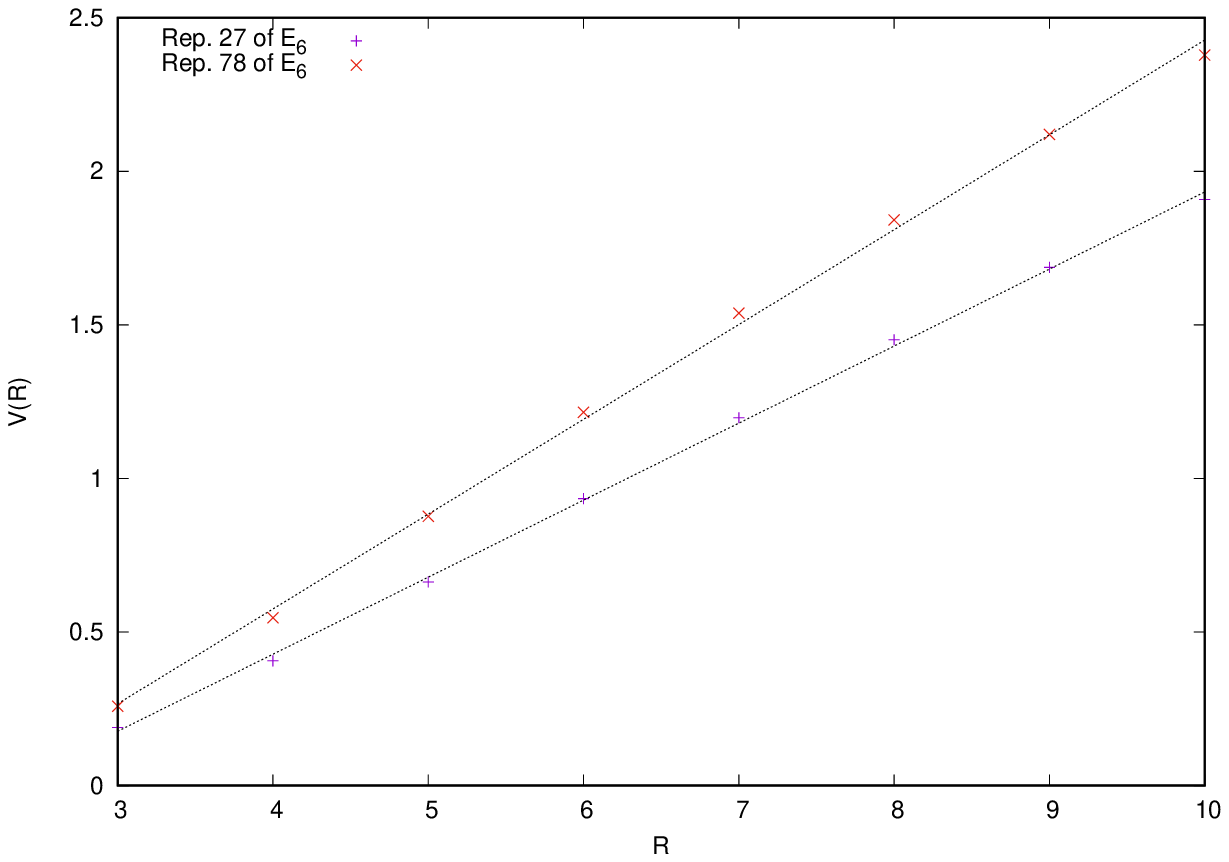}
\caption{Upper diagram shows trivial static potentials of the $E_6$ exceptional group for the fundamental and adjoint representations in the range $R \in [1,100]$. The potentials are screened at large distances which is due to the absence of the non-trivial center element. At intermediate quark separations, the potentials are linear which have been presented in the lower diagram in the range $R \in [3,10]$. The ratio of the adjoint potential to the fundamental one is in agreement with Casimir scaling.}
\label{fig-18}       
\end{figure}

The adjoint representation might be decomposed as the following:
\begin{equation} \label{e68}
\begin{aligned}
 78= &(8 , 1 , 1) \oplus (1 , 8 , 1) \oplus  (1 , 1 , 8) \oplus (3 , 3 , \bar{3}) \oplus \\ &(\bar{3} , \bar{3} , 3)=16 (1) \oplus 8 \oplus 9(\bar 3) \oplus 9(3). 
\end{aligned}
\end{equation}
So, an $E_6$ ``gluon" has been decomposed to nine $SU(3)$ quarks, nine anti-quarks, one gluon and 16 singlets. Hence, the Cartan generators structured from the $SU(3)$ decomposition are
\begin{equation} \label{e69}
\begin{aligned}
H_a^{78} =\frac{1}{2 \sqrt{6}} \, \textrm{diag} \Big[& \overbrace{0 , \cdots , 0}^{ 8 \, \textrm{times}} , \overbrace{0 , \cdots , 0}^{ 8 \, \textrm{times}} ,  \lambda_a^8  ,\\
& \overbrace{- (\lambda_a^3)^{\ast} , \cdots , - (\lambda_a^3)^{\ast}}^{9 \, \textrm{times}}  ,\overbrace{\lambda_a^3 , \cdots , \lambda_a^3}^{9 \, \textrm{times}}  \Big], 
\end{aligned}
\end{equation}
Applying the maximum flux condition of Eq.~\eqref{e19} for these Cartan generators, we have
\begin{equation} \label{e70}
\begin{aligned}
& \alpha_{{max}_1}^{78}=4 \pi \sqrt{6}, \\
& \alpha_{{max}_2}^{78}=12 \pi \sqrt{2}. \\
\end{aligned}
\end{equation}
The potential between static color sources using the trivial center element of $E_6$ exceptional gauge group for the fundamental and adjoint representations has been plotted in Fig.~\ref{fig-18}. The screening is visible at large distances while the intermediate parts are linear. The lower diagram shows the linear parts of the potentials in the interval $R \in [3,10]$. we have fitted our data to equation $V(R)=aR+b$. The slope of the potentials have been found to be 0.251(3) and 0.309(5) for the fundamental and adjoint representations, respectively. Therefor, the ratio of the adjoint potential to the fundamental one in this range is 1.23(5). In fact, the ratio of the adjoint potential to the fundamental one starts from $1.384$ which is the Casimir scaling of the adjoint representation \cite{feifer}. But, similar to Fig.~\ref{fig-9}, the adjoint potential ratio differs from the exact value of Casimir scaling at intermediate distances. 

To find what accounts for the intermediate linear potential, one needs to construct a matrix consists of $SU(3)$ center elements with respect to Eqs.~\eqref{e65} and \eqref{e68}:
\begin{equation}  \label{e71}
\begin{aligned}
 \mathbb{Z}_{SU(3)}^{27} &=\textrm{diag} \Big[ 1 , 1 , 1 , 1 , 1 , 1 , 1 , 1 , 1 , 
 z \, \mathbb{I}_{3 \times 3} , z \, \mathbb{I}_{3 \times 3} , \\ &z \mathbb{I}_{3 \times 3} , 
 z^{\ast} \, \mathbb{I}_{3 \times 3} , z^{\ast} \, \mathbb{I}_{3 \times 3} , z^{\ast} \, \mathbb{I}_{3 \times 3} \Big],\\
 \mathbb{Z}_{SU(3)}^{78} &= \textrm{diag} \Big[ \overbrace{ 0 , \cdots , 0}^{8 \, \textrm{times}} , \overbrace{ 0 , \cdots , 0}^{8 \, \textrm{times}} , \mathbb{I}_{8 \times 8} ,  \\
& \overbrace{ z^{\ast} \, \mathbb{I}_{3 \times 3} , \cdots ,  z^{\ast} \, \mathbb{I}_{3 \times 3}}^{9 \, \textrm{times}} , \overbrace{ z\, \mathbb{I}_{3 \times 3} , \cdots ,  z\, \mathbb{I}_{3 \times 3}}^{9 \, \textrm{times}}  \Big].
\end{aligned}
\end{equation}
Similar to $F_4$, the number of $z$ and $z^{\ast}$ vortices are the same in the above matrices. Using Eq.~\eqref{e6}, we have
\begin{equation}  \label{e72}
\begin{aligned}
 \alpha_{{max}_1}^{27 \textrm{-non}}=2 \pi \sqrt{6}, \\
 \alpha_{{max}_2}^{27 \textrm{-non}}=2 \pi \sqrt{2}, \\
\\
\alpha_{{max}_1}^{78 \textrm{-non}}=4 \pi \sqrt{6}, \\
\alpha_{{max}_1}^{78 \textrm{-non}}=4 \pi \sqrt{2} .\\
\end{aligned}
\end{equation}

\begin{figure}
\centering
\includegraphics[width=\linewidth]{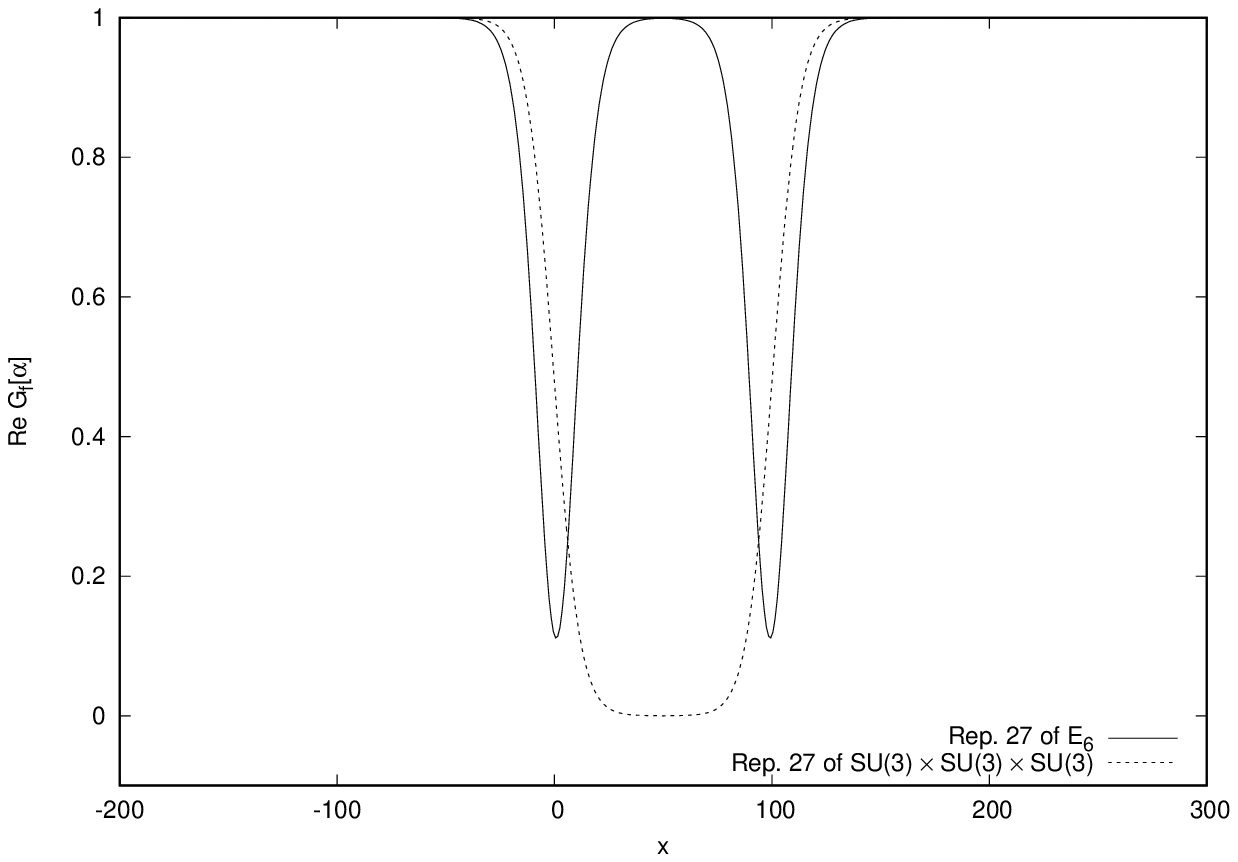}
\includegraphics[width=\linewidth]{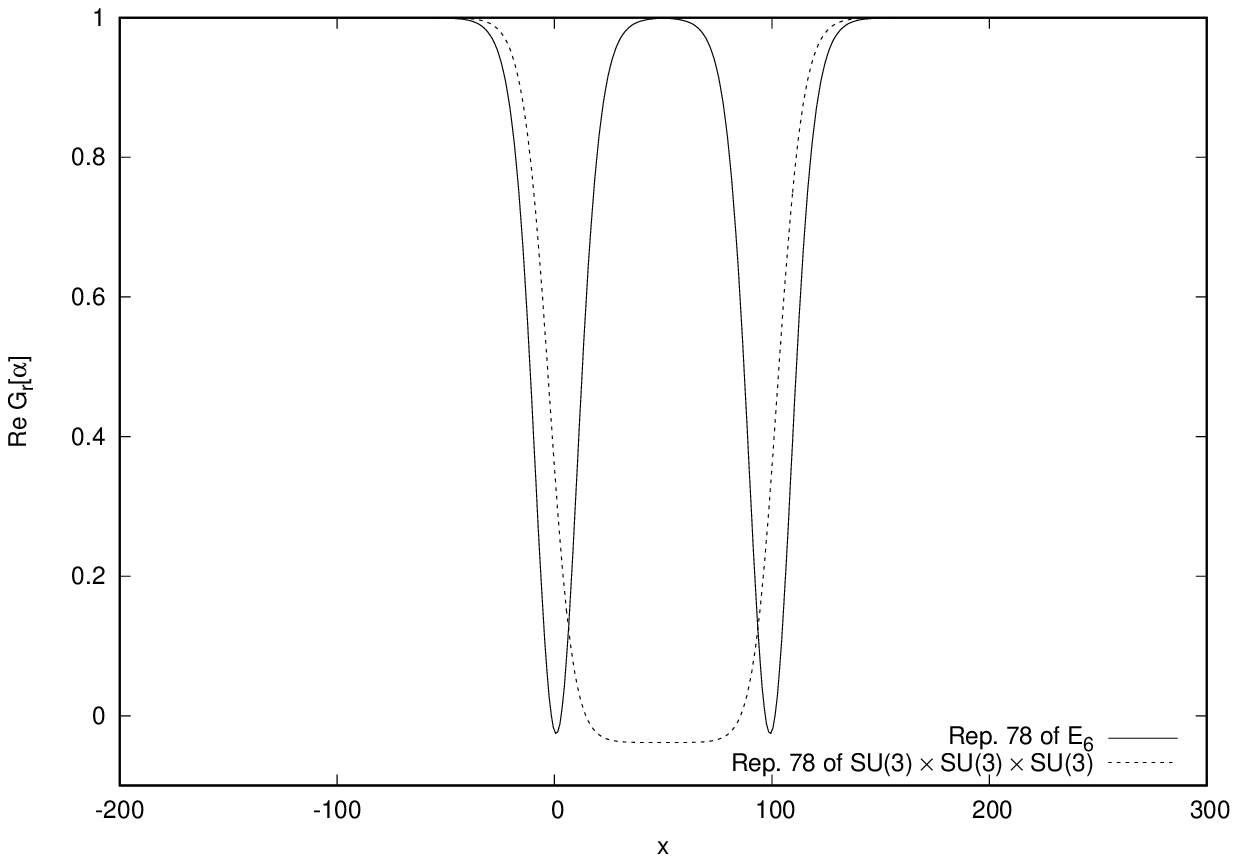}
\caption{ The real part of the group factor function versus the location $x$ of the vacuum domain midpoint, for $R=100$ and in the range $x \in [-200,300]$, for the fundamental and adjoint representation of the $E_6$ (solid lines) in comparison with the same function versus the location $x$ of the vortex midpoint obtained from $ SU(3) \times SU(3) \times SU(3)$ decomposition (dashed lines). The minimum points of the $E_6$ group factor which occur at $x=0$ and $x=100$ reach the amounts $0.111$ and $-0.025$ for the fundamental and adjoint representations, respectively. Whereas, these amounts are approximately $0$ and $-0.038$ for the $SU(3)$ subgroup.}
\label{fig:27-78-su3}       
\end{figure}

Using the above amounts, one might plot the group factor for the fundamental and adjoint representations in Fig.~\ref{fig:27-78-su3} and compare them with the corresponding ones for the $E_6$ . In this figure, we can observe that the minimum values are not the same, . Thus, non-trivial center elements of the $SU(3)$ subgroup are not in charge of confining part
in the trivial potential of $E_6$.
The third possibility to investigate the linearity of the $E_6$ trivial potentials in Fig.~\ref{fig-18}, is the non-trivial center element of the $SU(2)$ subgroups.

\hfill\\
\bm{$SU(2) \times SU(6)  $} \textbf{subgroup}\\
Now, we decompose $E_6$ to {$SU(2) \times SU(6)  $} subgroup and have
\begin{equation} \label{e73}
\begin{aligned}
27 &=(2 , \bar{6}) \oplus (1 , 15), \\
78 &=(3 , 1) \oplus (1 , 35) \oplus (2 , 20).
\end{aligned}
\end{equation}
Then, one can choose the following maximal subgroup of $SU(6)$ to decompose its representations:
\begin{equation} \label{e74}
\begin{aligned}
& SU(6) \supset SU(2) \times SU(4) \times  U(1) \quad (R)\\
& 6=(2 , 1) \oplus (1 , 4), \\
& 15=(1 , 1) \oplus (2 , 4) \oplus (1 , 6), \\
& 20=(1 , 4) \oplus (1 , \bar{4}) \oplus (2 , 6),  \\
& 35=(1 , 1) \oplus (3 , 1) \oplus (1 ,15) \oplus (2 , 4) \oplus (2 , \bar{4}). \\
\end{aligned}
\end{equation}
and for the $SU(4)$,
\begin{equation} \label{e75}
\begin{aligned}
& SU(4) \supset SU(2) \times SU(2) \times U(1) \quad (R)  \\
& 4=(2 , 1) \oplus (1 , 2),  \\
& 6=(1 , 1) \oplus (1 , 1) \oplus (2 , 2), \\
& 15= (1 , 1) \oplus (3 , 1) \oplus (1 , 3) \oplus (2 , 2) \oplus (2 , 2). \\
\end{aligned}
\end{equation}
Finally, we have:
\begin{equation}
\begin{aligned}
 27 =&(2 , 1) \oplus (2 , 1) \oplus (1 , 2) \oplus (2 , 1) \oplus (2 , 1) \oplus (1 , 2) \\
   \oplus &(1 , 1) \oplus (2 , 1) \oplus (1 , 2) \oplus  (2 , 1) \oplus (1 , 2) \oplus (1 , 1) \\
   \oplus &(1 , 1) \oplus (2 , 2),\nonumber
\end{aligned}
\end{equation}
\begin{equation} \label{e76}
\begin{aligned}
 78=&(3 , 1) \oplus (1 , 1) \oplus (3 , 1) \oplus (1 , 1) \oplus  (3 , 1) \oplus  (1 , 3) \\
  \oplus &(2 , 2) \oplus (2 , 2) \oplus (2 , 1)  \oplus  (1 , 2) \oplus  (2 , 1) \oplus (1 , 2)\\
  \oplus &(2 , 1) \oplus  (1 , 2) \oplus  (2 , 1) \oplus (1 , 2) \oplus (2 , 1) \oplus (1 , 2) \\
  \oplus & (2 , 1) \oplus  (1 , 2) \oplus (1 , 1) \oplus (1 , 1) \oplus (2 , 2) \oplus (1 , 1) \\
  \oplus & (1 , 1) \oplus (2 , 2) \oplus (2 , 1) \oplus  (1 , 2) \oplus (2 , 1) \oplus (1 , 2) \\
  \oplus & (1 , 1) \oplus (1 , 1) \oplus  (2 , 2) \oplus (1 , 1) \oplus  (1 , 1) \oplus (2 , 2).  
\end{aligned}
\end{equation}
In Eqs.~\eqref{e74}-\eqref{e76}, the $U(1)$ factor has been ignored. 
Reconstruction of the Cartan generators for the $SU(2)$ subgroup of the $E_6$ and for fundamental and adjoint representations using the decompositions in Eq.~\eqref{e76} are as the following:
\begin{equation} \label{e77}
\begin{aligned}
 H_{SU(2)}^{27}=\frac{1}{\sqrt{6}} \, \textrm{diag} \Big[& 0 , 0 , 0 , 0 , \sigma_3^2 , 0 , 0 ,   0 , 0 , \sigma_3^2 ,  0 , 0 , 0 , \sigma_3^2 ,  \\
&0 , 0 , \sigma_3^2 , 0 , 0 , \sigma_3^2  , \sigma_3^2 \Big], \\
 H_{SU(2)}^{78}=\frac{1}{2 \sqrt{6}} \, \textrm{diag} \Big[& \overbrace{0 , \cdots , 0}^{35 \, \textrm{times}} , \sigma_3^3 ,  \overbrace{\sigma_3^2  , \cdots , \sigma_3^2}^{20 \, \textrm{times}} \Big] .
\end{aligned}
\end{equation}
It should be noted that, for the sake of simplicity, the components of the matrix $H_{SU(2)}^{78}$ are not in order because it does not have any effect on our calculations.

To make a comparison with the potentials obtained from the trivial center element of the $E_6$ exceptional group and its $SU(2)$ subgroup, one can utilize Eq.~\eqref{e77} in Eq.~\eqref{e19} and find:
\begin{equation} \label{e78}
\begin{aligned}
\alpha_{\textrm{max}}^{27}&=4 \pi \sqrt{6},\\
\alpha_{\textrm{max}}^{78} &=8 \pi \sqrt{6}.
\end{aligned}
\end{equation}
Static potentials obtained by these maximum flux values and Cartans of Eq.~\eqref{e77} are identical with the $E_6$ potentials in Fig.~\ref{fig-18}. These results were foreseeable due to the similarity of matrix components of Eq.~\eqref{e77} with the corresponding Cartan generators of $E_6$ in Eqs.~\eqref{e66} and \eqref{e69}.

\begin{figure}
\centering
\includegraphics[width=\linewidth]{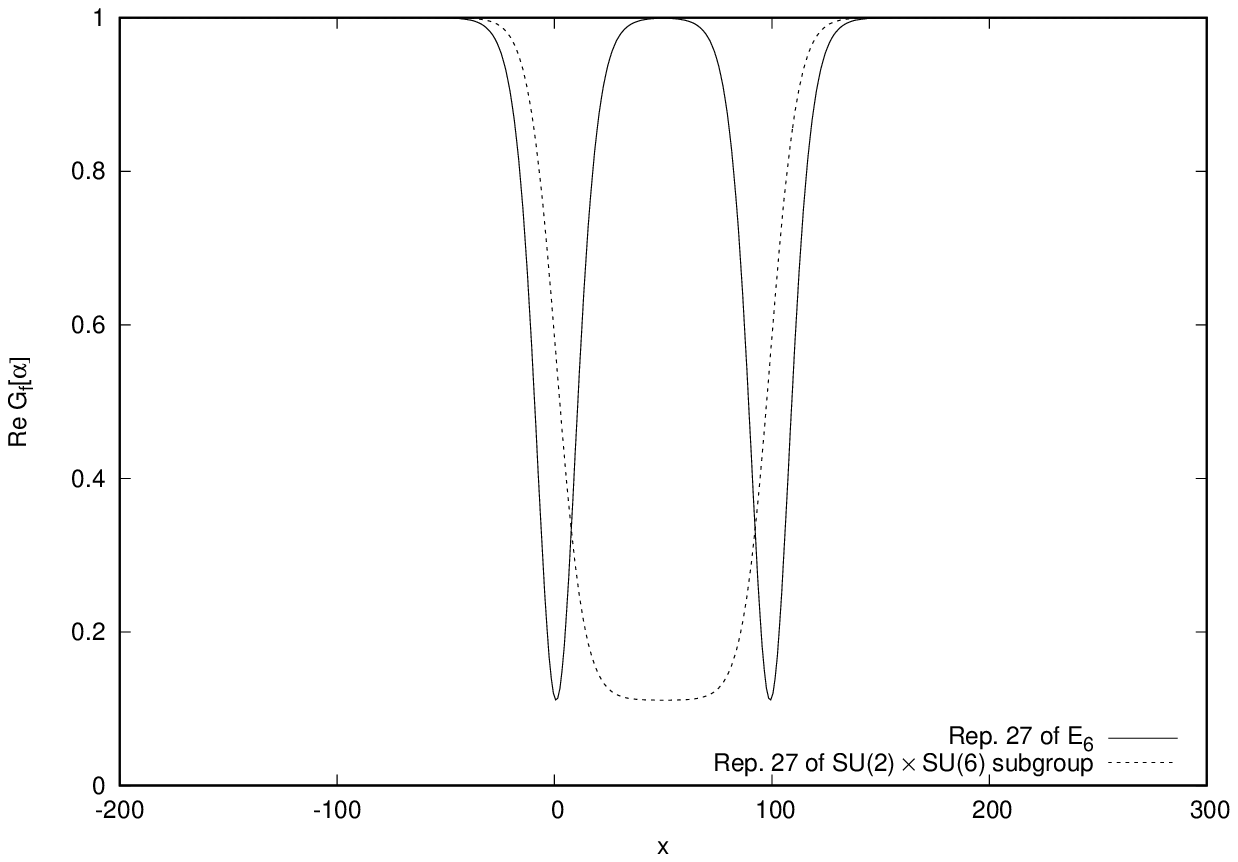}
\includegraphics[width=\linewidth]{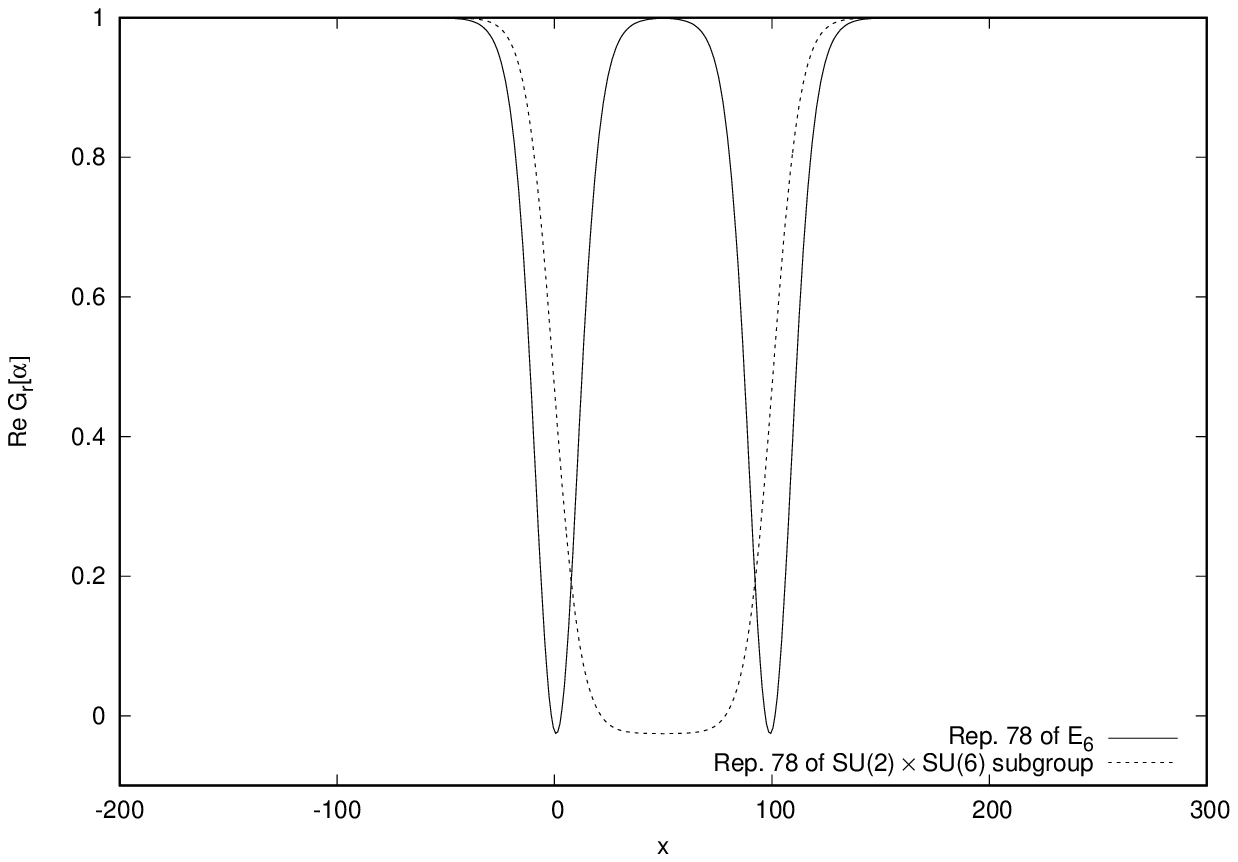}
\caption{The same as Fig.~\ref{fig:27-78-su3} but the dashed lines represent the group factor corresponding to the $SU(2) \times SU(6)$ subgroup of $E_6$. It is clear that in both diagrams, the minimum values are identical.}
\label{fig-19}       
\end{figure}

Matrices made of center elements of the $SU(2)$ subgroup considering Eq.~\eqref{e76} are
\begin{equation}  \label{e79}
\begin{aligned}
\mathbb{Z}_{SU(2)}^{27}= \Big[ &1 , 1 , 1 , 1 , z_1  \mathbb{I}_{2 \times 2} , 1 , 1 , 1 , 1 ,  z_1 , \mathbb{I}_{2 \times 2} ,  1 , 1 , 1 , \\
&z_1  \mathbb{I}_{2 \times 2} , 1 , 1 , z_1  \mathbb{I}_{2 \times 2} ,  1 , 1 , z_1  \mathbb{I}_{2 \times 2} , z_1  \mathbb{I}_{2 \times 2} \Big],   \\
 \mathbb{Z}_{SU(2)}^{78}= \Big[ &\overbrace{1 , \cdots , 1}^{35 \, \textrm{times}} ,  \mathbb{I}_{3 \times3} ,  \overbrace{z_1  \mathbb{I}_{2 \times 2} , \cdots , z_1  \mathbb{I}_{2 \times 2}}^{20 \, \textrm{times}} \Big].   
\end{aligned}
\end{equation}
So, using Eq.~\eqref{e6}, the flux maximum values are
\begin{equation} \label{e81}
\begin{aligned}
\alpha_{max}^{27 \textrm{-non}}&=2 \pi \sqrt{6},\\
\alpha_{max}^{78 \textrm{-non}}&=4 \pi \sqrt{6}.
\end{aligned}
\end{equation}
Using these values, we are able to plot the group factor function for the non-trivial center element of the $SU(2)$ subgroup and compare the results with the same function obtained by the trivial center element of the $E_6$ exceptional group or its $SU(3)$ subgroup. Fig.~\ref{fig-19} depicts this comparison.
It is clear that the minimum points are identical. So, one can conclude that the non-trivial center element of the $SU(2)$ subgroup is responsible for the linearity at intermediate distance scales.

\hfill\\
\bm{$F_4 $} \textbf{subgroup}\\
There is another way to decompose $E_6$ to its subgroup without having a $U(1)$ factor in the final result. For instance, the decomposition chain described below can satisfy our assumption and reconstruct matrices with the same components as in Eq.~\eqref{e77}:
\begin{equation} \label{e82}
\begin{aligned}
& E_6 \supset F_4  \qquad (S) \\
& F_4 \supset SU(2) \times Sp(6)  \quad (R) \\
& Sp(6) \supset SU(2) \times Sp(4)  \quad (R) \\
& Sp(4) \supset SU(2) \times SU(2)  \quad (R) \\
\end{aligned}
\end{equation}
Therefore, we expect the same result as $E_6 \supset SU(2) \times SU(6)$ decomposition.

\begin{figure}
\centering
\includegraphics[width=\linewidth]{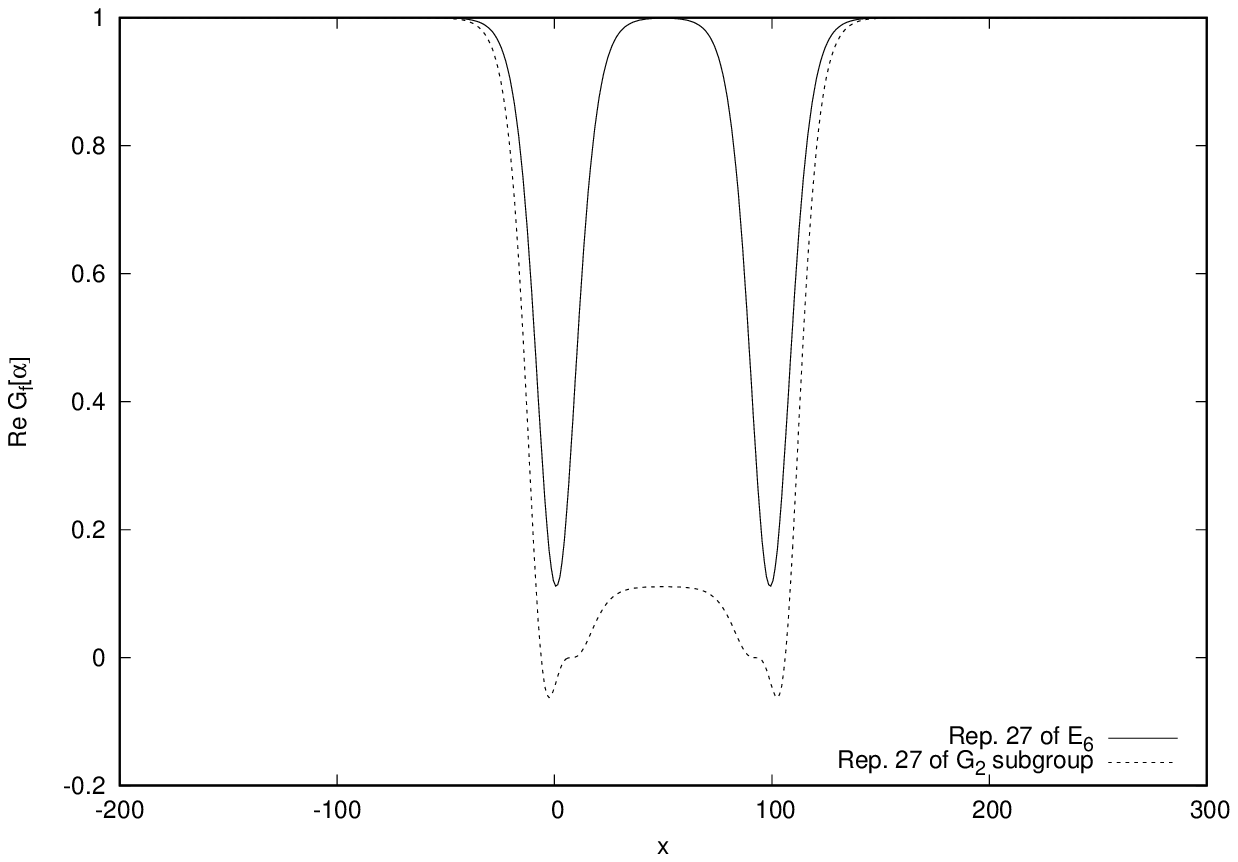}
\includegraphics[width=\linewidth]{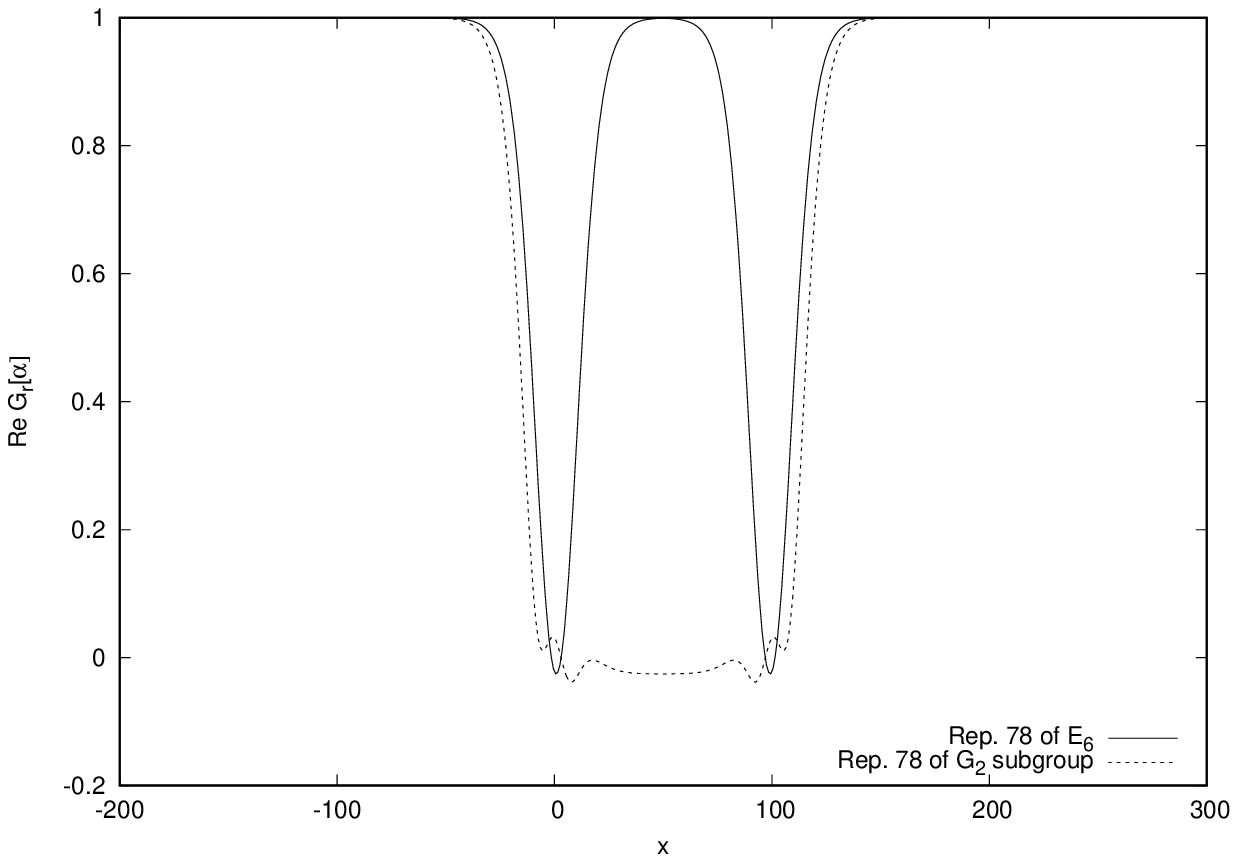}
\caption{The same as Fig.~\ref{fig:27-78-su3} but the dashed lines represent the group factor of the $G_2$ subgroup.}
\label{fig-22}       
\end{figure}

\hfill\\
\bm{$G_2 $} \textbf{subgroup}\\
In the $F_4$ exceptional group case, its singular maximal $SU(2) \times G_2$ subgroup has a property which could not produce the same potential as $F_4$ itself, because its reconstructed Cartan matrices consist of different components from $h_1$ and $h_2$ original Cartan generators of $F_4$. However, they are still able to
produce the same linear part as some of the other $SU(2)$ subgroups of the $F_4$. Here, for the $E_6$, branching $G_2$ singular maximal subgroup into $SU(2) \times SU(2)$ subgroup attributes the same. According to the branching rules, the decomposition for the fundamental representation of
the $E_6$ is as the following:
\begin{equation} \label{e83}
\begin{aligned}
& E_6 \supset G_2 \quad (S) \\
& 27=27.  \\
& 78=14 \oplus 64
\end{aligned}
\end{equation}
Then,
\begin{equation} \label{e84}
\begin{aligned}
 G_2 \supset & SU(2) \times SU(2) \quad (R)  \\
 27 = &(3 , 3) \oplus (2 , 4) \oplus (2 , 2) \oplus   (1 , 5) \oplus (1 , 1),  \\
 14 = &(1 , 3) \oplus (2 , 4) \oplus (3 , 1),\\
 64 = &(4 , 2) \oplus (3 , 5) \oplus (3 , 3) \oplus (2 , 6) \oplus  (2 , 4) \oplus (2 , 2) \oplus \\
&(1 , 5) \oplus (1 , 3).  \\
\end{aligned}
\end{equation}
Therefore,
\begin{equation} \label{e85}
\begin{aligned}
 H_{E_6 \supset G_2 }^{27}&=\frac{1}{3 \sqrt{6}} \, \textrm{diag} \Big[ \sigma_3^3 ,  \sigma_3^3 ,
 \sigma_3^3  ,  \sigma_3^4 , \sigma_3^4 , \sigma_3^2 , \sigma_3^2   ,  \sigma_3^5 , 0 \Big],\\
 H_{E_6 \supset G_2}^{78}&=\frac{1}{6 \sqrt{6}} \, \textrm{diag} \Big[ \sigma_3^3 , \sigma_3^4  , \sigma_3^4 , 0 , 0 , 0 ,  
 \sigma_3^2 ,  \sigma_3^2  , \sigma_3^2  ,  \sigma_3^2  ,  \sigma_3^5 , \\
& \sigma_3^5  ,  \sigma_3^5  , \sigma_3^3  ,  \sigma_3^3 , \sigma_3^3 , \sigma_3^6  ,  \sigma_3^6 , \sigma_3^4  , \sigma_3^4  ,  \sigma_3^2  , \sigma_3^2  ,  \sigma_3^5  , \sigma_3^3  \Big],
\end{aligned}
\end{equation}
and
\begin{equation} \label{e86}
\begin{aligned}
\mathbb{Z}_{E_6 \supset G_2}^{27}= \textrm{diag}\Big[& \mathbb{I}_{ 3  \times 3 } , \mathbb{I}_{ 3  \times 3 } , \mathbb{I}_{ 3  \times 3 }  , z_1 \mathbb{I}_{ 4  \times 4 } , z_1 \mathbb{I}_{ 4  \times 4 } ,\\
&   z_1 \mathbb{I}_{ 2  \times 2 } , z_1 \mathbb{I}_{ 2  \times 2 } , \mathbb{I}_{ 5  \times 5 } , 1 \Big],\\
 \mathbb{Z}_{E_6 \supset G_2 }^{78}= \textrm{diag} \Big[& \mathbb{I}_{3 \times 3} , z_1 \, \mathbb{I}_{4 \times 4}  ,   z_1 \, \mathbb{I}_{4 \times 4} ,  1  , 1  , 1  ,  \\
& z_1 \, \mathbb{I}_{2 \times 2} , z_1 \, \mathbb{I}_{2 \times 2} , z_1 \, \mathbb{I}_{2 \times 2} ,  
 z_1 \, \mathbb{I}_{2 \times 2} ,   \\
&\mathbb{I}_{5 \times 5} , \mathbb{I}_{5 \times 5} , \mathbb{I}_{5 \times 5} , \mathbb{I}_{3 \times 3}  ,  
 \mathbb{I}_{3 \times 3} ,  \mathbb{I}_{3 \times 3} ,\\
& z_1 \, \mathbb{I}_{6 \times 6} , z_1 \, \mathbb{I}_{6 \times 6} ,  z_1 \, \mathbb{I}_{4 \times 4} ,  
 z_1 \, \mathbb{I}_{4 \times 4}  ,\\
& z_1 \, \mathbb{I}_{2 \times 2} , z_1 \, \mathbb{I}_{2 \times 2}  ,  \mathbb{I}_{5 \times 5} ,  \mathbb{I}_{3 \times 3}  \Big],
\end{aligned}
\end{equation}
The flux condition of Eq.~\eqref{e19} gives:
\begin{equation}
\begin{aligned}
\alpha_{max}^{27 \textrm{-non}}= 6 \pi \sqrt{6} ,\\
\alpha_{SU(2) \textrm{-max}}^{78 \textrm{-non}}=12 \pi \sqrt{6}.
\end{aligned}
\label{e87}
\end{equation}
Fig.~\ref{fig-22} shows the group factor obtained from this decomposition versus the location $x$ of the vortex midpoint. It is observed that the amount of the group factor when the vortex is completely inside the Wilson loop, equals to the minimum values of the $E_6$ group factor. Therefore, this decomposition is able to describe $E_6$ temporary confinement.
It should be pointed out that the number of center elements emerge in the center element matrices of Eqs.~\eqref{e79} and \eqref{e85} are the same. So, the similar argument as $SU(2) \times G_2$ subgroup of $F_4$ could be applied here.

\subsection{\label{sec:g2}$G_2$ exceptional group}

$ G_2 $ is the simplest exceptional group with rank 2 likewise $SU(3)$. All of its representations are real and it is its own universal covering group. Despite $F_4$ and $E_6$ exceptional groups which do not have numerical supports yet, pending future investigations, there are lattice calculations are in favor of $G_2$ exceptional gauge group \cite{holland,pepe,elia,maas,olej,wipf1,wipf2}.

In references \cite{deldar1,deldar2,deldar3}, the static potentials of the $G_2$ exceptional gauge group has been investigated and the dominant role of non-trivial center elements of its $SU(2)$ and $SU(3)$ subgroups on the intermediate confinement has been studied. In this research, we are going to insert our
generalized method to calculate the potentials of higher representations of $G_2$, as well.

To begin, one needs the original Cartan generators of the $G_2$ exceptional group to simulate the static potential using the vacuum domain structure model. The Cartan generators of the $G_2$ in the fundamental 7-dimensional representation are as the following \cite{deldar1,ekins}:
\begin{equation}
\begin{aligned}
& h_1^7=\frac{1}{2 \sqrt{2}} \, \textrm{diag} \Big[ +1 , -1 , 0 , 0 , -1 , +1 , 0  \Big], \\
& h_2^7=\frac{1}{2 \sqrt{6}} \, \textrm{diag} \Big[ +1 , +1 , -2 , 0 , -1 ,   -1 , +2 \Big].  
\label{e88}
\end{aligned}
\end{equation}
These matrices are normalized with Eq.~\eqref{e17} condition. Plotting the potential of Eq.~\eqref{e10} requires the group factor in Eq.~\eqref{e2} and the flux profile in Eq.~\eqref{e7}. In order to compute the maximum value of the flux profile, one has to apply the trivial flux condition in Eq.~\eqref{e19} and solve three
independent equations:
\begin{equation}
\begin{aligned}
\exp[\frac{ \alpha_{\textrm{max}_1}^{7}}{2\sqrt{2}}+\frac{ \alpha_{\textrm{max}_2}^{7}}{2\sqrt{6}}]=\mathbb{I},\\
\exp[\frac{- \alpha_{\textrm{max}_1}^{7}}{2\sqrt{2}}+\frac{ \alpha_{\textrm{max}_2}^{7}}{2\sqrt{6}}]=\mathbb{I},\\
\exp[\frac{- \alpha_{\textrm{max}_2}^{7}}{\sqrt{6}}]=\mathbb{I},
\end{aligned}
\label{e89}
\end{equation}
to find
\begin{equation}
\begin{aligned}
  \alpha_{\textrm{max}_1}^{7}&=2 \pi \sqrt{2}, \\
  \alpha_{\textrm{max}_2}^{7}&=2 \pi \sqrt{6}. 
\end{aligned}
\label{e90}
\end{equation}
It can be easily shown that the first Cartan generator $ h_1^7$ in Eq.~\eqref{e88}, with the maximum flux value of $\alpha_1^{max}=4 \pi \sqrt{2} $ is capable to produce the whole $G_2$ potential individually, without using $ h_2^7$. In the next stage, we are going to calculate reconstructed Cartan generators in $7$, $14$, $27$, $64$, $77$ and $77^{\prime} $-dimensional representations from the decomposition of the $G_2$ to its $SU(3)$ subgroup. 

\hfill\\
\bm{$SU(3) $} \textbf{subgroup}\\
The decomposition of the fundamental representation to the $SU(3)$ subgroup is as the following \cite{sorba,slansky}:
\begin{equation}
\begin{aligned}
& G_2 \supset SU(3) \quad (R) \\
& 7=3 \oplus \bar{3} \oplus 1.  
\end{aligned}
\label{e91}
\end{equation}
The reconstructed Cartan generator with respect to this decomposition is,
\begin{equation}
\begin{aligned}
H_{a}^{7}= \frac{1}{\sqrt{2}} \, \textrm{diag} \Big[ \lambda_a^3  , -(\lambda_a^3)^{\ast}  , 0 \Big],
\end{aligned}
\label{e92}
\end{equation}
with $a=3, 8$. It is clear that the decomposition of this representation into the $SU(3)$ subgroup results in the same matrices as Eq.~\eqref{e88}. This fact is similar to the $F_4$ and $E_6$ exceptional groups which one could reproduce the group potentials by applying their $SU(3)$ subgroups. This matter
enables us to calculate the potentials of higher representations by taking the same procedure.
The decomposition of the higher representations of $G_2$ into the $SU(3)$ subgroup is \cite{slansky}
\begin{equation}
\begin{aligned}
14 &=3 \oplus \bar{3} \oplus 8,\\
27 &=8 \oplus 6 \oplus \bar{6} \oplus 3 \oplus \bar{3} \oplus 1,\\
64 &=15 \oplus \overline{15} \oplus 8 \oplus 8  \oplus 6 \oplus \bar{6} \oplus 3 \oplus \bar{3},\\
77 &=27 \oplus 15 \oplus \overline{15} \oplus 8 \oplus 6 \oplus \bar{6},\\
77^{\prime} &=15 \oplus \overline{15} \oplus 10 \oplus \overline{10} \oplus 8  \oplus 6 \oplus  \bar{6} \oplus 3 \oplus \bar{3}\oplus 1.\\
\end{aligned}
\label{e93}
\end{equation}
So, the corresponding Cartan generators are decomposed as the following:
\begin{equation}
\begin{aligned}
H_a^{14} = \frac{1}{\sqrt{8}} \, \textrm{diag} \Big[ &\lambda_a^3  , -(\lambda_a^3)^{\ast} , \lambda_a^8  \Big],\\
H_a^{27} =\frac{1}{\sqrt{18}} \, \textrm{diag} \Big[&\lambda_a^8 , \lambda_6^6 , -(\lambda_a^6)^{\ast} ,  \lambda_a^3 , -(\lambda_a^3)^{\ast} , 0 \Big],\\
 H_a^{64} =\frac{1}{8} \, \textrm{diag} \Big[&\lambda_a^{15} , -(\lambda_a^{15})^{\ast}  ,  \lambda_a^8  , 
 \lambda_a^8 , \lambda_a^6 , -(\lambda_a^6)^{\ast} ,  \lambda_a^3 ,\\ &-(\lambda_a^3)^{\ast}  \Big],\\
 H_a^{77} =\frac{1}{\sqrt{110}} \, \textrm{diag} \Big[ &\lambda_a^{27}  ,  \lambda_a^{15} ,   
 -(\lambda_a^{15})^{\ast}  ,  \lambda_a^8 , \lambda_a^6 , -(\lambda_a^6)^{\ast}  \Big],\\
 H_a^{77^{\prime}} =\frac{1}{2  \sqrt{22}} \, \textrm{diag} \Big[ &\lambda_a^{15} , -(\lambda_a^{15})^{\ast}  ,  \lambda_a^{10}  ,    -(\lambda_a^{10})^{ \ast}  , \lambda_a^8  ,  \\
&\lambda_a^6 , -(\lambda_a^6)^{\ast} ,  \lambda_a^3 , -(\lambda_a^3)^{\ast}  , 0 \Big],
\label{e94}
\end{aligned}
\end{equation}
where the upper indices of the $\lambda_a$s indicate the dimensions of the $SU(3)$ representations. The maximum flux values extracted from Eq.~\eqref{e19} are:
\begin{equation}
\begin{aligned}
& \alpha_{\textrm{max}_1}^{14}=2 \pi \sqrt{8}, ~~~~~\alpha_{\textrm{max}_2}^{14}=2 \pi \sqrt{24},\\
& \alpha_{\textrm{max}_1}^{27}=6 \pi \sqrt{2}, ~~~~~\alpha_{\textrm{max}_2}^{27}=6 \pi \sqrt{6},\\
& \alpha_{\textrm{max}_1}^{64}=16 \pi,~~~~~~~~ \alpha_{\textrm{max}_2}^{64}=16 \pi \sqrt{3},  \\
& \alpha_{\textrm{max}_1}^{77}=2 \pi \sqrt{110}, ~~ \alpha_{\textrm{max}_2}^{77}=2 \pi \sqrt{330},  \\
& \alpha_{\textrm{max}_1}^{77^{\prime}}=4 \pi \sqrt{22}, ~~~~\alpha_{\textrm{max}_2}^{77^{\prime}}=4 \pi \sqrt{66}.  \\
\end{aligned}
\label{e95}
\end{equation}

\begin{figure}
\centering
\includegraphics[width=\linewidth]{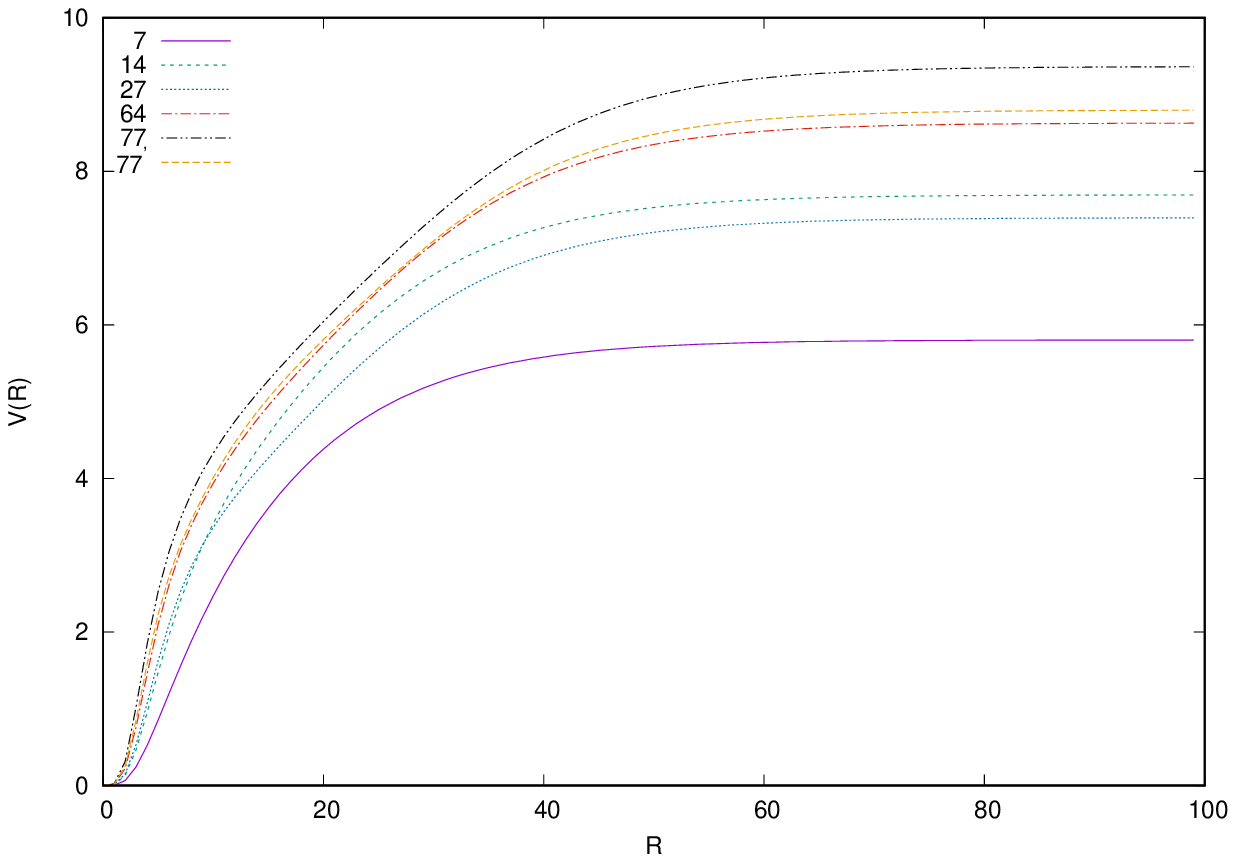}
\includegraphics[width=\linewidth]{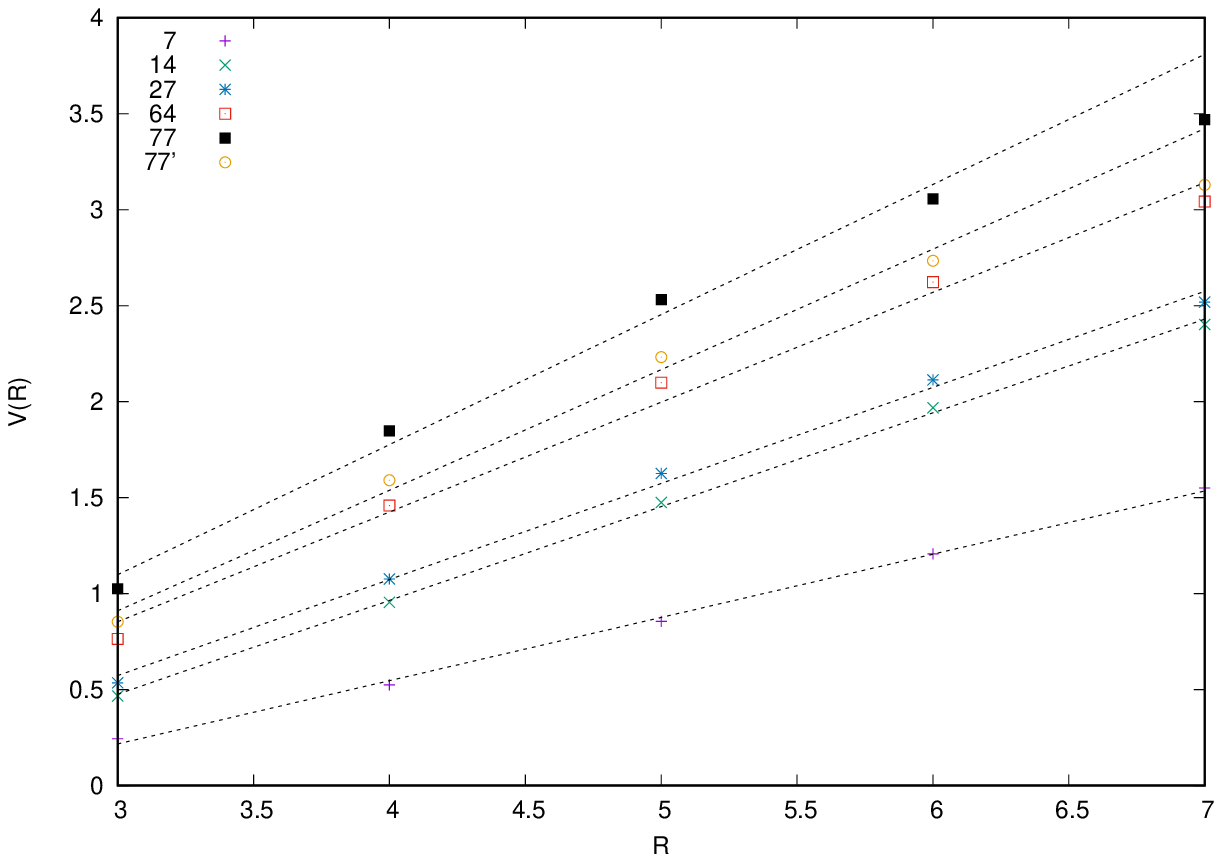}
\caption{Upper diagram: Static potentials of the $G_2$ exceptional group for the fundamental, adjoint, $27$, $64$, $77$ and $77^{\prime}$ representations in the range $R \in [1,100]$. All potentials are screened at far distances and are linear at intermediate distance scales. Lower diagram: Linear parts of the potentials in the range $[3,7]$. The slope of the potentials have been given in the forth column of Tab.~\ref{tab6}. }
\label{fig-23}      
\end{figure}

\begin{figure}
\centering
\includegraphics[width=\linewidth]{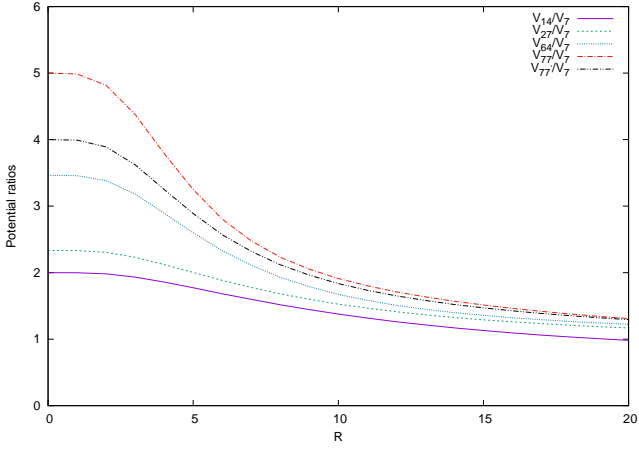}
\caption{Potential ratios in different representations of $G_2$ exceptional group which launch at exact Casimir ratios but deviate abruptly at farther distance ranges.}
\label{fig-24}       
\end{figure}

The trivial static potentials of Eq.~\eqref{e10} for the fundamental, adjoint, $27$, $64$, $77$ and $77^{\prime}$-dimensional representations have been plotted in Fig.~\ref{fig-23}.
Screening is observed for all representations, which is a consequence of adjoint gluons which pop out of the vacuum due to high energies and screen the initial color sources. The tensor products of all representations, when they create a singlet, is an implication of this phenomenon:
\begin{equation}
\begin{aligned}
&7 \times 14 \times 14 \times 14 = \mathbf{1} \oplus 10 (7) \oplus 6 (14) \oplus \cdots ,  \\
&14 \times 14 = \mathbf{1} \oplus 14 \oplus 27 \oplus \cdots   ,  \\
&27 \times 14 \times 14 = \mathbf{1} \oplus 3 (7) \oplus 3 (14) \oplus \cdots   ,  \\
&64 \times 14 \times 14 \times 14 = 2 (\mathbf{1}) \oplus 20 (7) \oplus \cdots  ,  \\
&77 \times 14 \times 14 = \mathbf{1} \oplus 2 (7)  \oplus 4 (14) \oplus \cdots   ,  \\
&77^{\prime} \times 14 \times 14 = \mathbf{1} \oplus 3 (14)  \oplus 3 (27) \oplus \cdots . 
\end{aligned}
\label{e96}
\end{equation}
Lower diagram of Fig.~\ref{fig-23} shows the linear parts of the potentials in the range $R \in [3,7]$. The slope of the linear potentials and the potential ratios (${\frac{k_r}{k_F}} $) have been depicted in the forth and fifth columns of tab.~\ref{tab6}, respectively. Comparing ${\frac{k_r}{k_F}} $ values with the values of  ${\frac{C_r}{C_F}} $ represents that potentials are in agreement with Casimir scaling qualitatively.
\begin{table}
\caption{Second column represents Casimir numbers of several representations of the $G_2$ exceptional group \cite{olej,feifer,wipf1,wipf2}. Casimir ratios, slope of the potentials obtained from the lower diagram of Fig.~\ref{fig-23} and the potential ratios have been given in the third, forth and fifth columns, respectively. The numbers in the parentheses indicate the fit error. }
\begin{center}
{ \setlength{\extrarowheight}{3pt}
\begin{tabular}{ | c | c | c | c | c|}
\hline
\footnotesize{ \textbf{Rep.}} &  \footnotesize{\textbf{Casimir Numbers}}  & $ \mathbf{\frac{C_r}{C_F}} $ & Potential slope & $ \mathbf{\frac{k_r}{k_F}} $\\
\hline
7  &  $ \frac{1}{2}  $  &   1  & 0.329(8) & 1\\ 
\hline
14  &  $ 1 $  &  2  & 0.488(8) & 1.48(1)\\
\hline
27 & $ \frac{7}{6} $  & 2.33 & 0.50(2) & 1.52(1)\\
\hline
64 & $ \frac{7}{4}  $ &   3.5 & 0.57(3) & 1.74(3)\\
\hline
$ 77^{\prime} $ & $2$ & 4 & 0.63(4) & 1.91(4)\\
\hline
77 & $ \frac{5}{2}  $  & 5 & 0.68(5) & 2.06(5)\\
\hline
\end{tabular}
}
\end{center}
\label{tab6}
\end{table} 
The point by point ratios of the Potentials have been plotted in Fig.~\ref{fig-24} in the range $R \in [1,20]$. The potential ratios start at accurate Casimir ratios,. However, they plummet considerably at larger distances of $R$. In fact, the potential ratios almost reach a plateau at $ R \to \infty $.
To investigate the reason why the potentials are linear at intermediate distances, we study the effects of the subgroups of $G_2$.

The $G_2$ exceptional group owns three direct maximal subgroups which have been presented in Tab.~\ref{tab2}. The same as $F_4$ and $E_6$ exceptional groups, the center elements of $SU(3)$ subgroup are not a direct cause of intermediate confining potentials of $G_2$ in several representations. Hence, we do not give the detailed calculation for this subgroup, as the results are the same as in $F_4$ and $E_6$. So, we study the other subgroup of $G_2$. It should be mentioned that $SU(2) \times SU(2)$ is a regular subgroup. Hence, it proves that these features do not appear exclusively for singular maximal subgroups.

\begin{figure} 
\centering
\includegraphics[width=\linewidth]{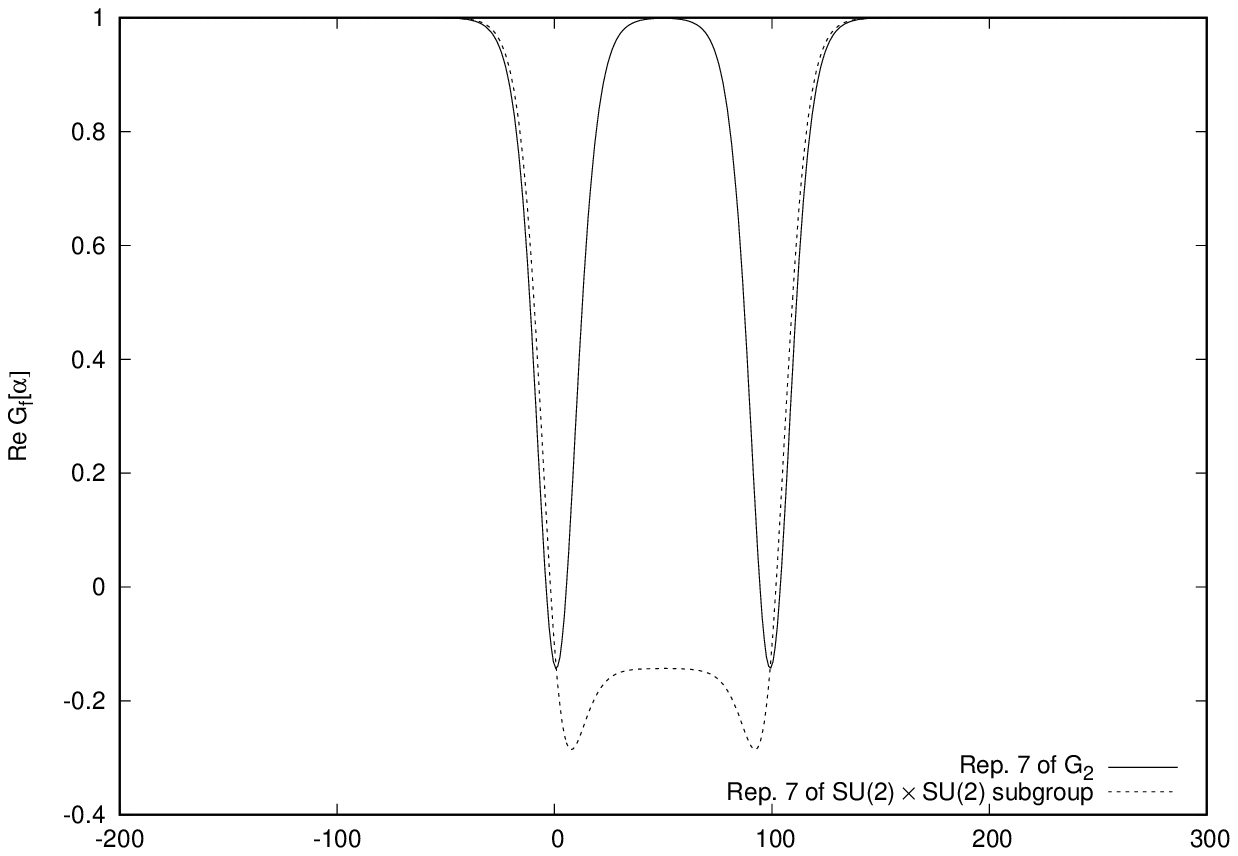}
\includegraphics[width=\linewidth]{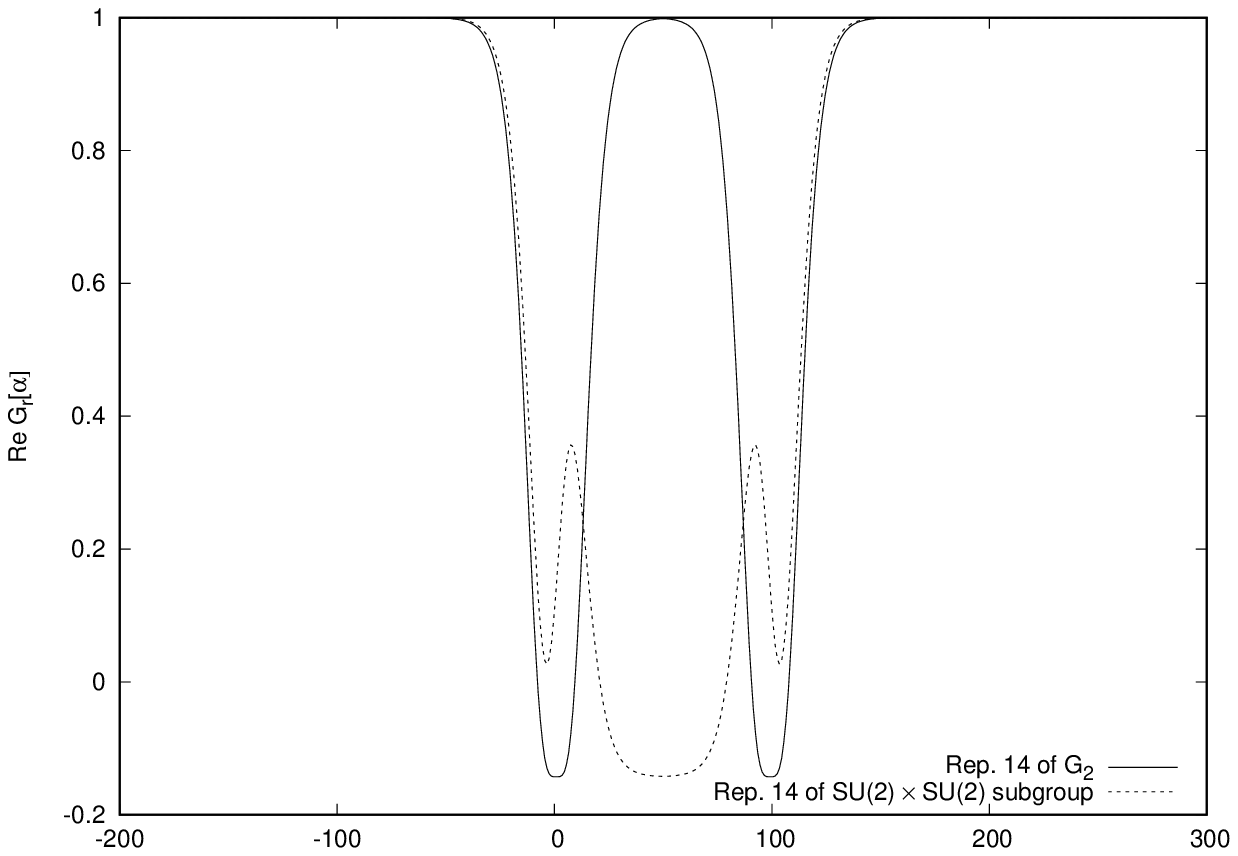}
\caption{ The real part of the group factor function versus the location $x$ of the vacuum domain midpoint, for $R=100$ and in the range $x \in [-200,300]$, for the fundamental and adjoint representation of the $G_2$ (solid lines) in comparison with the same function versus the location $x$ of the vortex midpoint obtained from $ SU(2) \times SU(2) $ decomposition (dashed lines). The minimum points of the $G_2$ group factor which occur at $x=0$ and $x=100$ reach the amounts $-0.142$ and $-0.143$ for the fundamental and adjoint representations, respectively. }  
\label{fig-26} 
\end{figure}

\hfill\\
\bm{$SU(2) \times SU(2) $} \textbf{subgroup}\\
Using this subgroup, the fundamental and adjoint representations could be decomposed as follows \cite{sorba,slansky}:
\begin{equation}
\begin{aligned}
G_2 & \supset SU(2) \times SU(2) \quad (R) \\
 7 &=(2 , 2) \oplus (1 , 3), \\
14 &=(1 , 3)  \oplus (3 , 1) \oplus (2 , 4).\\
\end{aligned}
\label{e97}
\end{equation}
Therefore, the reconstructed diagonal matrices for the fundamental and adjoint representations of the $G_2$ with respect to the $ SU(2) \times SU(2)$ subgroup are 
\begin{equation}\label{e98}
\begin{aligned}
H_{SU(2)}^{7} &=\frac{1}{\sqrt{6}} \, \textrm{diag} \Big[ \sigma_3^2  , \sigma_3^2 , \sigma_3^3  \Big],\\
H_{SU(2)}^{14}&= \frac{1}{2 \sqrt{6}} \, \textrm{diag} \Big[ \sigma_3^3 , 0 , 0 , 0 , \sigma_3^4  , \sigma_3^4 \Big].
\end{aligned}
\end{equation}
It is seen that the elements of these matrices are not identical with $H_3^7$ and $H_3^{14}$ in Eqs.~\eqref{e91} and \eqref{e93}, respectively. Therefore, the trivial potentials obtained from this decomposition are not the same as the original ones for the $G_2$. Nevertheless, the center element matrix of the $SU(2)$ subgroup is
\begin{equation}\label{e99}
\begin{aligned}
\mathbb{Z}_{SU(2)}^{7} &=\Big[ z_1 \, \mathbb{I}_{2 \times 2} ,  z_1 \, \mathbb{I}_{2 \times 2} , \mathbb{I}_{3 \times 3}  \Big],\\
\mathbb{Z}_{SU(2)}^{14}&= \Big[  \mathbb{I}_{3 \times 3}  , 1 , 1 , 1  , z_1 \, \mathbb{I}_{4 \times 4} ,  z_1 \, \mathbb{I}_{4 \times 4}   \Big].
\end{aligned}
\end{equation}
So, if one uses the non-trivial maximum flux condition in Eq.~\eqref{e6}, the non-trivial maximum flux values are
\begin{equation}
\begin{aligned}\label{e100}
\alpha_{\textrm{max-SU(2)}}^{7\textrm{-non}} &=2 \pi \sqrt{6},\\
\alpha_{\textrm{max-SU(2)}}^{14 \textrm{-non}}&=2 \pi \sqrt{24}.
\end{aligned}
\end{equation}
The group factor function of the fundamental and adjoint representations obtained from this decomposition have been illustrated in Fig.~\ref{fig-26} as well as the corresponding ones for the $G_2$. The detailed calculation for the higher representations has been given in Appendix~\ref{appc}.
In this figure, the minimum points of the $G_2$ group factor which occur at $x=0$ and $x=100$, reach the values $-0.142$ and $-0.143$ for the fundamental and adjoint representations, respectively.  The corresponding group factors of the $SU(2) \times SU(2)$ subgroup reach the same amounts at $x=50$ . Therefore, similar to $ F_4 $ and $ E_6 $ cases, the non-trivial center of the $ SU(2)$ subgroup induces temporary confinement in $ G_2 $ exceptional group. 

Now, we go one step further and decompose the $SU(3)$ subgroup into its $SU(2)$ subgroup. This decomposition enables us to give a comprehensive conclusion out of this work. 

\begin{figure} 
\centering
\includegraphics[width=\linewidth]{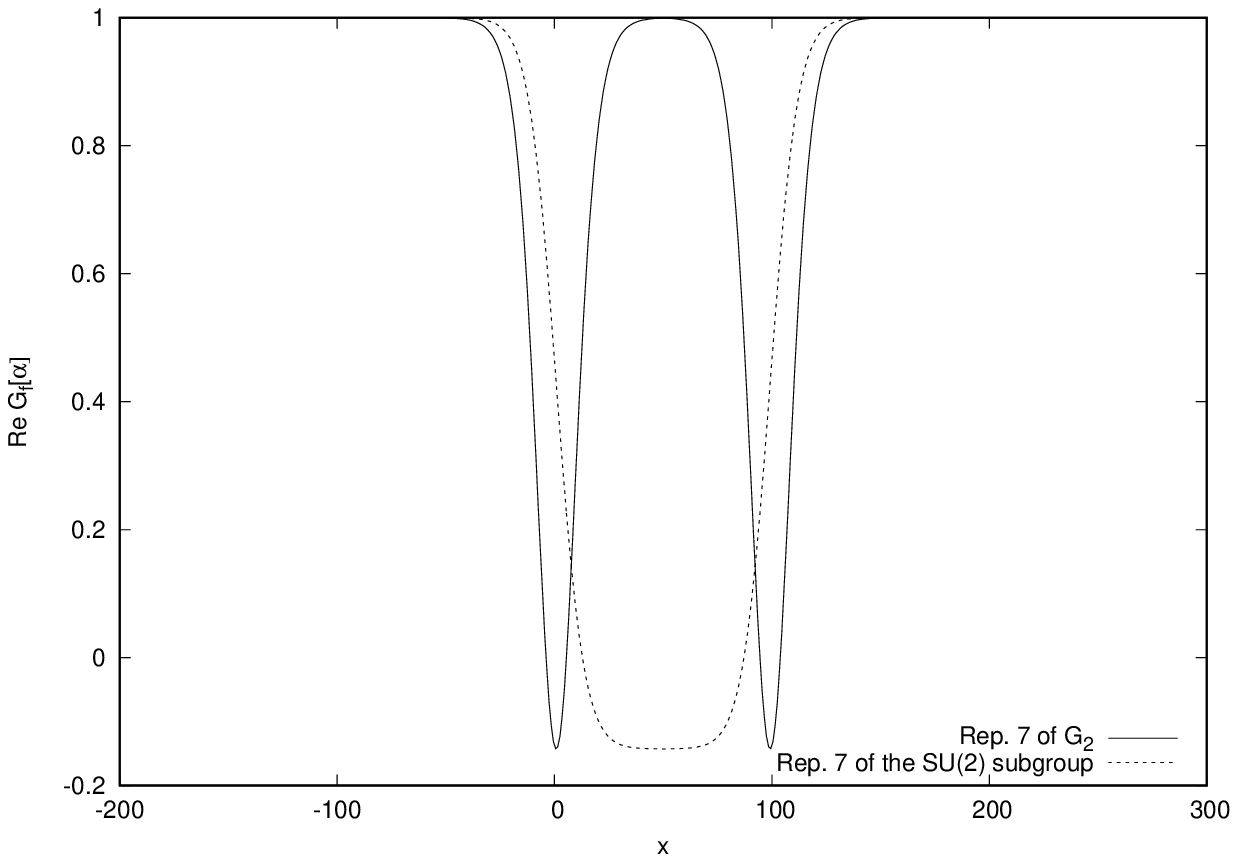}
\includegraphics[width=\linewidth]{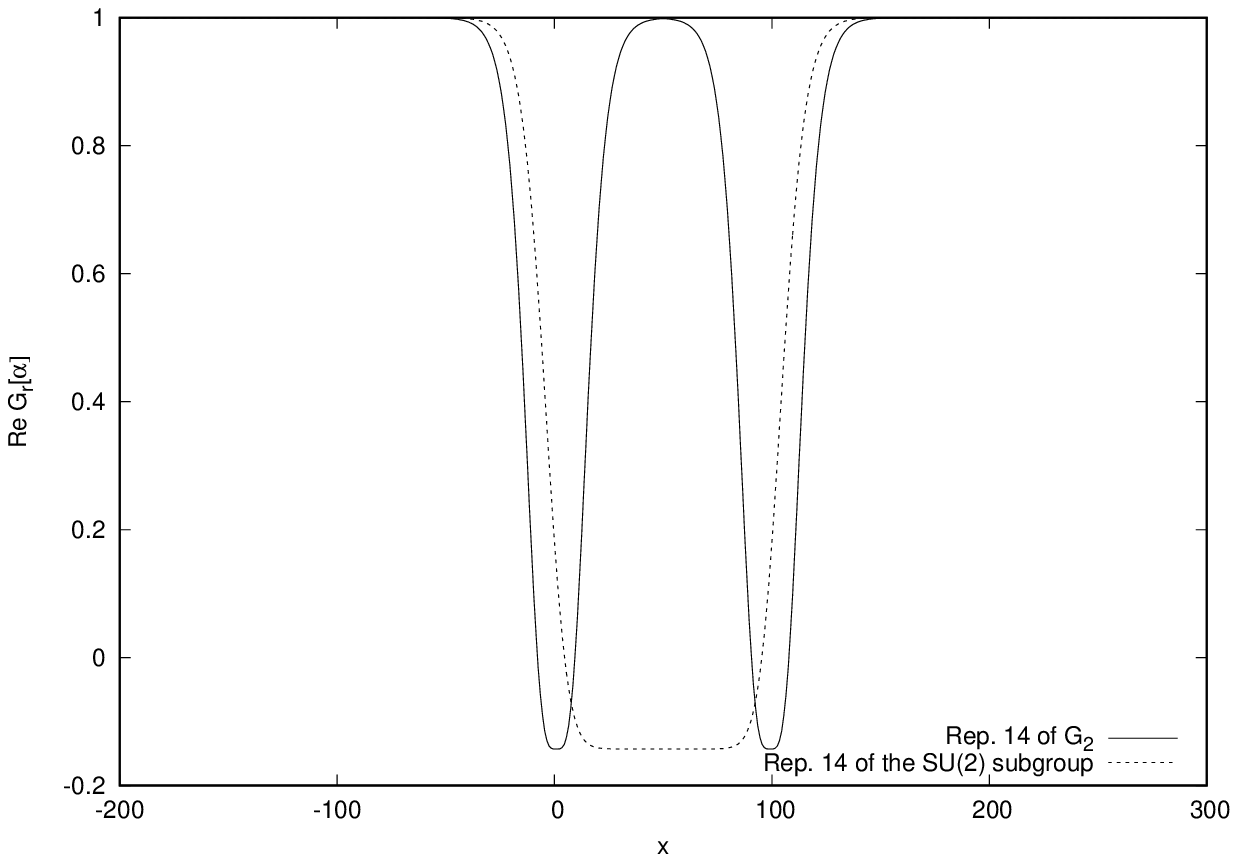}
\caption{The same as Fig.~\ref{fig-26} but the dashed lines represent the group factor for the $SU(3) \supset SU(2) \times U(1)$ decomposition. }  
\label{fig:g2-su3-su2} 
\end{figure}

\hfill\\
\bm{$SU(3) \supset SU(2) \times U(1)$} \textbf{subgroup}\\
The decomposition of the fundamental and adjoint representations of the $G_2$ to the $SU(3)$ subgroup are
\begin{equation}\label{e101}
\begin{aligned} 
& G_2 \supset SU(3) \\
& 7 =3 \oplus \bar{3} \oplus 1,  \\
& 14 =3 \oplus \bar{3} \oplus 8.
\end{aligned}
\end{equation}
In the next step:
\begin{equation}\label{e102}
\begin{aligned}
& SU(3) \supset SU(2)  \times U(1) \quad \textrm{(R)} \\
& 3 =2 \oplus 1, \\
& 8 = 3 \oplus 2 \oplus 2 \oplus 1. \\
\end{aligned}
\end{equation}
It should be recalled that the $U(1)$ factor has been ignored in these decompositions. Ultimately, one could have
\begin{equation}\label{e103}
\begin{aligned}
 7 &= 2 \oplus 1 \oplus 2 \oplus 1 \oplus 1 , \\
 14&=  2 \oplus 1 \oplus 2 \oplus 1 \oplus 3 \oplus 2 \oplus 2 \oplus 1.  \\
\end{aligned}
\end{equation}
Using the above decompositions, one is able to reconstruct the Cartan matrices for the fundamental and adjoint representations as the following:
\begin{equation}
\begin{aligned}
H_{SU(3) \supset SU(2)}^{7} &=\frac{1}{\sqrt{2}}\, \textrm{diag} \left[ \sigma_{3}^{2} , 0 , \sigma_{3}^{2} , 0 , 0 \right],  \\
H_{SU(3) \supset SU(2)}^{14}&=\frac{1}{2 \sqrt{2}}\, \textrm{diag} \left[ \sigma_{3}^{2} , 0 , \sigma_{3}^{2} , 0 , \sigma_{3}^{3} , \sigma_{3}^{2} , \sigma_{3}^{2} , 0 \right].
\end{aligned}
\label{e104}
\end{equation}
We are going to investigate the role of this decomposition in the intermediate linear potentials of $G_2$. So, the matrices of center elements are calculated:
\begin{equation}
\begin{aligned}
\mathbb{Z}_{SU(3) \supset SU(2)}^{7}=\textrm{diag} [ & z_1 \, \mathbb{I}_{2 \times 2} , 1 , z_1 \, \mathbb{I}_{2 \times 2} , 1 , 1 ],\\
\mathbb{Z}_{SU(3) \supset SU(2)}^{14}=\textrm{diag} [ &  z_1 \, \mathbb{I}_{2 \times 2} , 1 , z_1 \, \mathbb{I}_{2 \times 2} , 1 , \mathbb{I}_{3 \times 3}, z_1 \, \mathbb{I}_{2 \times 2} , \\ 
&z_1 \, \mathbb{I}_{2 \times 2} , 1 ].
\end{aligned}
\label{e105}
\end{equation}
Using the maximum flux condition in Eq.~\eqref{e6}, we find:
\begin{equation}
\begin{aligned}
\alpha_{max}^{7\textrm{-non}}&= 2 \pi \sqrt{2},\\
\alpha_{max}^{14 \textrm{-non}}&= 4  \pi \sqrt{2}.
\end{aligned}
\label{e106}
\end{equation}
The group factor function of the fundamental and adjoint representations obtained from this decomposition have been illustrated in Fig.~\ref{fig:g2-su3-su2} as well as the corresponding ones for the $G_2$. It is observed that in each diagram the minimum values of two graphs are identical. Hence, $SU(2)$ gauge group has a dominant role in the linear part of the trivial potentials of the exceptional gauge groups.

\section{conclusion}
\label{sec6}
In this article, we have presented a generalized scenario whereby the static potentials in different representations of exceptional gauge groups could be calculated by means of their unit center elements in the framework of the vacuum domain structure model. Although $G_2$ and $F_4$ exceptional groups do not possess any non-trivial center elements and confinement is
not expected, linear potential is observed for all representations at intermediate distances. This fact is also correct for the $E_6$ exceptional gauge group, when one uses only trivial center element in the calculation. In
addition, to calculate these types of trivial potentials, there is no need to use all Cartan generators of the gauge group. For example, concerning $G_2$, $F_4$ and $E_6$ exceptional groups, it seems adequate to consider only their first Cartan generator which are $h_1^7$, $h_1^{26}$ and $h_1^{27}$ in their fundamental
representations, respectively. On the other hand, if the Cartan generators reconstructed by the group decomposition to the maximal subgroups have the same elements as $h_1$, they are able to simulate the exact static potentials as the exceptional super-groups, themselves. Thus, one is able to apply these subgroup decompositions to gain the static potential of the higher representations of the exceptional groups. Hence, Casimir scaling of different representations of these groups is observed. This method is not applicable for the potentials obtained by the
non-trivial center elements of $E_6$ i.e. the potentials calculated by the non-trivial center elements in the thick center vortex model are not identical with the potentials of their subgroups. Hence, it seems that this method is just valid when we use only unit center element of the gauge groups to calculate the
static potentials.

To find the reason of the temporary confinement at intermediate distances, we have turned to the center elements of the $SU(N)$ subgroups by which their center vortices indirectly produce the intermediate linear part in the super-groups. So, the group factor function $ \textrm{Re} \, \mathcal{G}_r \left[
\vec{\alpha}(x) \right] $ has been investigated in different representations of $G_2$, $F_4$ and $E_6$ exceptional gauge groups using the unit center element only. Comparison of this function with the corresponding one obtained from the non-trivial center elements of the $SU(N)$ subgroups, results that the center vortices of
the $ SU(3) $ subgroups in non of these exceptional groups could be responsible for the intermediate linear potential, since the group factor functions reach different minimum amounts. However, by means of the trivial center element of this subgroup, the same potential as the exceptional group itself is produced.

Any regular or singular decomposition to the $SU(2)$ subgroup which produces a Cartan generator with the same elements as $h_1$, gives rise to the linear intermediate parts in the potentials
of the super-groups. In fact, the extremums of $ \textrm{Re} \, \mathcal{G}_r \left[\vec{\alpha}(x) \right] $ which occur at the points where $50\%$ of the vacuum domain flux enter the Wilson loop, are responsible for the intermediate linear potential. When the center element obtained from these $SU(2)$ decompositions lies entirely inside the Wilson loop, the corresponding group factor reaches a value which equals to the extremum values of $ \textrm{Re} \, \mathcal{G}_r \left[\vec{\alpha}(x) \right] $ for the given exceptional group.

Furthermore, there are some subgroups such as $SU(2) \times SU(2)$ for $G_2$, $SU(2) \times G_2$ for $F_4$ and $G_2$ singular subgroup of $E_6$ which produce different potentials from their super-group. Yet, they are responsible for the temporary confinement in different
representations. In fact, if the number of center elements or center vortices in the matrix of center elements obtained from two different decompositions is the same, the corresponding group factors reach the same value when the vortex is located completely inside the Wilson loop. The dominant role of the $ SU(2) $ subgroup in observing the temporary confinement obtained by the unit center element, is not exclusive to the exceptional gauge groups. In the next work, we argue that this dominant role of the $SU(2)$ subgroup is seen for the trivial potentials of the $SU(N)$ gauge
groups as well. We should mention that, due to the over-simplicity of the model, the results which have been presented in this paper, seem to be restricted in the framework of vacuum domain structure and thick center vortex models. 

\appendix
\section{\label{appa}$F_4 \supset SU(3) \times SU(3)$}
\begin{figure}
\centering
\includegraphics[width=\linewidth]{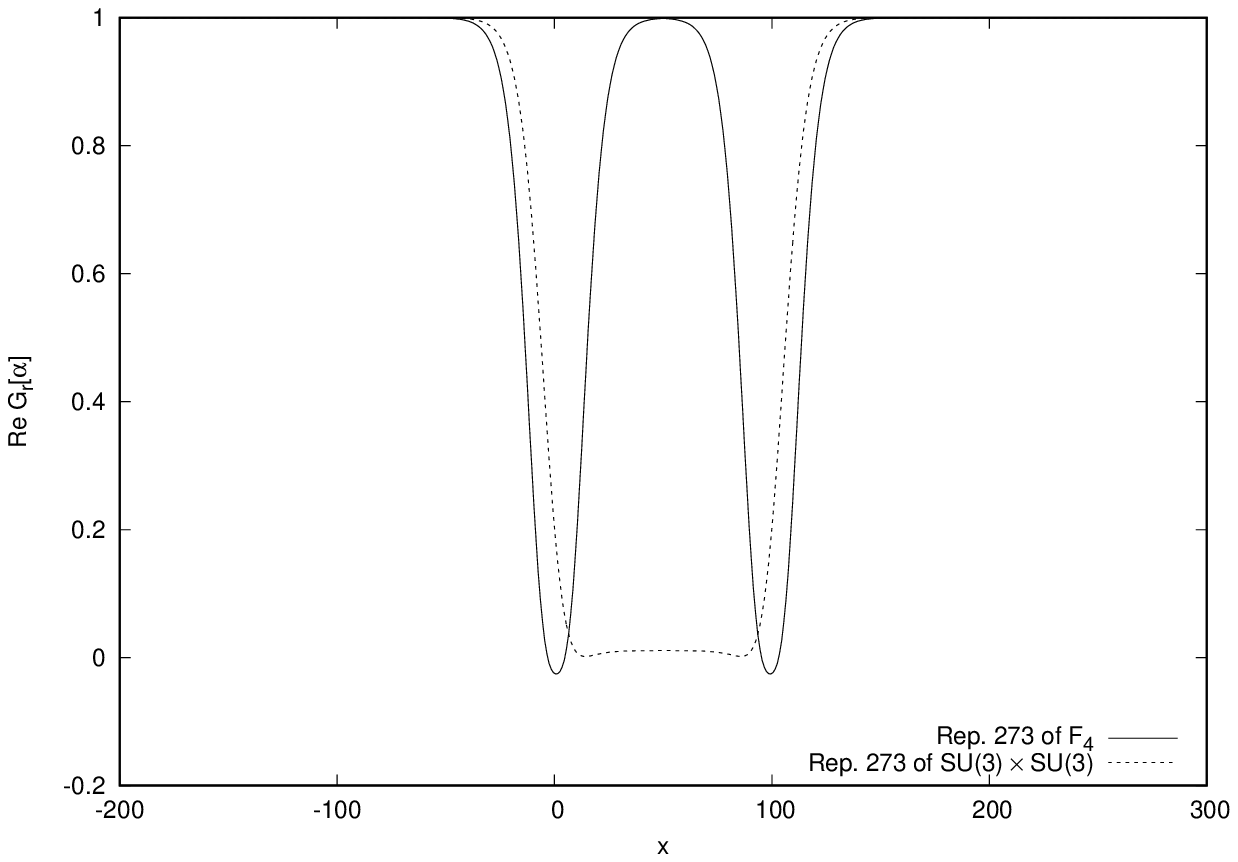}
\includegraphics[width=\linewidth]{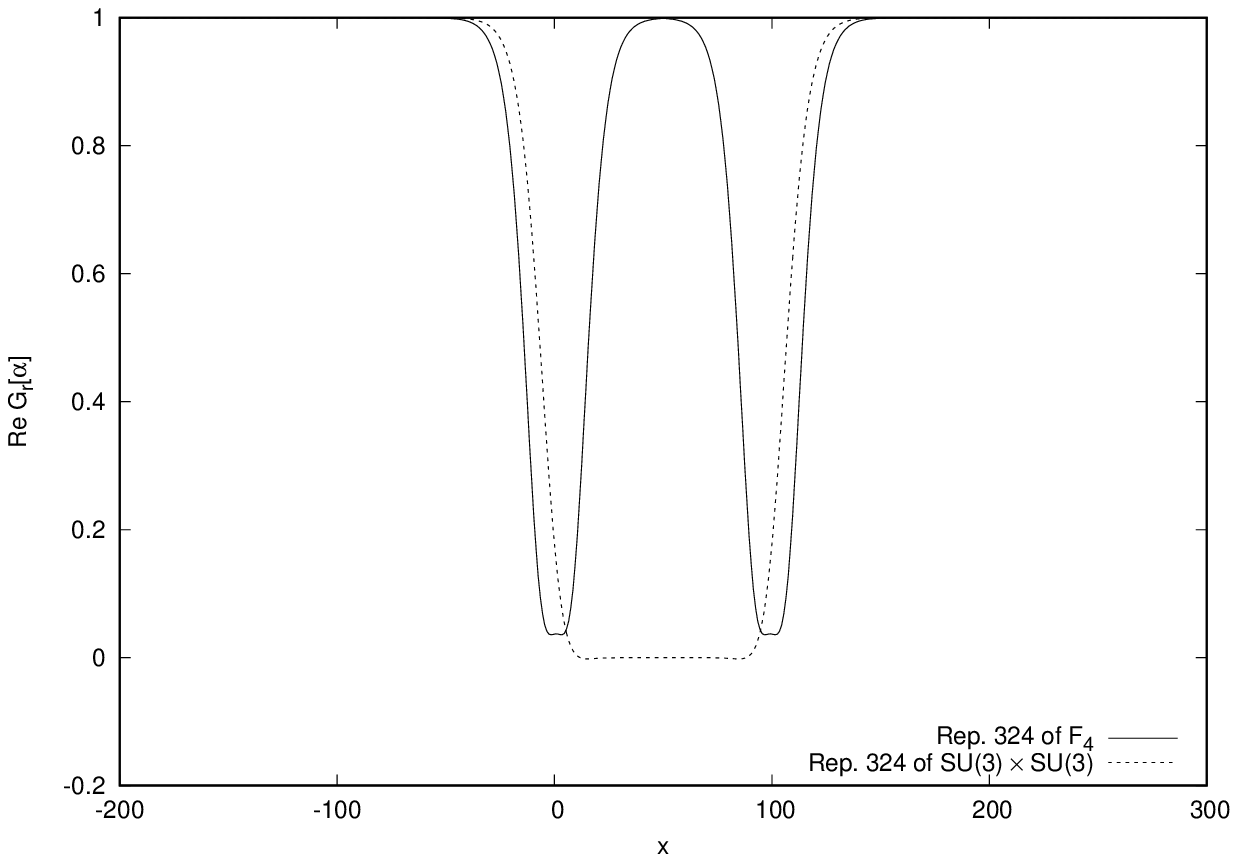}
\caption{The same as Fig.~\ref{fig:f4-su3} but for representations $273$ and $324$. The minimum values of the $F_4$ group factor for representations $273$ and $324$ are $-0.025$ and $0.037$, respectively. It is seen that the minimum values of the group factors for the $SU(3) \times SU(3)$ decomposition are not identical with the corresponding ones for the $F_4$.}
\label{fig:273-324}       
\end{figure}

The decompositions of $273$ and $324$-dimensional irreps of the $ F_4 $ to the irreps of the $ SU(3) \times SU(3) $ subgroup are as following \cite{slansky,sorba}
\begin{equation} \label{a1}
\begin{aligned}
273 =&(1 , 1) \oplus (8 , 1) \oplus (3 , 3) \oplus (\bar{3} , \bar{3}) \oplus  (10 , 1) \oplus (\overline{10} , 1) \oplus \\
    &(6 , \bar{3}) \oplus (\bar{6} , 3) \oplus  (3 , \bar{6})  \oplus  (\bar{3} , 6) \oplus (15 , 3) \oplus (\overline{15} , \bar{3}) \oplus \\ &(8 , 8),
\end{aligned}
\end{equation}
\begin{equation} \label{a2}
\begin{aligned}324 =&(1 , 1) \oplus (8 , 1) \oplus (1 , 8) \oplus (\bar{3} , \bar{3}) \oplus 
 (3 , 3) \oplus (6 , \bar{3}) \oplus \\
&(\bar{6} , 3) \oplus  (27 , 1) \oplus (\bar{6} , \bar{6})
\oplus  (6 , 6) \oplus (15 , 3) \oplus (\overline{15} , \bar{3}) \oplus \\&(8 , 8).
\end{aligned}
\end{equation}
Thus, Cartan diagonal generators reconstructed by taking advantage of Eqs.~\eqref{a1} and \eqref{a2}are
\begin{equation}  \label{a3}
\begin{aligned}
& H_a^{273}= \frac{1}{3 \sqrt{14}} \textrm{diag} \Big[ 0 , \overbrace{0 , \cdots , 0}^{8 \, \textrm{times}}, 
  \lambda_a^3 ,  \lambda_a^3 ,  \lambda_a^3  , \\
& -(\lambda_a^3)^{\ast} ,   -(\lambda_a^3)^{\ast} ,  -(\lambda_a^3)^{\ast} , \overbrace{0 , \cdots , 0}^{10 \, \textrm{times}}, 
\overbrace{0 , \cdots , 0}^{10 \, \textrm{times}} ,   \\
& \overbrace{-(\lambda_a^3)^{\ast}, \cdots , -(\lambda_a^3)^{\ast}}^{6 \, \textrm{times}}, ,  \overbrace{\lambda_a^3, \cdots , \lambda_a^3}^{6 \, \textrm{times}},
 -(\lambda_a^6)^{\ast} ,  -(\lambda_a^6)^{\ast},  -(\lambda_a^6)^{\ast}, \\
& \lambda_a^6 ,  \lambda_a^6 , \lambda_a^6 ,  \overbrace{\lambda_a^3, \cdots , \lambda_a^3}^{15 \, \textrm{times}}, ,   \overbrace{-(\lambda_a^3)^{\ast}, \cdots , -(\lambda_a^3)^{\ast}}^{15 \, \textrm{times}},
 \overbrace{\lambda_a^8 , \cdots , \lambda_a^8}^{8 \, \textrm{times}}   \Big],\\
\end{aligned}
\end{equation}
\begin{equation}  
\begin{aligned}\label{a4}
&H_{a}^{324} = \frac{1}{9 \sqrt{2}} \,  \textrm{diag} \Big[ 0 ,  \overbrace{0 , \cdots , 0}^{8 \, \textrm{times}} , \lambda_{a}^8 , 
 \overbrace{\lambda_a^3 , \cdots , \lambda_a^3}^{3 \, \textrm{times}} , \\
&\overbrace{-(\lambda_a^3)^{\ast} , \cdots , -(\lambda_a^3)^{\ast}}^{3 \, \textrm{times}} ,
 \overbrace{-(\lambda_a^3)^{\ast} , \cdots , -(\lambda_a^3)^{\ast}}^{6 \, \textrm{times}} ,  \overbrace{\lambda_a^3 , \cdots , \lambda_a^3}^{6 \, \textrm{times}} , \\
&\overbrace{0 , \cdots , 0}^{27 \, \textrm{times}} , \overbrace{-(\lambda_a^6)^{\ast} , \cdots , -(\lambda_a^6)^{\ast}}^{6 \, \textrm{times}} ,
 \overbrace{\lambda_a^6 , \cdots , \lambda_a^6}^{6 \, \textrm{times}} ,  \overbrace{\lambda_a^3 , \cdots , \lambda_a^3}^{15 \, \textrm{times}} , \\
& \overbrace{-(\lambda_a^3)^{\ast} , \cdots , -(\lambda_a^3)^{\ast}}^{15 \, \textrm{times}} , \overbrace{\lambda_a^8 , \cdots , \lambda_a^8}^{8 \, \textrm{times}} \Big].
\end{aligned}
\end{equation}
Using the trivial flux condition in Eq.~\eqref{e19}, the maximum flux values are calculated as the following:
\begin{equation} \label{a5}
\begin{aligned}
& \alpha_{max_1}^{273}= 6 \pi \sqrt{14}, \\
& \alpha_{max_2}^{273}= 6 \pi \sqrt{42}, \\
\\
& \alpha_{max_1}^{324}= 18 \pi \sqrt{2} ,\\
& \alpha_{max_2}^{324}= 18 \pi \sqrt{6}. \\
\end{aligned}
\end{equation}
the static potential calculated by means of the above equations, has been given in Fig.~\ref{fig-8}. We can built up the center element matrices:
\begin{equation}\label{a6}
\begin{aligned}
& \mathbb{Z}_{SU(3)}^{273}= \textrm{diag} \Big[ 1 , \overbrace{1 , \cdots , 1}^{8 \, \textrm{times}} , \overbrace{z_n \, \mathbb{I}_{3 \times 3} , \cdots , z_n \, \mathbb{I}_{3 \times 3}}^{3 \, \textrm{times}} ,  \\
& \overbrace{z_n^{\ast} \, \mathbb{I}_{3 \times 3} , \cdots , z_n^{\ast} \, \mathbb{I}_{3 \times 3}}^{3 \, \textrm{times}} , \overbrace{1 , \cdots , 1}^{10 \, \textrm{times}} ,  \overbrace{1 , \cdots , 1}^{10 \, \textrm{times}} , \\
& \overbrace{z_n \, \mathbb{I}_{3 \times 3} , \cdots ,z_n \, \mathbb{I}_{3 \times 3}}^{6 \, \textrm{times}} ,  \overbrace{z_n^{\ast} \, \mathbb{I}_{3 \times 3} , \cdots , z_n^{\ast} \, \mathbb{I}_{3 \times 3}}^{6 \, \textrm{times}}  , \\
&  \overbrace{z_n^{\ast} \, \mathbb{I}_{6 \times 6} , \cdots , z_n^{\ast} \, \mathbb{I}_{6 \times 6}}^{3 \, \textrm{times}}  ,   
 \overbrace{z_n \, \mathbb{I}_{6 \times 6} , \cdots , z_n \, \mathbb{I}_{6 \times 6}}^{3 \, \textrm{times}} , \\
&\overbrace{z_n \, \mathbb{I}_{3 \times 3} , \cdots ,  z_n \, \mathbb{I}_{3 \times 3}}^{15 \, \textrm{times}}  ,  
 \overbrace{z_n^{\ast} \, \mathbb{I}_{3 \times 3} , \cdots , z_n^{\ast} \, \mathbb{I}_{3 \times 3}}^{15 \, \textrm{times}} , \\
& \overbrace{\mathbb{I}_{8 \times 8} , \cdots , \mathbb{I}_{8 \times 8}}^{8 \, \textrm{times}}  \Big],\\
\end{aligned}
\end{equation}
\begin{equation}\label{a7}
\begin{aligned}
& \mathbb{Z}_{SU(3)}^{324}= \textrm{diag} \Big[ 1 , \overbrace{1 , \cdots , 1}^{8 \, \textrm{times}} , \mathbb{I}_{8 \times 8} \\
& \overbrace{z_n^{\ast} \, \mathbb{I}_{3 \times 3} , \cdots , z_n^{\ast} \, \mathbb{I}_{3 \times 3}}^{3 \, \textrm{times}} ,   \overbrace{z_n \, \mathbb{I}_{3 \times 3} , \cdots , z_n \, \mathbb{I}_{3 \times 3}}^{3 \, \textrm{times}} ,   \\
& \overbrace{z_n^{\ast} \, \mathbb{I}_{3 \times 3} , \cdots , z_n^{\ast} \, \mathbb{I}_{3 \times 3}}^{6 \, \textrm{times}} ,  \overbrace{z_n \, \mathbb{I}_{3 \times 3} , \cdots , z_n \, \mathbb{I}_{3 \times 3}}^{6 \, \textrm{times}}  ,  \\
& \overbrace{1 , \cdots , 1}^{27 \, \textrm{times}} ,  \overbrace{z_n^{\ast} \, \mathbb{I}_{6 \times 6} , \cdots , z_n^{\ast} \, \mathbb{I}_{6 \times 6}}^{6 \, \textrm{times}}  ,   \\
& \overbrace{z_n \, \mathbb{I}_{6 \times 6} , \cdots , z_n \, \mathbb{I}_{6 \times 6}}^{6 \, \textrm{times}} , \overbrace{z_n \, \mathbb{I}_{3 \times 3} , \cdots , z_n \, \mathbb{I}_{3 \times 3}}^{15 \, \textrm{times}}  ,  \\
& \overbrace{z_n^{\ast} \, \mathbb{I}_{3 \times 3} , \cdots , z_n^{\ast} \, \mathbb{I}_{3 \times 3}}^{15 \, \textrm{times}} ,  \overbrace{\mathbb{I}_{8 \times 8} , \cdots , \mathbb{I}_{8 \times 8}}^{8 \, \textrm{times}}  \Big].
\end{aligned}
\end{equation}
Then, we use non-trivial flux profile condition of Eq.~\eqref{e6} to estimate the maximum flux values for these representations:
\begin{equation}\label{a8}
\begin{aligned}
& \alpha_{max_1}^{273-\textrm{non}}= 6 \pi \sqrt{14} \\
& \alpha_{max_2}^{273-\textrm{non}}= 2 \pi \sqrt{42} \\
\\
& \alpha_{max_1}^{324-\textrm{non}}= 18 \pi \sqrt{2} \\
& \alpha_{max_2}^{324-\textrm{non}}= 6 \pi \sqrt{6} \\
\end{aligned}
\end{equation}
Accordingly, using Eqs.~\eqref{a2}-\eqref{a4} and \eqref{a8}, the group factor functions could be plotted in Fig.~\ref{fig:273-324}.

\section{\label{appb}$F_4 \supset SO(9)\supset SU(2) \times SU(4)$}
\begin{equation} \label{b1}
\begin{aligned}
F_4 &\supset SO(9) \\
273 &=9 \oplus 16 \oplus 36 \oplus 84 \oplus 128, \\
324 &=1 \oplus 9 \oplus 16 \oplus 44 \oplus 126 \oplus 128,\\
\end{aligned}
\end{equation}
In the next step,
\begin{equation}\label{b2}
\begin{aligned}
SO(9) &\supset SU(2) \times SU(4)\\
9 &=(3 , 1) \oplus (1 , 6),\\
16 &=(2 , 4) \oplus (2 , \bar{4}),\\
36 &=(3 , 1) \oplus (1 , 15) \oplus (3 , 6),\\
44 &=(1 , 1) \oplus (5 , 1) \oplus (3 , 6) \oplus (1 , 20^{\prime}),\\
84 &=(1 , 1) \oplus (1 , 10) \oplus (1 , \overline{10}) \oplus  (3 , 6) \oplus (3 , 15),\\
126 &=(1 , 6) \oplus (3 , 10) \oplus (3 , \overline{10}) \oplus   (1 , 15) \oplus (3 , 15),\\
128 &=(2 , 4) \oplus (2 , \bar{4}) \oplus (4 , 4) \, \oplus (4 , \bar{4}) \oplus  (2 , 20) \oplus (2 , \overline{20}),  \\
\end{aligned}
\end{equation}
Now, we decompose $ SU(4) $ into its $ SU(2) $ subgroup:
\begin{equation}\label{b3}
\begin{aligned}
SU(4) &\supset SU(2) \times SU(2) \times U(1),\\   
4 &=(2 , 1) \oplus (1 , 2),\\
6 &=(1 , 1) \oplus (1 , 1) \oplus (2 , 2),\\
10 &=(2 , 2) \oplus (3 , 1) \oplus (1 , 3),\\ 
15 &=(1 , 1) \oplus (2 , 2) \oplus (2 , 2) \oplus (3 , 1) \oplus  (1 , 3),\\
20 &=(2 , 1) \oplus (2 , 1) \oplus (1 , 2) \oplus (1 , 2) \oplus  (3 , 2) \oplus (2 , 3),\\
20^{\prime} &=(1 , 1) \oplus (1 , 1) \oplus (1 , 1) \oplus (2 , 2) \oplus   (2 , 2) \oplus (3 , 3),\\
\end{aligned}
\end{equation}
Ultimately we have,
\begin{equation}\label{b4}
\begin{aligned}
& 273=\overbrace{(1 , 1) \oplus \cdots \oplus (1 , 1)}^{19 \, \textrm{times}} \oplus \overbrace{(3 , 1) \oplus \cdots \oplus (3 , 1)}^{8 \, \textrm{times}} \oplus  \\
& \overbrace{(2 , 2) \oplus \cdots \oplus (2 , 2)}^{17 \, \textrm{times}} \oplus  \overbrace{(2 , 1) \oplus \cdots \oplus (2 , 1)}^{24 \, \textrm{times}} \oplus    \\
& \overbrace{(1 , 2) \oplus \cdots \oplus (1 , 2)}^{24 \, \textrm{times}} \oplus \overbrace{(1 , 3) \oplus \cdots \oplus (1 , 3)}^{6 \, \textrm{times}} \oplus   \\
& \overbrace{(2 , 3) \oplus \cdots \oplus (2 , 3)}^{4 \, \textrm{times}} \oplus \overbrace{(3 , 2) \oplus \cdots \oplus (3 , 2)}^{4 \, \textrm{times}}   \\
\end{aligned}
\end{equation}
\begin{equation}\label{b5}
\begin{aligned}
& 324=1 \oplus \overbrace{(1 , 1) \oplus \cdots \oplus (1 , 1)}^{18 \, \textrm{times}} \oplus  \\
& \overbrace{(3 , 1) \oplus \cdots \oplus (3 , 1)}^{11 \, \textrm{times}} \oplus  \overbrace{(2 , 2) \oplus \cdots \oplus (2 , 2)}^{21 \, \textrm{times}} \oplus  \\
& \overbrace{(2 , 1) \oplus \cdots \oplus (2 , 1)}^{24 \, \textrm{times}} \oplus   \overbrace{(1 , 2) \oplus \cdots \oplus (1 , 2)}^{24 \, \textrm{times}} \oplus  \\
& \overbrace{(1 , 3) \oplus \cdots \oplus (1 , 3)}^{10 \, \textrm{times}} \oplus  \overbrace{(2 , 3) \oplus \cdots \oplus (2 , 3)}^{4 \, \textrm{times}} \oplus  \\
& \overbrace{(3 , 2) \oplus \cdots \oplus (3 , 2)}^{4 \, \textrm{times}} \oplus  (3 , 3) \oplus (5 , 1)  \\
\end{aligned}
\end{equation}
and,
\begin{equation}\label{b6}
\begin{aligned}
H_{SU(2)}^{273}&=\frac{1}{3 \sqrt{14}} \, \textrm{diag} \Big[ \overbrace{0 , \cdots , 0}^{91 \, \textrm{times}} , \overbrace{\sigma_3^2 , \cdots , \sigma_3^2}^{70 \, \textrm{times}} ,  \overbrace{\sigma_3^3 , \cdots , \sigma_3^3}^{14 \, \textrm{times}} \Big],\\
H_{SU(2)}^{324}&=\frac{1}{9 \sqrt{2}} \, \textrm{diag} \Big[ \overbrace{0 , \cdots , 0}^{105 \, \textrm{times}} ,  \overbrace{\sigma_3^2 , \cdots , \sigma_3^2}^{78 \, \textrm{times}} ,  \overbrace{\sigma_3^3 , \cdots , \sigma_3^3}^{21 \, \textrm{times}} \Big].
\end{aligned}
\end{equation}

\begin{figure}
\centering
\includegraphics[width=\linewidth]{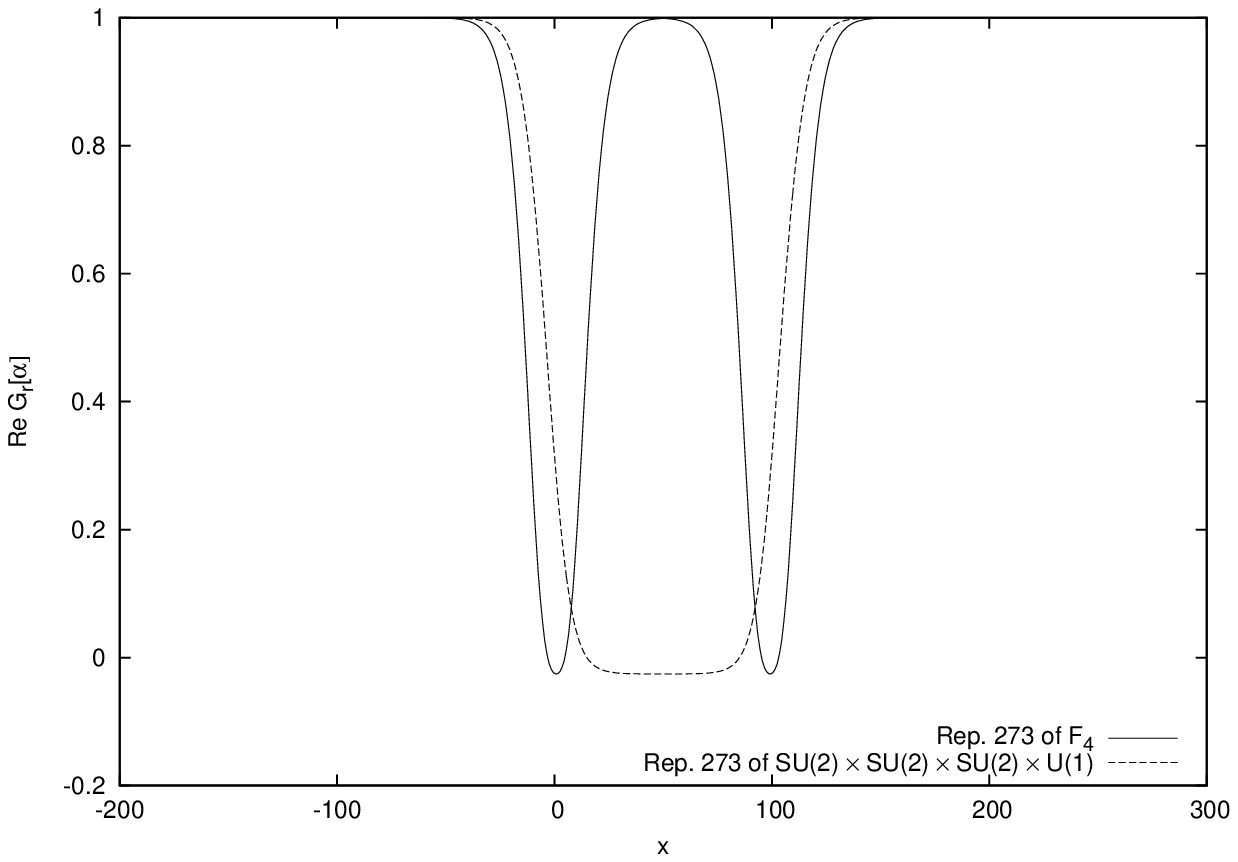}
\includegraphics[width=\linewidth]{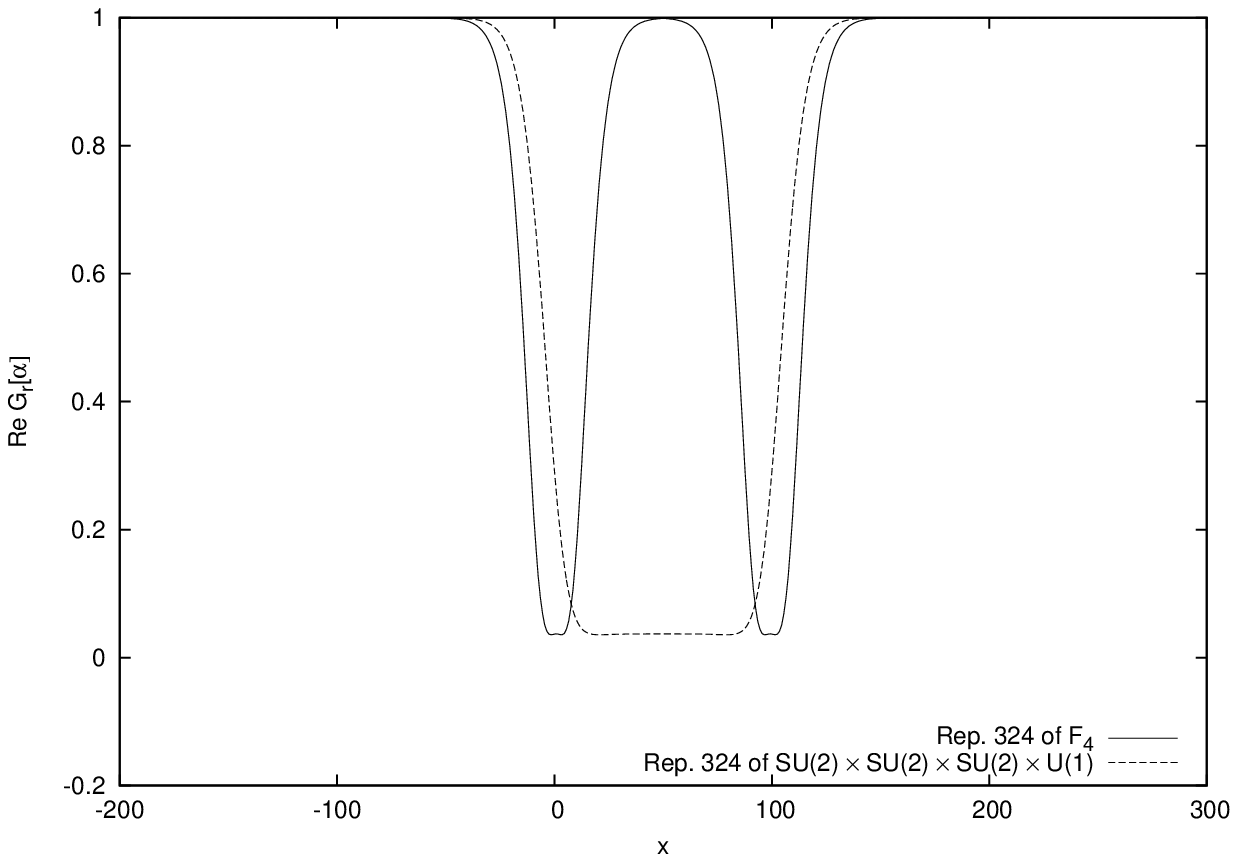}
\caption{The real part of the group factor versus the location $x$ of the vacuum domain midpoint, for $R=100$ and in the range $x \in [-200,300]$, for representations $273$ and $324$ of the $F_4$ (solid lines) in comparison with the one obtained from $ SO(9)\supset SU(2) \times SU(4) $ decomposition using its non-trivial center elements (dashed lines). In each diagram, the minimum values are identical. }
\label{fig:273-324-su2}       
\end{figure}

The matrices made of the center element of $SU(2)$ gauge group corresponding to the duality of its representation with respect to Eqs.~\eqref{b4} and \eqref{b5} are as the following:
\begin{equation}\label{b8}
\begin{aligned}
 \mathbb{Z}_{SU(2)}^{273}= \textrm{diag} \Big[&  \overbrace{1 , \cdots , 1}^{91 \, \textrm{times}} ,  
 \overbrace{z_1 \, \mathbb{I}_{2 \times 2}  , \cdots , z_1 \, \mathbb{I}_{2 \times 2}}^{70 \, \textrm{times}} , \\ &\overbrace{\mathbb{I}_{3 \times 3} , \cdots , \mathbb{I}_{3 \times 3}}^{14 \, \textrm{times}}  \Big],\\
 \mathbb{Z}_{SU(2)}^{324} = \textrm{diag} \Big[& \overbrace{1 , \cdots , 1}^{105 \, \textrm{times}} ,  
 \overbrace{z_1 \, \mathbb{I}_{2 \times 2}  , \cdots , z_1 \, \mathbb{I}_{2 \times 2}}^{78 \, \textrm{times}} , \\ &\overbrace{\mathbb{I}_{3 \times 3} , \cdots , \mathbb{I}_{3 \times 3}}^{21 \, \textrm{times}}  \Big].
\end{aligned}
\end{equation}
The maximum flux values could be calculated from the non-trivial flux condition of Eq.~\ref{e6}:
\begin{equation}\label{b9}
\begin{aligned}
\alpha_{SU(2) \textrm{-max}}^{273 \textrm{-non}}&=6 \pi \sqrt{14},\\
\alpha_{SU(2) \textrm{-max}}^{324 \textrm{-non}}&=18 \pi \sqrt{2}.
\end{aligned}
\end{equation}
The group factor functions of these representations have been given in Fig.~\ref{fig:273-324-su2}. 

\section{\label{appc}$ G_2 \supset SU(2) \times SU(2)$}
\begin{figure}
\centering
\includegraphics[width=\linewidth]{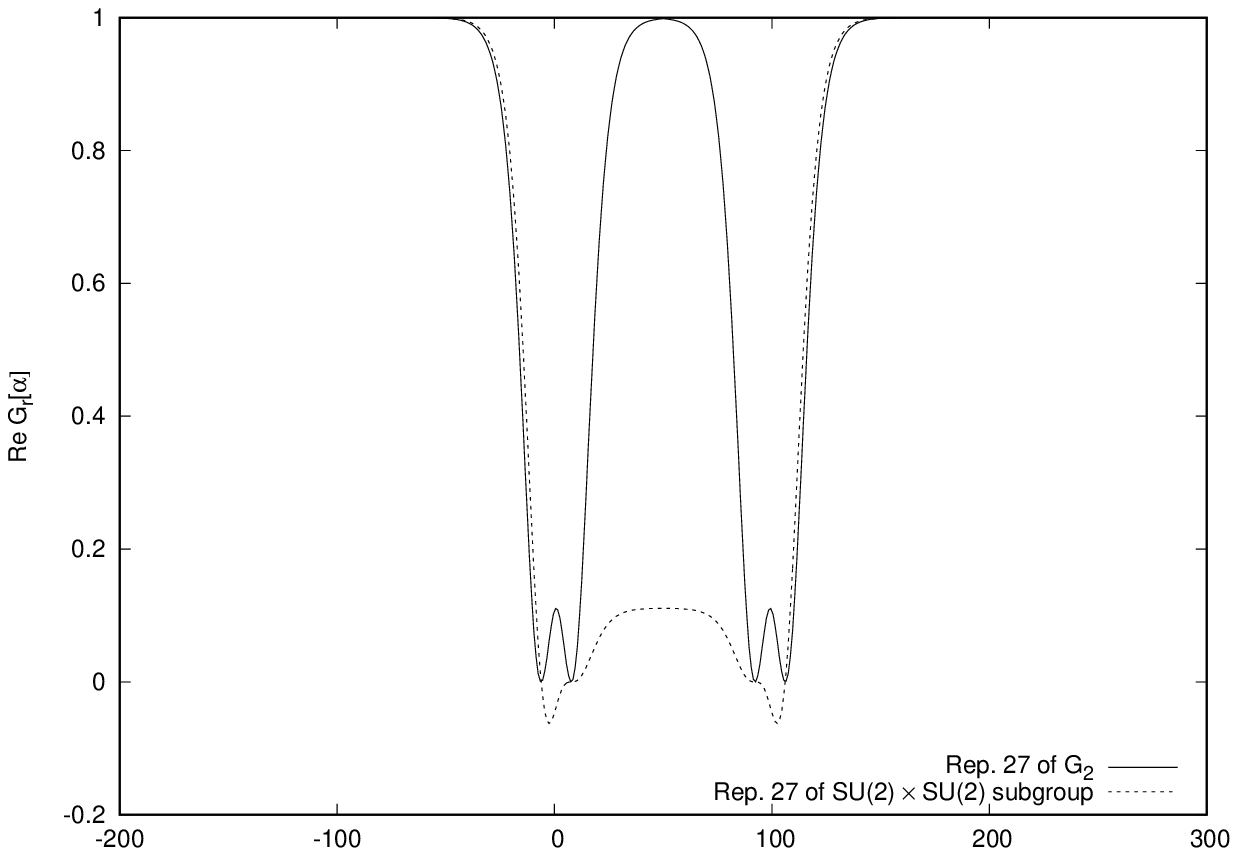}
\caption{The same as Fig.~\ref{fig-26} but for representation $27$. The extremum values of the $F_4$ group factor at $x=0$ and $x=100$ approximately equal to $0.111$.}
\label{27-g2}       
\end{figure}

\begin{figure}
\centering
\includegraphics[width=\linewidth]{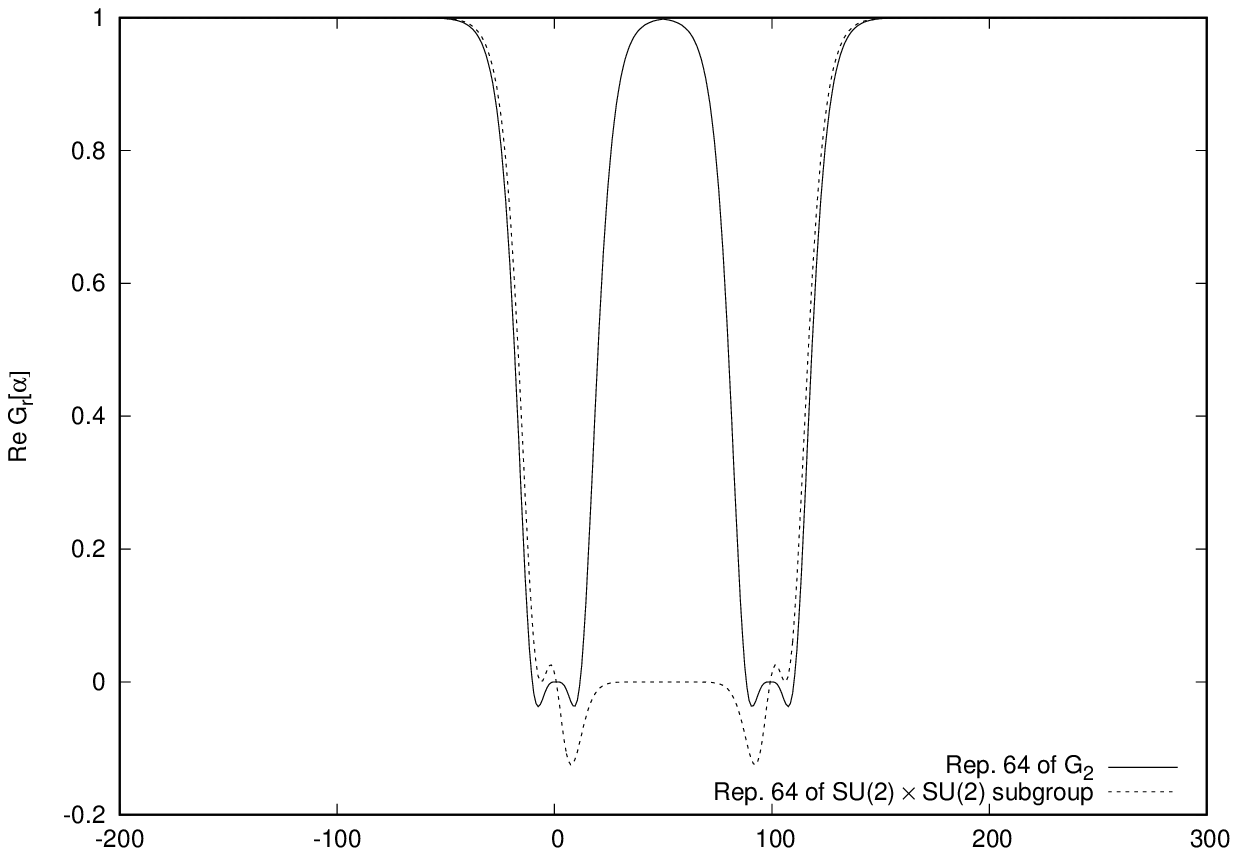}
\caption{The same as Fig.~\ref{fig-26} but for representation $64$. The extremum values of the $F_4$ group factor at $x=0$ and $x=100$ approximately equal to $0$.}
\label{64-g2}       
\end{figure}

\begin{figure}
\centering
\includegraphics[width=\linewidth]{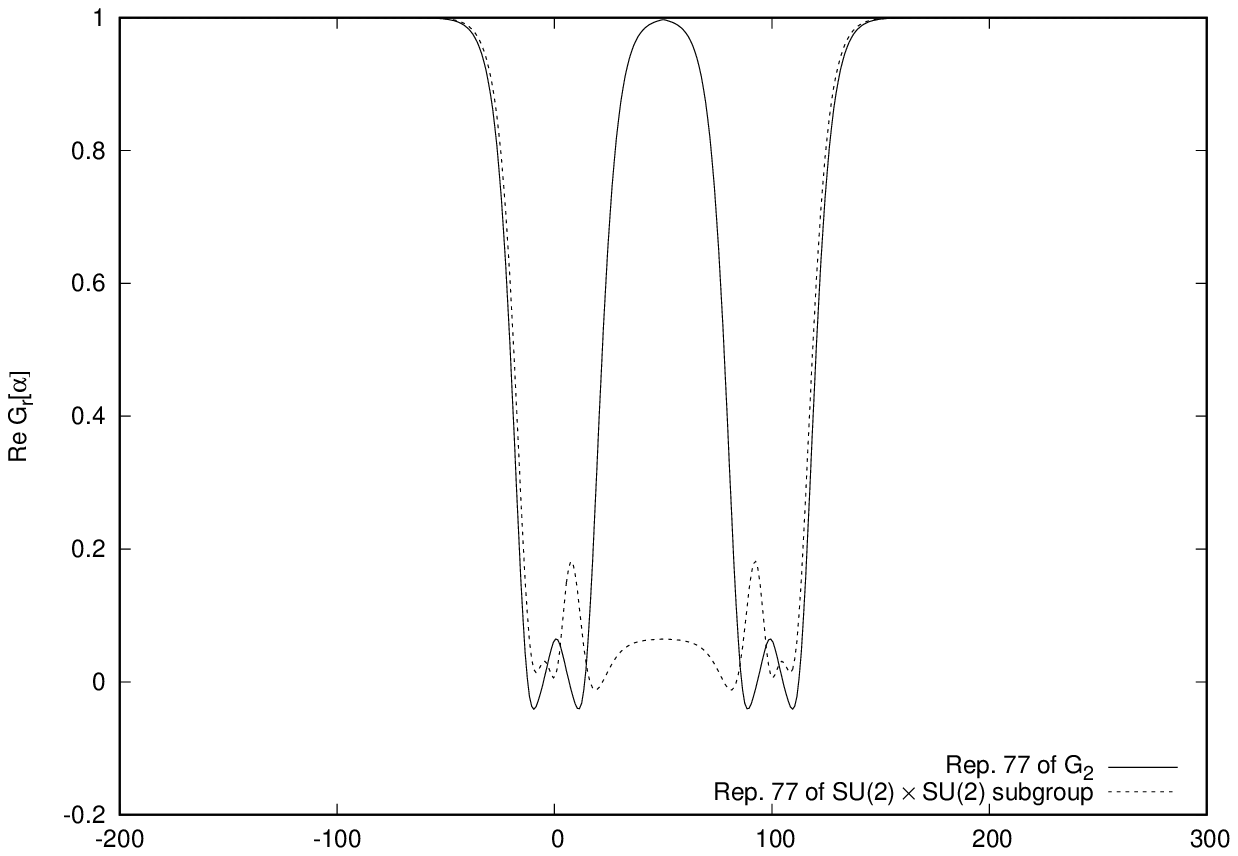}
\caption{The same as Fig.~\ref{fig-26} but for representation $77$. The extremum values of the $F_4$ group factor at $x=0$ and $x=100$ approximately equal to $0.064$.}
\label{77-g2}       
\end{figure}

\begin{figure}
\centering
\includegraphics[width=\linewidth]{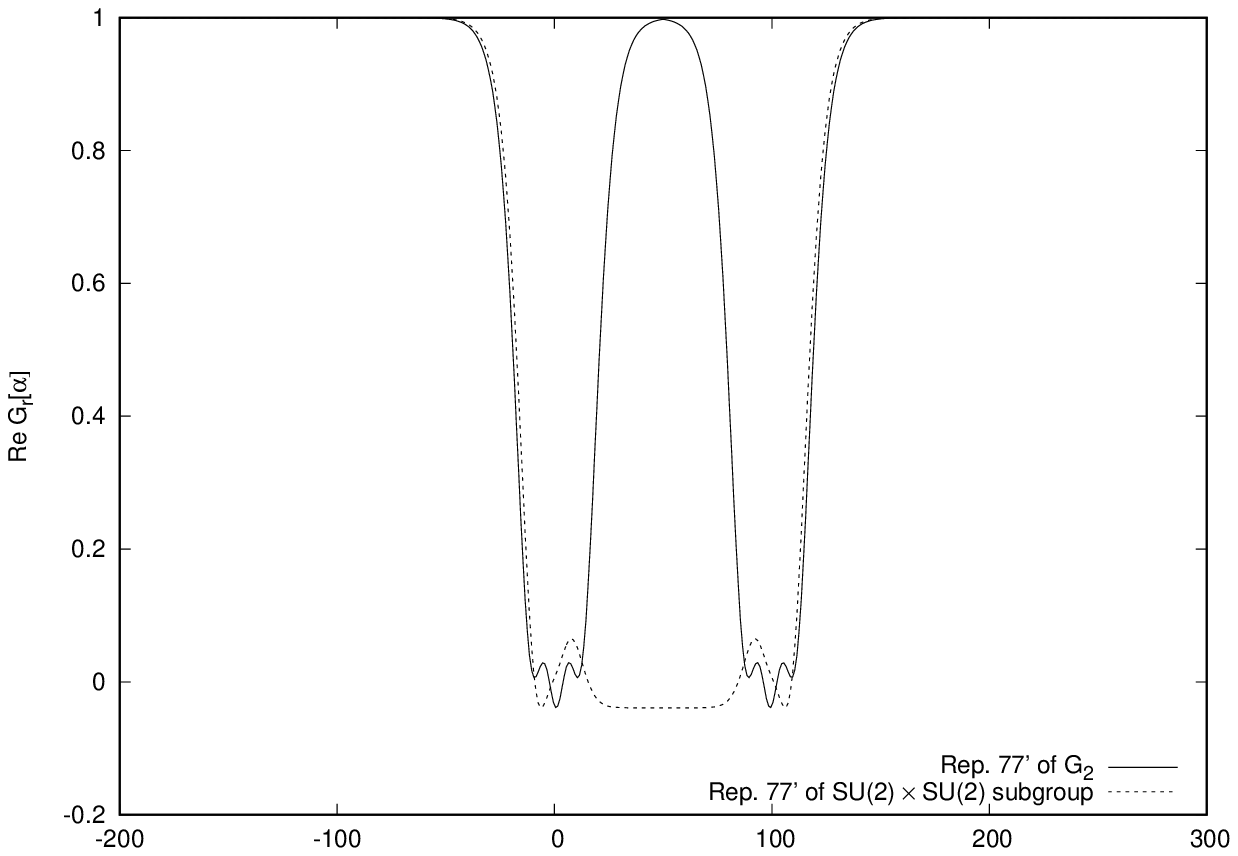}
\caption{The same as Fig.~\ref{fig-26} but for representation $77^\prime$. The extremum values of the $F_4$ group factor at $x=0$ and $x=100$ approximately equal to $-0.038$.}
\label{77p-g2}       
\end{figure}

Decompositions of the $ G_2 $ representations into  $ SU(2) \times SU(2) $ regular subgroup representations are
\begin{equation}\label{c1}
\begin{aligned}
27 = &(3 , 3) \oplus (2 , 4) \oplus (2 , 2) \oplus (1 , 5) \oplus (1 , 1),\\
64 = &(4 , 2)\oplus(3 , 5)\oplus(3 , 3)\oplus(2 , 6)\oplus(2 , 4)\oplus(2 , 2)\oplus\\
     &(1 , 5)\oplus(1 , 3),\\
77 = &(5 , 1)\oplus(4 , 4)\oplus(3 , 7)\oplus(3 , 3)\oplus(2 , 6)\oplus \\
     &(2 , 4)\oplus(1 , 5)\oplus(1 , 1),\\
77^\prime = &(4 , 4)\oplus(3 , 5)\oplus(3 , 3)\oplus(3 , 1)\oplus(2 , 6)\oplus(2 , 4)\oplus\\
     &(2 , 2)\oplus(1 , 7)\oplus(1 , 3).
\end{aligned}
\end{equation}
The Cartan generators could be reconstructed as the following:
\begin{equation}\label{c2}
\begin{aligned}
 H_{SU(2)}^{27} =\frac{1}{3 \sqrt{6}} \, \textrm{diag} \Big[& \sigma_3^3 , \sigma_3^3 , \sigma_3^3  ,  \sigma_3^4  ,   
 \sigma_3^4  ,  \sigma_3^2  ,  \sigma_3^2  , \sigma_3^5 , 0 \Big],\nonumber\\
\end{aligned}
\end{equation}
\begin{equation}
\begin{aligned}
 H_{SU(2)}^{64} =\frac{1}{8 \sqrt{3}}\, \textrm{diag} \Big[& \sigma_3^2 ,  \sigma_3^2  ,  \sigma_3^2  ,  \sigma_3^2  ,  
 \sigma_3^5  ,  \sigma_3^5 ,  \sigma_3^5  , \sigma_3^3 , \sigma_3^3 , \sigma_3^3  ,\nonumber\\
\end{aligned}
\end{equation}
\begin{equation}
\begin{aligned}
& \sigma_3^6 , \sigma_3^6 , \sigma_3^4 , \sigma_3^4  ,  \sigma_3^2  ,  \sigma_3^2  , \sigma_3^5  , \sigma_3^3  \Big],\\
 H_{SU(2)}^{77}=\frac{1}{\sqrt{330}} \, \textrm{diag} \Big[& 0 , 0 , 0 , 0 , 0 , \sigma_3^4  ,    \sigma_3^4  , \sigma_3^4  , \sigma_3^4   ,  \sigma_3^7  , \sigma_3^7  , \sigma_3^7  ,\nonumber \\
\end{aligned}
\end{equation}
\begin{equation}
\begin{aligned}
& \sigma_3^3  , \sigma_3^3  , \sigma_3^3  , \sigma_3^6  ,  \sigma_3^6  ,  \sigma_3^4  ,  \sigma_3^4  ,   \sigma_3^5  , 0  \Big],\\
 H_{SU(2)}^{{77}^{\prime}}=\frac{1}{2 \sqrt{66}} \, \textrm{diag} \Big[& \sigma_3^4  ,  \sigma_3^4  , \sigma_3^4  , \sigma_3^4  ,    \sigma_3^5  , \sigma_3^5  , \sigma_3^5  ,  \sigma_3^3  , \sigma_3^3  , \sigma_3^3  , \\
&0  , 0  , 0  , \sigma_3^6  , \sigma_3^6  ,  \sigma_3^4  ,  \sigma_3^4  ,  \sigma_3^2  ,   \sigma_3^2  , \sigma_3^7  , \sigma_3^3   \Big].
\end{aligned}
\label{c3}
\end{equation}
The center element matrices of the $SU(2) \times SU(2)$ subgroup are:
\begin{equation}
\begin{aligned}
\mathbb{Z}_{SU(2)}^{27}= \textrm{diag} \Big[& \mathbb{I}_{3 \times 3} , \mathbb{I}_{3 \times 3} , \mathbb{I}_{3 \times 3} ,   z_1 \, \mathbb{I}_{4 \times 4} ,  z_1 \, \mathbb{I}_{4 \times 4}  , z_1 \, \mathbb{I}_{2 \times 2}  ,  \\
& z_1 \, \mathbb{I}_{2 \times 2} , \mathbb{I}_{5 \times 5} , 1 \Big],\\
 \mathbb{Z}_{SU(2)}^{64}=\textrm{diag} \Big[& z_1 \, \mathbb{I}_{2 \times 2} , z_1 \, \mathbb{I}_{2 \times 2} , z_1 \, \mathbb{I}_{2 \times 2}  ,  
 z_1 \, \mathbb{I}_{2 \times 2} ,  \mathbb{I}_{5 \times 5} , \\
&\mathbb{I}_{5 \times 5} ,  \mathbb{I}_{5 \times 5} , \mathbb{I}_{3 \times 3} , \mathbb{I}_{3 \times 3} , \mathbb{I}_{3 \times 3} ,  
 z_1 \, \mathbb{I}_{6 \times 6} , z_1 \, \mathbb{I}_{6 \times 6} ,\\
& z_1 \, \mathbb{I}_{4 \times 4} , z_1 \, \mathbb{I}_{4 \times 4} ,  
 z_1 \, \mathbb{I}_{2 \times 2} , z_1 \, \mathbb{I}_{2 \times 2} ,  \mathbb{I}_{5 \times 5} , \mathbb{I}_{3 \times 3} \Big],\\
 \mathbb{Z}_{SU(2)}^{77}=\textrm{diag} \Big[& 1 , 1 , 1 , 1 , 1  ,  z_1 \, \mathbb{I}_{4 \times 4}  ,   
 z_1 \, \mathbb{I}_{4 \times 4} , z_1 \, \mathbb{I}_{4 \times 4}  ,  z_1 \, \mathbb{I}_{4 \times 4} ,  \\
& \mathbb{I}_{7 \times 7} , \mathbb{I}_{7 \times 7}  ,  
 \mathbb{I}_{7 \times 7}  ,   \mathbb{I}_{3 \times 3} , \mathbb{I}_{3 \times 3} , \mathbb{I}_{3 \times 3} ,  z_1 \, \mathbb{I}_{6 \times 6}  ,\\   
& z_1 \, \mathbb{I}_{6 \times 6} , z_1 \, \mathbb{I}_{4 \times 4} , z_1 \, \mathbb{I}_{4 \times 4}  ,  \mathbb{I}_{5 \times 5} , 1  \Big],\nonumber
\end{aligned}
\end{equation}
\begin{equation}\label{c4}
\begin{aligned}
 \mathbb{Z}_{SU(2)}^{{77}^{\prime}}=\textrm{diag} \Big[&  z_1 \, \mathbb{I}_{4 \times 4}  ,  z_1 \, \mathbb{I}_{4 \times 4} , z_1 \, \mathbb{I}_{4 \times 4}   ,  z_1 \, \mathbb{I}_{4 \times 4} ,  \mathbb{I}_{5 \times 5} , \mathbb{I}_{5 \times 5}  , \\
& \mathbb{I}_{5 \times 5}  ,   \mathbb{I}_{3 \times 3} , \mathbb{I}_{3 \times 3} ,   \mathbb{I}_{3 \times 3} , 1 , 1 , 1 ,  z_1 \, \mathbb{I}_{6 \times 6}  ,   z_1 \, \mathbb{I}_{6 \times 6}  , \\
& z_1 \, \mathbb{I}_{4 \times 4}  ,  
 z_1 \, \mathbb{I}_{4 \times 4}  ,  z_1 \, \mathbb{I}_{2 \times 2} ,  z_1 \, \mathbb{I}_{2 \times 2} , \mathbb{I}_{7 \times 7}  , \mathbb{I}_{3 \times 3}   \Big].
\end{aligned}
\end{equation}
Using non-trivial maximum flux condition of Eq.~\eqref{e6}, we find:
\begin{equation}\label{c5}
\begin{aligned}
\alpha_{SU(2) \textrm{-max}}^{27 \textrm{-non}}&=6 \pi \sqrt{6},\\
\alpha_{SU(2) \textrm{-max}}^{64 \textrm{-non}}&=16 \pi \sqrt{3},\\
\alpha_{SU(2) \textrm{-max}}^{77 \textrm{-non}}&=2 \pi \sqrt{330},\\
\alpha_{SU(2) \textrm{-max}}^{77^{\prime} \textrm{-non}}&=4 \pi \sqrt{66}.
\end{aligned}
\end{equation}
The group factor of representations $27$, $64$, $77$ and $77^\prime$ have been presented in Figs.~\ref{27-g2}-\ref{77p-g2}.

\bibliography{new}

\end{document}